\documentclass{elsart}

\usepackage{latexsym}
\usepackage{amssymb}
\usepackage{epsfig} 

\journal{Physics Reports}

\newcommand\eqn[1]{(\ref{#1})}      
\newcommand\Eqn[1]{Eq.~(\ref{#1})}  
\newcommand\cref[1]{Ref.~\cite{#1}}  
\newcommand\Fig[1]{Fig.~\ref{#1}}  
\newcommand\Sec[1]{Sec.~\ref{#1}}  
\newcommand{\beq}{\begin{equation}}
\newcommand{\eeq}{\end{equation}}
\newcommand{\bea}{\begin{eqnarray}}
\newcommand{\eea}{\end{eqnarray}}
\newcommand{\nn}{\nonumber\\}
\newcommand{\ie}{i.e.\ }
\newcommand{\eg}{e.g.\ }

\newcommand{\bk}{{\bf k}}
\newcommand{\bp}{{\bf p}}
\newcommand{\bq}{{\bf q}}

\newcommand{\bs}{{\bf s}}
\newcommand{\bx}{{\bf x}}
\newcommand{\by}{{\bf y}}

\newcommand{\vA}{{\vec A}}

\newcommand{\vJ}{{\vec J}}
\newcommand{\vK}{{\vec K}}
\newcommand{\vV}{{\vec V}}
\newcommand{\va}{{\vec a}}
\newcommand{\vb}{{\vec b}}
\newcommand{\vc}{{\vec c}}
\newcommand{\ve}{{\vec e}}

\newcommand{\vpi}{{\vec\pi}}

\newcommand{\vtau}{{\vec\tau}}

\renewcommand{\e}{ {\rm e} }

\newcommand{\p}{\partial}
\newcommand{\bra}{\langle}
\newcommand{\ket}{\rangle}

\newcommand{\T}{{_T}}
\newcommand{\B}{{_{\rm B}}}
\newcommand{\Tr}{{\rm Tr}}

\newcommand{\fpi}{f_\pi}
\newcommand{\mpi}{m_\pi}
\newcommand{\msi}{m_\sigma}
\newcommand{\dk}{\frac{d^3k}{(2\pi)^3}}

\newcommand{\tdp}{\tilde{dp}}
\newcommand{\tdq}{\tilde{dq}}
\newcommand{\tdr}{\tilde{dr}}
\newcommand{\tpi}{\tilde{\pi}}

\newcommand{\dvphi}{{\delta\varphi}}
\newcommand{\tomega}{\tilde{\omega}}
\newcommand{\dd}{{\rm d}}
\newcommand{\cY}{{\mathcal Y}}
\newcommand{\D}{{\mathcal D}}
\newcommand{\cl}{{\rm cl}}
\newcommand{\obs}{{\rm obs}}

\newcommand{\nc}{{n_{\rm ch}}}
\renewcommand{\ng}{{n_\gamma}}
\newcommand{\Nc}{{N_{\rm ch}}}
\newcommand{\Ng}{{N_\gamma}}
\newcommand{\Ngl}{{N_{\gamma-{\rm like}}}}
\newcommand{\zc}{{z_{\rm ch}}}
\newcommand{\zg}{{z_\gamma}}

\begin{document}

\begin{frontmatter}

\title{DISORIENTED CHIRAL CONDENSATE: THEORY AND EXPERIMENT}

\author[Bedanga]{B. Mohanty}
\ead{bmohanty@veccal.ernet.in}
and 
\author[Julien]{J. Serreau\corauthref{cor}}
\corauth[cor]{Corresponding Author. Tel: +33 1 69 15 70 36, Fax: +33 1 69 15 82 87}
\ead{Julien.Serreau@th.u-psud.fr}

\address[Bedanga]{Variable Energy Cyclotron Center, 1/AF, Bidhan Nagar
Kolkata - 700064, India}
\address[Julien]{Astro-Particule et Cosmologie,\\
11, place Marcelin Berthelot, F-75231 Paris Cedex 05, France\\
and\\
Laboratoire de Physique Th\'eorique,\\
B\^atiment 210, Universit\'e Paris-Sud 11, 91405 Orsay Cedex, France\thanksref{Lab}}
\thanks[Lab]{APC is unit\'e mixte de recherche UMR 7164 (CNRS, 
Universit\'e Paris 7, CEA, Observatoire de Paris). LPT is UMR 8627
(CNRS, Universtit\'e Paris-Sud~11).}

\begin{abstract}
It is thought that a region of pseudo-vacuum, where the chiral order 
parameter is misaligned from its vacuum orientation in isospin space, 
might occasionally form in high energy hadronic or nuclear collisions. 
The possible detection of such disoriented chiral condensate (DCC) would 
provide useful information about the chiral structure of the QCD vacuum 
and/or the chiral phase transition of strong interactions at high  
temperature. We review the theoretical developments concerning the possible
DCC formation in high-energy collisions as well as the various experimental 
searches that have been performed so far. We discuss future prospects for 
upcoming DCC searches, \eg in high-energy heavy-ion collision 
experiments at RHIC and LHC.
\end{abstract}


\end{frontmatter}


\section{Introduction}

In very high energy hadronic and/or nuclear collisions highly excited states are 
produced and subsequently decay toward vacuum via incoherent multi-particle emission. 
Due to the approximate $SU_R(2)\times SU_L(2)$ chiral 
symmetry of strong interactions, there exists a continuum of nearly 
degenerate low-energy (pseudo-vacuum) states. These correspond to 
collective excitations where the chiral quark condensate is rotated 
from its vacuum orientation in chiral space $\sigma\sim\bra\bar q q\ket$ and 
can be seen as semi-classical configurations of the pion field 
$\vpi\sim\bra \bar q\vtau\gamma_5 q\ket$, where $q\equiv(u,d)$ denotes the 
light-quark doublet and $\vec\tau$ are the usual Pauli matrices. An interesting 
possibility is that the decay of highly excited states produced in high-energy 
collisions may proceed via one of these collective states characterized by a 
disoriented chiral condensate (DCC). The latter would subsequently decay 
toward ordinary vacuum through coherent emission of low-momentum pions. 
Due to the semi-classical nature of the corresponding emission process,
this may lead to specific signatures, such as anomalously large event-by-event
fluctuations of the charged-to-neutral ratio of produced pions. If the 
space-time region where this happens is large enough the phenomenon might 
be experimentally observable, thereby providing an interesting opportunity 
to study the chiral structure of QCD. 

Although very speculative, the idea that a DCC may form in high-energy collisions
is quite appealing. Since it has been proposed in the early $1990$'s 
\cite{Bjorken:1991sg,Blaizot:1992at} (see also 
\cite{Nelson:1987dg,Anselm:1989pk,Anselm:1991pi})
it has attracted a lot of interest and has generated an intense theoretical and 
experimental activity. One of the main motivations -- beyond its appealing simplicity 
-- has probably been the existence of exotic, so-called Centauro events reported in 
the cosmic ray literature \cite{jacee,pamir,Gladysz-Dziadus:2001cq}, where clusters 
consisting of almost exclusively charged pions and no neutrals have been observed. 
The DCC would indeed provide a simple explanation for such phenomenon. 

On the theory side, a plausible mechanism for DCC formation in the context
of high-energy heavy-ion collisions has been identified 
\cite{Rajagopal:1992qz,Rajagopal:1993ah}: Due to the
large energy deposit in the collision zone, a hot, chirally symmetric state (quark-gluon 
plasma) is formed. Due to the fast expansion at early times, the system is 
suddenly quenched down to the low-temperature phase, where chiral symmetry is 
spontaneously broken. The subsequent far-from-equilibrium evolution triggers 
an exponential growth of long-wavelength pion modes, resulting in the formation 
of a strong (semi-classical) pion field configuration. This ``quench scenario'' 
has become a paradigm for DCC formation in heavy-ion collisions and has been
widely used to investigate the phenomenological aspects of DCC production. 
As a side effect, this has led people to think more about non-equilibrium dynamics
in the context of high-energy physics. In particular, this has triggered a number 
of theoretical developments concerning the description of far-from-equilibrium 
quantum fields,\footnote{For a recent review see \eg \cref{Berges:2003pc}.} in 
connection with other areas of research, such as condensed-matter physics, or 
early-time cosmology (see \eg \cref{Boyanovsky:1999wd}).

Numerous experimental searches have been performed in parallel with the development 
of theoretical ideas. These include the analysis of various cosmic ray experiments
\cite{jacee,pamir}, nucleon-nucleon collisions at CERN \cite{ua1,ua5_540,ua5_900} and 
Fermilab \cite{cdf}, with, in particular, the dedicated MiniMAX experiment \cite{minimax}, 
as well as nucleus-nucleus collisions at the CERN SPS 
\cite{wa98_global,na49,wa98_local,wa98_local_cen} 
and presently at RHIC \cite{phenixqm,starnim}. The search for DCCs 
and other exotic events is part of the heavy-ions physics program to be performed by the 
multi-purpose detector ALICE at the LHC \cite{alicetp,Angelis:2003ap}. 
These investigations have led, in particular, to the development of powerful experimental 
tools to search for non-statistical fluctuations and/or to detect non-trivial structures 
in high-multiplicity events. No clear positive signal has been reported so far and upper 
bounds have been put on the likelihood of DCC formation, in particular, in heavy-ion 
collisions at SPS energies \cite{wa98_global,na49,wa98_local_cen}. However, at the same 
time, it has been understood that the present experimental limit actually is consistent with 
theoretical expectations based on the quench scenario \cite{Krzywicki:1998sc,Serreau:2003gj}.

This paper presents a status report of this field of research. In the first part,
we review the main theoretical ideas and developments relative to DCC physics, 
including DCC formation and evolution in the multi-particle environment as well as 
phenomenological aspects. The second part is devoted to the numerous experimental 
studies which have been performed to search for DCC signals in high-energy hadronic 
and/or nuclear collisions. We emphasize that this report is, by no means, an exhaustive 
review of the extensive literature on this topics. Our main aim is, instead, to 
provide a comprehensive synthesis of what we believe are the most 
relevant aspects of DCC physics, both theoretically and experimentally, in view of 
further investigations, in particular, with upcoming DCC searches at RHIC and LHC.
We mention that earlier reviews can be found in 
\cite{Bjorken:1993cz,Rajagopal:1995bc,Blaizot:1996js,Bjorken:1997re}

To end this introduction and open our review, let us cite Bjorken's words in
his ``DCC trouble list'' \cite{Bjorken:trento}, which illustrate very well the 
spirit of the present report:

\hspace*{1.cm}Existence of DCC:

\begin{center}
\begin{tabular}{ll}
Must it exist?      &     NO                          \\
Should it exist?    &     MAYBE                       \\
Might it exist?     &     YES                         \\
Does it exist?      &     IT'S WORTH HAVE A LOOK      \\
\end{tabular}
\end{center}

\section{Theory}

\subsection{\label{sec:DCC} The disoriented chiral condensate: basic ideas}

The theory of strong interactions exhibits an approximate chiral symmetry
$SU_R(2)\times SU_L(2)$, which is spontaneously broken in the vacuum -- or 
in thermal equilibrium at sufficiently low temperatures. The associated order
parameter, namely the quark condensate $\bra\bar q_R q_L\ket$, can be represented
as a four-component vector $\varphi_a\equiv(\vpi,\sigma)$ transforming under the 
$O(4)$ subgroup of $SU_R(2)\times SU_L(2)$. In the vacuum, the order parameter 
points in the $\sigma$-direction. However, due to the approximate chiral 
symmetry, one might expect that under appropriate conditions, there could exist 
a region of space, separated from the physical vacuum for some period of time, 
where the order parameter develops a non-trivial pionic component: This is the 
disoriented chiral condensate.\footnote{It is worth emphasizing that there
is no contradiction with the well-known Vafa-Witten theorem \cite{Vafa:1983tf}, 
which states that no pionic component of the chiral condensate can develop in 
a stationary state: The DCC is, by essence, a transient phenomenon. 
Similarly, it should be stressed that the production of a DCC state does
not contradict usual conservation laws, such as \eg parity, isospin, or 
charge conservation, as one eventually has to average over all possible 
equivalent directions in isospin space to obtain physical results (see 
subsection \Sec{sec:coherent} below).}

\subsubsection{Baked-Alaska}

Bjorken and collaborators \cite{Bjorken:1991sg,Bjorken:1993cz,Bjorken:1994ug} have 
put forward a very simple and intuitive physical picture of the possible formation of a 
disoriented pseudo-vacuum state in hadronic collisions. We start our report by
reviewing this so-called baked-Alaska scenario, which provides a useful guide for 
physical intuition.
Consider a high multiplicity collision event with a large transverse
energy release but no high-$p_T$ jets. In this situation, the hadronization time 
can be rather long, as large as a few fm/c. Prior to hadronization, 
the primary partons carry most of the released energy away from the 
collision point at essentially the speed of light. One imagines a thin, 
``hot'' expanding shell which isolates the relatively ``cold'' interior 
from the outer vacuum. If the energy density left behind is low enough, 
the interior of the fireball should look very similar to the vacuum,
with an associated quark condensate. However, if the time it takes to cook 
this baked-Alaska is short enough, the quark condensate in the interior might 
be rotated from its usual orientation since the energy density associated 
with the explicit breaking of chiral symmetry is small.\footnote{A rough order
of magnitude is: $\epsilon \sim \fpi^2\mpi^2 \approx 20$~MeV/fm$^3$.}
When the hot shell hadronizes, the disoriented interior comes into contact 
with the true vacuum and radiates away its pionic orientation, resulting
in coherent emission of soft pions, with strong isospin correlations.

As a simple realization of these ideas \cite{Amelino-Camelia:1997in}, one might 
represent the hot debris located on the surface of the fireball as a source for 
the long-wavelength pionic excitation associated with the disoriented interior. 
In the linear approximation, the dynamics of the latter can be described by the 
following equation:
\beq
\label{freeprop}
 \left(\square+\mpi^2\right)\vpi(x)=\vJ(x)\,,
\eeq
where the arrows denote vectors in isospin-space. A simple example is, 
for instance, $\vJ(x)=\vJ(t)\,\Theta(t-t_0)\,\delta(t-r)$, corresponding
to a spherically expanding shell with initial radius $r_0=t_0$, where $t_0$ 
is the time where the expansion starts. After a 
typical decoupling (hadronization) time, the source $\vJ(t)$ vanishes and 
the field excitation $\vpi(x)$ decays into freely propagating pions.
In this simple model, pion emission is characterized by the coherent 
state:
\beq
\label{coherentstate}
 |\vJ\,\rangle={\mathcal N}\,\exp\left(\sum_{a=1}^3\int 
 \dk\,J_a(\bk)\,a_a^\dagger(\bk)\right)|0\rangle
\eeq
where $\mathcal N$ is a normalization factor and the subscript $a=1,2,3$ 
denotes Cartesian isospin orientations. Here, $a_a^\dagger(\bk)$ is the 
creation operator of a pion with momentum $\bk$ and isospin component $a$
and $J_a(\bk)$ is related to the on-shell $4$-dimensional Fourier transform 
of the source component $J_a(x)$ through: 
\beq
 J_a(\bk)=\frac{1}{\sqrt{2\omega_k}}\int d^4x\,\e^{ikx}J_a(x)\,
 \Big|_{k^0=\omega_k}\,,
\eeq
with $\omega_k=\sqrt{k^2+\mpi^2}$. Equivalently, it can be directly related 
to the spatial Fourier components of the asymptotic out-going field 
configuration $\pi^{out}_a(\bk,t)$ and its time derivative 
$\dot\pi^{out}_a(\bk,t)$:\footnote{It is easy to check that the 
RHS of \Eqn{outfield} does not depend on time by using the fact 
that $\pi^{out}_a(x)$ satisfies \Eqn{freeprop} with vanishing sources.}
\beq
\label{outfield}
 J_a(\bk)=-i\,\frac{\e^{-i\omega_k t}}{\sqrt{2\omega_k}}\,
 \left[i\dot\pi^{out}_a(\bk,t)+\omega_k\pi^{out}_a(\bk,t)\right]\,.
\eeq
For a given realization of the 
source, particles are produced independently and follow a Poisson 
distribution characterized by the average:
\beq
\label{avnum}
 \bar{n}_a(\bk)
 =\langle \vJ\,|a^\dagger_a(\bk)\,a_a(\bk)|\vJ\,\rangle
 =\left|J_a(\bk)\right|^2\,.
\eeq
When the latter is large enough -- which is required for the present
classical description to make sense at all, one can neglect the quantum
fluctuations, of relative order $\sim 1/\sqrt{\bar{n}}$. The number of 
pions produced per unit phase space in the collision event corresponding 
to the source $\vJ$ is then approximately given by:
\beq
 \frac{dN^{(J)}_a}{d^3k}\approx\frac{\left|J_a(\bk)\right|^2}
 {(2\pi)^3}\,.
\eeq

The magnitude and chiral orientation of the source $J_a(\bk)$ for 
each mode $\bk$ fluctuate from event to event. The above picture of a 
large region of space where the chiral condensate is coherently 
misaligned ideally corresponds to all relevant (soft) modes pointing 
in a given direction $\ve$ in isospin space:
\beq
\label{dccpol}
 \vJ_{\rm DCC}(\bk)=J(\bk)\,\ve\,.
\eeq
Notice from \Eqn{outfield} that this implies that both the field
and its time-derivative be aligned in the same direction $\ve$. 
This can be viewed as an out-going wave linearly polarized in 
isospin space. Clearly, the source \eqn{dccpol} induces non-trivial 
correlations between emitted pions with different isospin orientations. 
This can be nicely illustrated by means of the neutral fraction 
of emitted pions:
\beq
 f=\frac{N_{\pi^0}}{N_{\pi^0}+N_{\pi^\pm}}=\frac{N_3}{N_1+N_2+N_3}\,,
\eeq
where $N_a$ is the total number of pions with isospin component $a$
in a given event. For the DCC state \eqn{dccpol} one has:
\beq
 f_{\rm DCC}=\cos^2\hat\theta\,,
\eeq
where $\hat\theta$ is the angle between the unit vector $\ve$ and the third
axis in isospin space. Assuming that there is no privileged isospin
direction, all possible orientations of the unit vector $\ve$ are 
equally probable and one finds that the event-by-event distribution 
of the neutral fraction $f$ is given by \cite{Bjorken:1991sg,Blaizot:1992at}:
\beq
\label{DCCsign}
 \left.\frac{dP(f)}{df}\right|_{\rm DCC}=\frac{1}{2\sqrt f}\,.
\eeq
The latter exhibits striking fluctuations around the mean value
$\bar f=1/3$, which are a direct consequence of the coherence of 
the DCC state. Equation \eqn{DCCsign} is to be contrasted with the narrow Gaussian 
distribution predicted by statistical arguments for incoherent 
pion production.\footnote{For a binomial distribution, the typical 
fluctuations around the average value $\bar f=1/3$ are 
$\sim 1/\sqrt{N_{tot}}$, where $N_{tot}$ is the total multiplicity.} 
For instance, the probability 
that less than $10$~\% of the DCC pions be neutral is predicted to be as 
large as $30$~\%. This is what makes the DCC an interesting candidate to 
explain the Centauro events in cosmic ray showers. The property \eqn{DCCsign}
is also at the basis of most existing strategies for experimental searches.

\subsubsection{A dynamical perspective}

The idea that multiple-pion emission in high-energy hadronic and/or nuclear
collisions might be associated with classical radiation is rather old (see \eg 
Refs.~\cite{Horn:1971,Botke:1974ra,Botke:1974bs,Andreev:1981}) and can actually be 
traced back to some old papers by Heisenberg \cite{Heisenberg:1952}. This idea has, 
however, received only marginal attention until the early 1990's, where it 
has been rediscovered and further developed in the modern context of low-energy 
effective theories \cite{Anselm:1989pk,Anselm:1991pi,Blaizot:1992at,Bjorken:1993cz}. 
In \cref{Blaizot:1992at}, Blaizot and Krzywicki have investigated 
the question of soft pion emission in high energy nuclear collisions by studying 
classical solutions of the nonlinear $\sigma$ model, which describes the dynamics 
of low-energy pion fields. They adopted Heisenberg's ideal boundary
conditions \cite{Heisenberg:1952} (see also \cref{Bjorken:1982qr}) to 
model the expanding geometry of the collision. The picture is very close 
to the one described above, with the non-trivial pion field configuration 
generated by classical sources localized on the light-cone, that is receding
from each other at essentially the speed of light. 
Remarkably enough, there exist solutions which correspond to the DCC 
configuration \eqn{dccpol}, that is where the pion field oscillates in 
a given direction in isospin space. The analysis can be extended to the
linear $\sigma$ model \cite{Blaizot:1994ih} and provides an instructive 
dynamical realization of the qualitative ideas described previously. 

Written in terms of the quadruplet of scalar fields $\varphi_a\equiv(\vpi,\sigma)$, 
the classical action of the linear $\sigma$-model with the standard 
chiral-symmetry--breaking term reads:
\beq
 \label{action}
 {\mathcal S}=\int d^4x\left\{\frac{1}{2}\,\p_{\mu}\varphi_a\,\p^{\mu}\varphi_a
 -\frac{\lambda}{4}\,\left(\varphi_a\varphi_a-v^2\right)^2+H_a\varphi_a\right\}\,,
\eeq
where $H_a\equiv(\vec 0,H)$ points in the $\sigma$-direction in chiral space.
The parameters $v$, $\lambda$ and $H$ are related to physical quantities 
through:
\bea
 H&=&\fpi\mpi^2\nn
\label{parameters}
 \mpi^2&=&\lambda\left(\fpi^2-v^2\right)\\
 \msi^2&=&2\lambda\fpi^2+\mpi^2\nonumber
\eea
Note that in the phenomenologically relevant limit, namely $\msi\gg\mpi$ or,
equivalently, $\lambda\gg 1$, one has $v\approx\fpi$ and 
$\msi\approx\sqrt{2\lambda}\fpi$. 
In the following, we set the scale $v=1$ for simplicity. 
The classical equations of motion read:
\beq
 \label{eom}
 \left[\square+\lambda(\varphi^2-v^2)\right]\varphi_a=H_a\,.
\eeq
In the original treatment of \cite{Blaizot:1992at,Blaizot:1994ih}, 
the longitudinal expansion is modeled by viewing the colliding nuclei 
in the center of mass frame as two infinitesimally thin (Lorentz 
contracted) pancakes of infinite transverse extent (see also 
\cite{Bjorken:1982qr}). The symmetry of the problem then implies 
that the classical field $\varphi_a$ is a function of the proper time 
$\tau=\sqrt{t^2-z^2}$ only.\footnote{Notice
that strictly speaking, this assumes that the expansion never stops, or, 
in terms of the previous baked-Alaska description, that the sources of 
the pion field never decouple. This reflects the fact that the initial 
energy density is infinite in the present idealization. We stress 
however that, due to expansion, the energy density decreases inside the 
light-cone, and the pion dynamics eventually freezes out at a time $\tau_f$ 
(see \Eqn{pib} below). 
Therefore, the present boost-invariant idealization should provide a 
reasonable description if the decoupling (hadronization) time $t_h\gtrsim\tau_f$.} 
The field equations \eqn{eom} become ordinary differential equations and 
can be solved analytically. It is not difficult to extend Blaizot and 
Krzywicki's treatment to the case of a symmetric $d$-dimensional 
expansion. Here, we present the main results of such an analysis and 
essentially follow the presentation of \cref{Blaizot:1996js}.
The relevant proper-time variable is $\tau=\sqrt{t^2-r^2}$ 
with $r^2=\sum_{i=1}^dx_i^2$, and the four-dimensional Laplacian becomes:
\beq
\label{Lap}
 \square\varphi_a(\tau)=\tau^{-d}\p_\tau(\tau^d\p_\tau\varphi_a)
 =\ddot{\varphi_a}+\frac{d}{\tau}\dot{\varphi_a}\,,
\eeq
where the dot denotes derivative with respect to $\tau$. Hence, the
equations of motion involve a friction term $\sim \dot\varphi$, which
simply reflects the decrease of the energy density due to expansion.
The initial conditions are to be specified on the surface $\tau=\tau_0$.

From Eqs.~\eqn{eom} and \eqn{Lap} one easily obtains that:
\beq
\label{isovec}
 \vpi\times\dot{\vpi}=\frac{\va}{\tau^d}
\eeq
and
\beq
\label{axial}
 \vpi\dot{\sigma}-\sigma\dot{\vpi}=\frac{\vb}{\tau^d}
 +\frac{H}{\tau^d}\int^\tau \vpi \tau^d d\tau\,.
\eeq
The first of these equations is a consequence of the conservation
of the iso-vector current $\vV^\mu=\vpi\times\p^\mu\vpi$ and the second one
reflects the partial conservation of the corresponding axial-vector current
$\vA^\mu=\vpi\p^\mu\sigma-\sigma\p^\mu\vpi$. The orthogonal iso-vectors 
$\va$ and $\vb$ are integration constants. Their lengths measure the
initial strength of the respective currents. We shall focus here 
on the regime where $a\ll b$, which is the relevant one for our
present purpose. From \eqn{isovec}, one sees that the motion is 
planar in isospin space: $\pi_a\equiv\vpi\cdot\va=0$. 
At short enough time, the pion mass
is irrelevant and one can neglect the second term on the RHS of 
\Eqn{axial}. Following \cite{Blaizot:1994ih}, one can show that, 
for times $\tau\lesssim (b/\mpi)^{1/d}$:
\bea
 \pi_b&\approx&-r\sin\hat\theta\\
 \pi_c&\approx&\frac{a}{\sqrt{a^2+b^2}}\,r\cos\hat\theta\\
 \sigma&\approx&\frac{b}{\sqrt{a^2+b^2}}\,r\cos\hat\theta\,,
\eea
where $\vc=\va\times\vb$. The motion is approximately planar in the
$4$-dimensional chiral space. Notice that for $a\ll b$, the component
$\pi_c$ is very small and the pion field oscillates along the (random)
direction defined by the iso-vector $\vb$. This precisely corresponds
to the linearly polarized DCC configuration described in the previous 
subsection. 

For times $(b/\msi)^{1/d}\lesssim\tau\lesssim (b/\mpi)^{1/d}$, the 
length of the chiral field $r=\sqrt{\sigma^2+\pi^2}$ undergoes rapid, 
damped oscillations around the approximately degenerate minimum of
the potential. For instance, for large enough $\tau$, one has:
\beq
 r\approx1+\frac{C\cos(\msi\tau+\delta)}{(\msi\tau)^{d\over2}}\,,
\eeq
where $C$ and $\delta$ are integration constants. In this regime
one can replace $r$ by its time-averaged value $\bar r=1$ for all practical
purposes and the motion essentially takes place near the minimum of 
the mexican hat potential. One therefore obtains that the angle $\hat\theta$ 
is approximately given by:
\beq
 \hat\theta\approx b\,\int^\tau\frac{d\tau}{\tau^d}\,,
\eeq
corresponding to a circular motion in chiral space. In this regime 
the energy density is approximately given by: 
\beq
\label{endens}
 \epsilon\approx\frac{\dot\sigma^2+\dot\pi^2}{2}\approx\frac{b^2}{2\tau^{2d}}\,,
\eeq
and is still high enough so that the system does not feel the explicit 
symmetry breaking due to the pion mass, hence the circular motion. 

At a time $\tau\approx (b/\mpi)^{1/d}$, the damping produces a cross-over
from the circular to an oscillatory motion around the true minimum of the 
potential. This is also the time at which the energy density \eqn{endens}
becomes comparable to the mass of a pion, divided by the cube of its 
Compton wavelength $\sim \mpi^4$. Therefore, the non-linear dynamics
freezes out and one is left with freely propagating pions. Indeed, for
times $\tau\gtrsim (b/\mpi)^{1/d}$, one finds:
\beq
\label{pib}
 \pi_b\approx\frac{\sqrt{b}\cos(\mpi\tau+\delta')}{(\mpi\tau)^{d\over2}}
\eeq
and
\beq
 \pi_c\approx\frac{a^2}{b^2}\,\pi_b\,,
\eeq
which describes free propagation of pions in the expanding geometry.
These correspond to the DCC decay products. As already noticed, 
when the relative strengths of the initial vector versus axial-vector
current is small ($a\ll b$), the pion field is mainly polarized along 
the random direction $\vb$, which results in the event-by-event 
distribution \eqn{DCCsign} for the neutral fraction~$f$.

We see from \Eqn{pib} that the parameter $b$ controls the amplitude
of the out-going pion field, that is in turn, of the amount of radiated
energy. It is interesting to compute the probability $P(a,b)$ that $a$ 
and $b$ take particular values in a simple model characterizing 
the high energy, chirally symmetric initial state produced in 
the collision. Assuming that the values of the fields $(\vpi,\sigma)$ 
and their proper-time derivative at $\tau=\tau_0$ are Gaussian 
random numbers of zero mean and of variance $\sigma_1$ and 
$\sigma_2$ respectively, one easily obtains \cite{Blaizot:1994ih}:
\beq
\label{probab}
 \frac{d^2P(a,b)}{dadb}
 =\frac{K_1\left(\sqrt{a^2+b^2}/\kappa_0\right)}{2\pi^2\kappa_0^2} 
 \sim\frac{\e^{-b/\kappa_0}}{\sqrt b}
\eeq
where $K_1(z)$ is the second modified Bessel function and 
$\kappa_0=\sigma_1\sigma_2\tau_0^d$. The second term on the RHS
shows the behavior at $b\gg a,\kappa_0$: The probability of
a significant signal appears to be exponentially suppressed in
this simple model.

\subsubsection{\label{sec:coherent} Coherent state descriptions}

It is instructive to see how the qualitative argument leading to the
prediction \eqn{DCCsign} is modified when one takes the quantum nature 
of the emission process into account. This can be done in the
previous coherent state picture, Eqs. \eqn{coherentstate} and 
\eqn{dccpol},\footnote{Squeezed quantum states have also been considered 
in the literature, see \eg 
\cite{Amado:1994fu,Dremin:1995nd,Hiro-Oka:1999xk,Bambah:2004tq}.}
where the DCC state is characterized by a given orientation $\ve$ in isospin
space and a complex source $J$. For each given source $J$, isospin symmetry 
is ensured by averaging over all possible orientations with equal weight.
This can be described by the following density matrix:
\beq
\label{mixed}
 \rho(J,J^*)=\int\frac{d^2\Omega}{4\pi}\,
 |\vJ\,\ket\bra \vJ\,|\,,
\eeq
where $|\vJ\,\ket$ denotes the DCC state, Eqs.~\eqn{coherentstate}
and \eqn{dccpol}, and the integral is over all possible orientations
of the unit iso-vector $\ve$ characterized by the solid angle $\Omega$. 
Alternatively, one can imagine \cite{Horn:1971} a coherent superposition of states 
$|\vJ\,\ket$ with all possible orientations, corresponding to the following 
zero-isospin pure state:\footnote{Notice that this state is not normalized.
One has: $\bra J|J\ket=\prod_\bk {\mathcal N}(|J(\bk)|^2)$, with 
${\mathcal N}(x)\equiv4\pi\,\e^{-x}\sinh x/x$.}
\beq
\label{pure}
 |J\ket=\int d^2\Omega\, Y_0^0(\Omega)\,|\vJ\,\ket\,,
\eeq
with the spherical harmonic $Y_0^0(\Omega)=1/\sqrt{4\pi}$. It is easy
to show that the two descriptions \eqn{mixed} and \eqn{pure} are 
actually equivalent in the limit of large particle numbers 
$|J(\bk)|^2\gg1$. In that case, one can write:\footnote{The precise 
meaning of this relation is that correlation functions -- or multiplicity 
distributions considered below -- computed with either the mixed state 
\eqn{mixed} or the pure state \eqn{pure} agree in the (classical) limit of 
large particle numbers. This can be seen by computing the generating functional 
$Z(\vK,\vK^*)=\bra J|{\mathcal O}(\vK,\vK^*)|J\ket$, where
${\mathcal O}(\vK,\vK^*)= \exp\int\dk[\vK(\bk)\cdot\va_\bk
+\vK^*(\bk)\cdot\va_\bk^\dagger]$, from which one can obtain all
correlation functions by derivatives with respect to the sources
$\vK(\bk)$ and $\vK^*(\bk)$. The calculation involves an integration
over two directions, $\int d^2\Omega d^2\Omega'$. By employing a saddle 
point approximation in the limit of large particle numbers, one can show
that $Z(\vK,\vK^*)\simeq\int\frac{d^2\Omega}{4\pi}\bra \vJ\,| 
{\mathcal O}(\vK,\vK^*)|\vJ\,\ket={\rm Tr}\{\rho\,
{\mathcal O}(\vK,\vK^*)\}$, with $\rho$ given by \Eqn{mixed}.}
\beq
 \rho(J,J^*)\approx|J\ket\bra J|\,.
\eeq

To completely characterize the emission process, one has to consider all 
possible realizations of the complex source $J$, with appropriate weight 
${\mathcal P}(J,J^*)$ (the latter reflects the detailed dynamics of DCC 
formation, to be discussed later in this review). This can be described by 
the following density matrix:
\beq
\label{model}
 \rho_{\rm DCC}=\int\D J\D J^*\,{\mathcal P}(J,J^*)\,\rho(J,J^*)
\eeq
where the integration runs over all modes in the DCC state: 
$\D J\D J^*\equiv\Pi_\bk dJ(\bk)dJ^*(\bk)$.

The multiplicity distribution of emitted pions can be obtained as 
(we consider here a single mode $\bk$ for simplicity and we omit 
the explicit $\bk$-dependence):
\beq
\label{multdist}
 p(n_0,n_+,n_-)=\int\D J\D J^*\,{\mathcal P}(J,J^*)\,
 \int\frac{d^2\Omega}{4\pi}\,|\bra n_0,n_+,n_-|\vJ\,\ket|^2
\eeq
where $n_0$ and $n_\pm$ denote the actual multiplicity of neutral
and positive/ne\-ga\-ti\-ve pions respectively. The corresponding state 
is given by:
\beq
 |n_0,n_+,n_-\ket=\frac{{(a_0^\dagger)}^{n_0}}{\sqrt{n_0!}}\,
 \frac{{(a_+^\dagger)}^{n_+}}{\sqrt{n_+!}}\,
 \frac{{(a_-^\dagger)}^{n_-}}{\sqrt{n_-!}}\,|0\ket
\eeq
where the creation operators $a_0^\dagger$ and $a_\pm^\dagger$ for 
neutral and charged pions respectively are related to the 
Cartesian-coordinates creation operators $a_a^\dagger$ as: 
$a_0^\dagger=a_3^\dagger$ and $a_\pm^\dagger=\mp(a_1^\dagger\pm
ia_2^\dagger)/\sqrt 2$.
Correspondingly, it is convenient to
decompose the unit iso-vector $\ve$ by: $e_0=e_3=\cos\hat\theta$
and $e_\pm=\mp(e_1\pm ie_2)/\sqrt 2=\mp\sin\hat\theta e^{\pm i\hat\varphi}/\sqrt 2$,
where $\hat\theta$ and $\hat\varphi$ denote respectively the azimuthal and 
polar angles in isospin-space. 
The probability distribution $p(\nc,n_0)$ for charged and neutral  
pions is easily obtained from \eqn{multdist} and can be written as 
($N=\nc+n_0$):
\beq
 p(\nc,n_0)=P(N)\,p_N(\nc,n_0)\,,
\eeq
where $P(N)$ is the total multiplicity distribution and
$p_N(\nc,n_0)$ denotes the conditional probability that,
among $N$ produced pions, $\nc$ be charged and $n_0$ be 
neutral. In the present model, one gets:
\beq
 P(N)=\int\D J\D J^*\,{\mathcal P}(J,J^*)\,\frac{|J|^{2N}}{N!}\,
 \e^{-|J|^2}
\eeq
and
\bea
 p_N(\nc,n_0)&=&\frac{N!}{\nc!n_0!}\,\int\frac{d^2\Omega}{4\pi}\,|e_0|^{2n_0}
 (|e_+|^2+|e_-|^2)^\nc\nn
\label{DCCmodel}
 &=&\frac{N!}{\nc!n_0!}\int_0^1 \frac{df}{2\sqrt f}\,f^{n_0}\,(1-f)^\nc\,,
\eea 
where we introduced the variable $f=\cos^2\hat\theta$ in the second line.
The last integral in \Eqn{DCCmodel} can be expressed in terms of the Euler 
gamma function $\Gamma(z)$ and one gets:
\beq
 p_N(\nc,n_0)=\frac{\Gamma(N+1)\Gamma(n_0+\frac{1}{2})}{2\Gamma(N+\frac{3}{2})\Gamma(n_0+1)}\,,
\eeq
For large mulitplicities $n_0,\nc\gg1$, one finally obtains (see also 
\cite{Andreev:1981,Kowalski:1992xq}):
\beq
\label{dccquantum}
 p_N(\nc,n_0)\simeq\frac{1}{N}\frac{1}{2{\hat f}^{1/2}}
\eeq
where $\hat f=n_0/N$ is the fraction of neutral pions.
Thus one recovers the inverse square-root law obtained previously by 
a simple geometrical argument, neglecting quantum fluctuations.
As expected, this is a valid approximation when the number of produced
particles is large.

Another possibility for the DCC state has been extensively discussed in the literature 
\cite{Horn:1971,Kowalski:1992xq,Bjorken:1993wj,Greiner:1993jn,Biyajima:1998yh,Nakamura:1999ai}, 
namely the zero-isospin state with fixed 
number $N$ of pions (note that a state of total isospin zero can only contain 
an even number of pions) \cite{Horn:1971}:
\beq
\label{zeroiso}
 |I=0,N=2m\ket=\frac{1}{\sqrt{(2m+1)!}}
 \left[(a_0^\dagger)^2-2a_+^\dagger a_-^\dagger\right]^m|0\ket\,.
\eeq
To make link with the present considerations, we note that the iso\-spin-\-ave\-ra\-ged 
coherent state \eqn{pure} can in fact be written as a coherent superposition of 
the states \eqn{zeroiso} with arbitrary (even) number of pions. Indeed, one easily 
gets, after some calculations:
\beq
 |J\ket=\sqrt{4\pi}\,\e^{-{1\over2}|J|^2}\sum_{m\ge0} \frac{J^{2m}}{\sqrt{(2m+1)!}}\,
 |I=0,N=2m\ket\,.
\eeq
It is clearly seen on this expression that production of a DCC is consistent
with usual conservation laws, such as charge ($n_+=n_-$), isospin, or parity
(even number of pions) conservation.

It is interesting to characterize the correlations between the different charge
states \cite{Greiner:1993jn,Biyajima:1998yh,Nakamura:1999ai}. 
A convenient way is to introduce the following normalized 
correlation function ($i,j=+,0,-$) \cite{Gyulassy:1979yi,Nakamura:1999ai}:
\beq
\label{normcorrel}
 R_{ij}=\frac{\bra N\ket^2}{\bra N(N-1)\ket}\frac{\bra n_in_j\ket}
 {\bra n_i\ket\bra n_j\ket}-1
\eeq
where brackets denote an average with respect to the multiplicity distribution
\eqn{multdist}. Here, the factor $\bra N\ket^2/\bra N(N-1)\ket$ is introduced 
to correct for possible non-Poissonian multiplicity fluctuations.
For the model described above, one obviously has:
\beq
 \bra n_i\ket=\frac{1}{3}\bra N\ket\,,
\eeq
The correlators $\bra n_in_j\ket$ are most conveniently calculated by means of the formula:
$\bra n_in_j\ket={\rm Tr}\{\rho_{\rm DCC}\,a_i^\dagger a_i\,a_j^\dagger a_j\}$,
where $\rho_{\rm DCC}$ is given by Eqs.~\eqn{model} and \eqn{mixed}. One finds, 
neglecting terms of relative order $1/\bra N\ket$:
\beq
 \bra n_in_j\ket\simeq r_{ij}\bra N(N-1)\ket\,,
\eeq
where
\beq
 r_{ij}=\int\frac{d^2\Omega}{4\pi}\,|e_i|^2|e_j|^2\,.
\eeq
Finally, one obtains the following correlations \cite{Greiner:1993jn,Nakamura:1999ai}:
\bea
 R_{00}&\simeq&\frac{4}{5}\\
\label{n-ch}
 R_{0+}&\simeq&-\frac{2}{5}\\
 R_{++}&\simeq&\frac{1}{5}\,.
\eea
This is to be contrasted with the corresponding result $R_{ij}\propto \delta_{ij}/\bra N\ket$, 
expected from statistical arguments for incoherent pion emission. Clearly, the 
large correlations obtained here reflect the large event-by-event fluctuations 
of the neutral fraction of emitted pions, see \eqn{dccquantum}. Notice, in particular,
the negative neutral-charged correlation \eqn{n-ch}, characteristic of DCC 
emission.

\subsection{\label{sec:chiralPT} Dynamics of DCC formation: the out-of-equilibrium
chiral phase transition in heavy-ion collisions}

We have seen that the DCC field configuration is a solution of the low energy 
dynamics of strong interactions. The approaches described in the previous sections 
assume the existence of a classical pion field -- or, equivalently, of a classical
source for the latter. To go beyond this level of description, 
one needs to understand the dynamical origin of such a classical field.
An important progress has been made in this respect in \cref{Rajagopal:1993ah},
where Rajagopal and Wilczek have realized that a strong long-wavelength pion 
field configuration could indeed be formed during the out-of-equilibrium chiral 
phase transition in the context of heavy-ion collisions: The rapid expansion of 
the system results in a sudden suppression of initial fluctuations (quenching) 
which in turn triggers a dramatic amplification of soft pion modes~\cite{Rajagopal:1993ah}.

\subsubsection{\label{sec:quench} The quench scenario: amplification of long wavelength modes}

To model the dynamics of the far-from-equilibrium chiral phase transition,
Rajagopal and Wilczek \cite{Rajagopal:1993ah} consider the $O(4)$ linear sigma
model for the chiral quadruplet $\varphi_a=(\vec\pi,\sigma)$ with action 
given in \Eqn{action} above, where the parameters \eqn{parameters} are chosen 
so that $\mpi=135$~MeV, $\msi=600$~MeV and $\fpi=92.5$~MeV. 
Anticipating the fact that the relevant field configurations correspond 
to large field amplitudes (that is large occupation numbers), it is 
justified to employ the classical statistical field approximation. In practice, 
the possible initial field configurations at time $t=0$, characterized by the 
values of the fields $\varphi_0^a(\bx)\equiv\varphi_a(\bx,t=0)$ and their 
time-derivatives $\dot\varphi_0^a(\bx)\equiv\dot\varphi_a(\bx,t=0)$ at each point 
of space, are sampled from a given statistical ensemble reflecting the initial density 
matrix of the corresponding quantum system. Each such field configuration is 
then evolved in time according to the classical equations of motion \eqn{eom} 
-- supplemented by appropriate boundary conditions. The time-evolution 
of a given physical observable ${\mathcal O}\equiv{\mathcal O}[\varphi,\dot\varphi]$ 
is obtained by averaging over all possible field configurations. This is summarized 
in the following formula, where we omit the explicit spatial dependence as well 
as chiral indices for simplicity:
\beq
\label{statistical}
 \bra {\mathcal O}(t)\ket=\frac{\int d\varphi_0d\dot\varphi_0\,
 {\mathcal P}[\varphi_0,\dot\varphi_0]\,{\mathcal O}[\varphi(t),\dot\varphi(t)]}
 {\int d\varphi_0d\dot\varphi_0\,{\mathcal P}[\varphi_0,\dot\varphi_0]}\,,
\eeq
where $\int d\varphi_0 d\dot\varphi_0$ represents an integral over possible 
initial field configurations $\{\varphi_0,\dot\varphi_0\}$ with appropriate
weight ${\mathcal P}[\varphi_0,\dot\varphi_0]$ and $\{\varphi(t),\dot\varphi(t)\}$ 
is the corresponding time-evolved configuration at time $t$. In the strong field 
regime, where quantum fluctuations are suppressed compared to statistical fluctuations, 
the previous procedure provides a good approximation to the full (quantum) dynamics 
(see \eg \cite{Aarts:1997kp,Aarts:2001yn}).

The physical picture of the Rajagopal and Wilczek scenario is as follows: 
One assumes that a large amount 
of energy has been deposited in the collision zone, corresponding to 
a very high temperature, presumably well above the chiral transition 
temperature. Due to the rapid expansion, this hot system experiences
rapid cooling, which results in a strong suppression of the initial 
(thermal) fluctuations. To model this effect, Rajagopal and Wilczek 
assume an instantaneous quench from above to below the critical 
temperature at the initial time. The initial field configurations are 
sampled from a chirally symmetric probability distribution, characteristic 
of the high-temperature phase, but where the fluctuations are frozen by 
hand. In the simplest realization of this scenario \cite{Rajagopal:1993ah}, 
the values of the field $\varphi_0^a(\bx)$ and its time-derivative $\dot\varphi_0^a(\bx)$ 
at the initial time are chosen as independent random variables 
on each site of a cubic lattice.\footnote{The lattice spacing has 
therefore the physical meaning of the correlation length $\lambda_c$ in 
the high-temperature phase.} They are sampled from a Gaussian distribution 
with respective means $\bra\varphi_0^a\ket=0$ and $\bra\dot\varphi_0^a\ket=0$, 
variances $\bra\varphi_0^a\varphi_0^b\ket=\sigma_1^2\,\delta_{ab}$ and 
$\bra\dot\varphi_0^a\dot\varphi_0^b\ket=\sigma_2^2\,\delta_{ab}$, 
and covariance $\bra\varphi_0^a\dot\varphi_0^b\ket=0$, where 
$\sigma_1^2$ and $\sigma_2^2$ are chosen to be smaller than what 
they would be in the high-temperature phase, thereby describing 
quenched fluctuations. The authors of \cref{Rajagopal:1993ah} choose 
$\sigma_1^2=v^2/16$ and $\sigma_2^2=v^2/4\lambda_c^2$, where 
$\lambda_c=a$ is the correlation length in the initial high-temperature
phase, here identified with the lattice spacing $a$.

\begin{figure}[t]
\begin{center}
 \epsfig{file=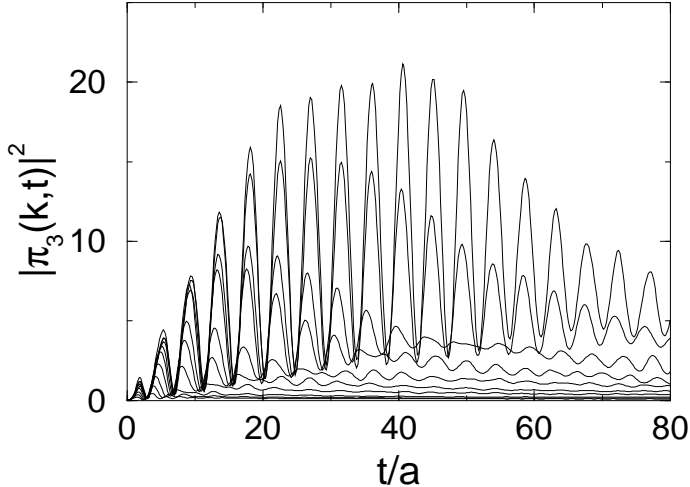,width=9.cm}
\end{center}
\caption{\label{fig:quench}
 \small The squared amplitude of the pion field in Fourier space $|\pi_3(\bk,t)|^2$ 
 as a function of time (in lattice units) for a given initial field configuration in 
 the quenched ensemble (see text), as in \cref{Rajagopal:1993ah}. Low-momentum modes 
 experience dramatic amplification, the softer the mode the larger the amplification, 
 and exhibit coherent oscillations with period $\sim 2\pi/m_\pi$.}
\end{figure}

The main result of \cref{Rajagopal:1993ah} is reproduced in \Fig{fig:quench}, 
which shows the squared amplitude of different field modes as a function of 
time, starting from a given field configuration in the ensemble described above: 
Due to the far-from-equilibrium initial condition, one observes a dramatic 
amplification of low momentum modes of the pion field at intermediate times. 
This phenomenon is analogous to that of domain formation after quenching a 
ferromagnet below the critical temperature. It is also operative \eg in the 
physics of particle creation in the early universe in so-called new
inflationary scenarios (see \eg \cref{Boyanovsky:1996sq}).

\begin{figure}[t]
\begin{center}
 \epsfig{file=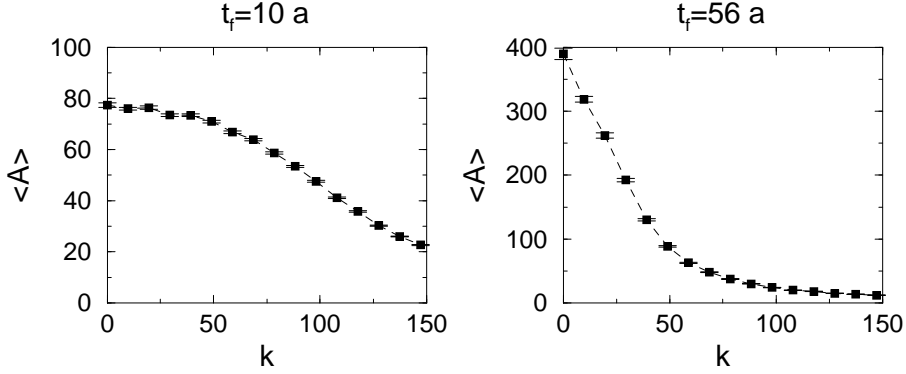,width=12.cm}
\end{center}
\caption{\label{fig:amplifquench}
 \small Momentum dependence of the amplification factor \eqn{ampliffac} 
 averaged over the initial ensemble in the quench scenario at $t_f=10a$
 ($21.0\times10^3$ events) and $t_f=56a$ ($10.9\times10^3$ events), where
 $a$ is the lattice spacing. Error bars are statistical. The dashed lines 
 are just guide for the eyes. From Ref. \cite{Serreau:2003gj}.} 
\end{figure}

It is worth emphasizing that the dramatic growth of field amplitudes for 
low-momentum modes at intermediate times is independent of the particular 
initial configuration chosen here. It is a generic feature of the typical 
field configurations in the statistical ensemble described previously. 
To illustrate this, we define an amplification factor as:
\beq 
\label{ampliffac}
 A(\bk,t)=\frac{P(\bk,t)}{P(\bk,0)}\,,
\eeq
where 
\beq
 P(\bk,t)=\omega_k\sum_{a=1}^3\bar n_a(\bk,t)
\eeq
is the pion power spectrum in mode $\bk$ at time $t$, with 
the average occupation number in mode $\bk$ defined by 
($\omega_k=\sqrt{k^2+\mpi^2}$):
\beq
\label{number1}
 \bar n_a(\bk,t)=\left|\frac{i\dot\varphi_a(\bk,t)+
   \omega_k\,\varphi_a(\bk,t)}{\sqrt{2\omega_k}}\right|^2\,,
\eeq
in analogy with Eqs.~\eqn{outfield}-\eqn{avnum}.\footnote{This is 
inspired by the fact that each classical field configuration in the 
statistical ensemble might be viewed as a particular coherent state
of the corresponding quantum system.} Here, $\varphi_a(\bk,t)$ and $\dot\varphi_a(\bk,t)$ 
are the spatial Fourier transforms of the field $\varphi_a(\bx,t)$ and 
its time-derivative $\dot\varphi_a(\bx,t)$ respectively. Figure~\ref{fig:amplifquench} 
shows the amplification factor averaged over a large number of field 
configuration as a function of momentum at two different times \cite{Serreau:2003gj} 
(see below). 
One observes that, starting from a quenched initial ensemble, the 
typical field configurations indeed experience dramatic amplification
at low-momentum, the softer the mode the stronger the amplification.

This can be qualitatively understood using the following mean-field 
type approximation~\cite{Rajagopal:1993ah}: In the 
classical equation of motion \eqn{eom}, replace the quadratic term
in the square brackets on the the LHS by its ensemble average value:
$\varphi^2\to\bra\varphi^2\ket$. Assuming a spatially homogeneous ensemble
for simplicity, the latter average is a function of $t$ only and the 
equation of motion for the pion field can be written, in Fourier space:\footnote{The 
approximation described here corresponds to the so-called large-$N$ approximation.}
\beq
\label{MFeom}
 \left[\frac{d^2}{dt^2}+k^2+m_{\rm eff}^2(t)\right]\vpi(\bk,t)={\vec 0}\,,
\eeq
where the effective time-dependent mass
\beq 
\label{effmass} 
 m_{\rm eff}^2(t)=\lambda\left[\bra\varphi^2\ket(t)-v^2\right]
\eeq
The problem is analogous to that of an ensemble of oscillators with
time dependent frequencies. The mass \eqn{effmass} can be viewed as the 
instantaneous curvature of the effective potential seen by the field 
modes at time $t$. 
One immediately sees that whenever the effective mass squared 
\eqn{effmass} becomes negative, soft modes with $k^2 < |m_{\rm eff}^2(t)|$
undergo exponential growth.
This is analogous to the so-called spinodal instability.
This happens when the field fluctuations get small enough: 
$\bra\varphi^2\ket<v^2$, which is the case in particular at early 
times in the quench scenario. 
\begin{figure}[t]
\begin{center}
 \epsfig{file=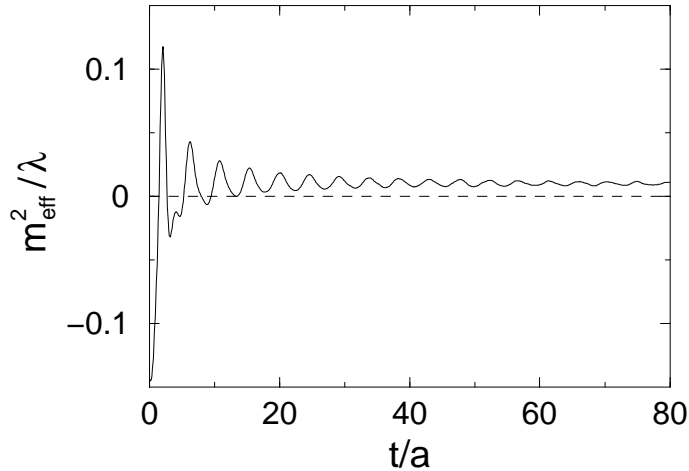,width=9.cm}
\end{center}
\caption{\label{fig:effmass}
 \small Effective mass squared $m_{\rm eff}^2(t)$ as a function 
  of time (in units of the lattice spacing $a$) in the quench 
  scenario \cite{Rajagopal:1993ah}.}
\end{figure}
Figure \ref{fig:effmass} shows the actual time evolution of the effective 
mass squared defined in \Eqn{effmass}, as obtained from the exact classical
Monte-Carlo simulation described previously \cite{Rajagopal:1993ah}. We see that the 
amplification of low-momentum modes at early times ($t\lesssim 10a$) is 
indeed due to the phenomenon of spinodal instability.
The corresponding amplification factor is shown on the left panel of 
\Fig{fig:amplifquench}.

It is clear from \Fig{fig:effmass} that the spinodal instability alone
cannot explain the amplification observed on \Fig{fig:quench} after times
$t\sim10a$, where the effective mass squared \eqn{effmass} is always 
positive. In fact, this intermediate-time amplification for $10a\lesssim t
\lesssim60a$, can be understood as resulting from the quasi-periodic 
oscillations of $m_{\rm eff}^2(t)$ around its asymptotic late time value 
\cite{Mrowczynski:1995at,Kaiser:1998hf}. This phenomenon is
known as parametric resonance and plays an important role \eg in describing
the (p)reheating of the early universe in chaotic inflationary scenarios 
\cite{Traschen:1990sw,Kofman:1994rk,Boyanovsky:1996sq}. Figure \ref{fig:amplifquench}
shows the momentum dependence of the average amplification factor
\eqn{ampliffac} at two different times, characterizing the two
amplification mechanisms at work. We see that most of the observed
amplification is due to the spinodal instability at early-times, leading
to an amplification $\sim 80$ for the most amplified mode, whereas the
parametric resonance phenomenon at intermediate-times only gives an
extra factor $\sim 5$ amplification. Moreover, it is important to 
emphasize that the regular oscillations of the effective mass are
in fact suppressed when expansion is explicitly taken into account (see \eg 
\Fig{fig:chi} below) and, therefore, so is the intermediate-time 
parametric amplification.

Thus we see that, due to the very unstable initial state, which was
assumed to be formed as a consequence of rapid expansion, typical pion field
configurations undergo a dramatic amplification and a strong long-wavelength 
pion field develops rapidly.\footnote{This justifies {\it a posteriori}
the use of classical statistical field theory.} Expansion causes the energy 
density to drop and the dynamics eventually linearizes as the modes stop interacting. 
If this freeze-out happens short enough after the collision, the system may be 
left in such a strong field configuration and subsequently decay through 
coherent pion emission.\footnote{One may wonder whether a similar phenomenon
could lead to enhanced production of strange mesons due to the approximate 
restoration of the $SU_R(3)\times SU_L(3)$ symmetry at very high temperatures. 
This has been investigated in \cref{Schaffner-Bielich:1998zi} (see also 
\cite{Gavin:2001uk} for a phenomenological study), with somewhat negative 
conclusions (which can be partly understood as being due to the too large 
mass of the strange quark).}  This provides a microscopic 
scenario for the formation of a strong (semi-classical) pion field configuration 
in heavy-ion collisions.\footnote{It is of common use in the literature to interpret 
each representative of the statistical ensemble described here as a possible 
realization of the pion field formed in a given collision event. It is worth 
emphasizing, however, that physical results, which require averaging over the 
statistical ensemble, do not depend on this interpretation.}
It is worth mentioning that this phenomenon has been demonstrated in other models
as well \cite{Bedaque:1993fa,Barducci:1996eu,Abada:1997mb}.\footnote{We mention that
a slightly different scenario for the growth of long-wavelength pion modes has been 
considered in \cref{Gavin:1993px}. In this so-called annealing scenario, soft pion 
modes evolve in an effective thermal potential with slowly decreasing temperature.
This essentially differs from the quench scenario in that the system evolves close
to equilibrium, which leads to less efficient amplification \cite{Asakawa:1994wk}.}

\subsubsection{\label{sec:exp} Expansion and initial conditions}

A crucial assumption in the previous treatment concerns the role of 
expansion. Including expansion explicitly 
in the description is needed e.g. in order to assess whether the 
decoupling of modes (freeze-out) occurs soon enough before rescatterings 
destroy the large field amplitude and thermalize the system (see e.g. 
\cref{Bialas:1994du}). But the main aspect of implementing expansion is to 
relax the drastic quench assumption made in \cref{Rajagopal:1993ah}. 
There are two points to this assumption: First, the time-scale of the initial 
energy drop must be sufficiently short compared to the typical time characterizing 
the interactions between long-wavelength modes; Second, the amplitude of the 
initial energy drop must be large enough so that, starting in the high-temperature
chirally symmetric phase, the system is indeed cooled down to the unstable region.
Clearly, the possible occurrence of an instability will depend both on the efficiency 
of the expansion in quenching initial fluctuations and on the actual initial state 
of the system.

These aspects have been investigated in details by Randrup in \cref{Randrup:1996ay}
in the context of classical statistical field theory. We review this work here. 
The effect of expansion is modeled by adding a cooling term by hand in the equations 
of motion \eqn{eom}. Although not a rigorous treatment of an expanding system, this 
captures the physics of the cooling process in a simple way.\footnote{For
descriptions in actual expanding geometries, see \eg 
\cite{Bialas:1994du,Cooper:1994ji,Lampert:1996qw,Petersen:1999jc} 
(see also subsection \ref{sec:expansion} below).}  
For a symmetric $d$-dimensional expansion, one replaces the $4$-dimensional
Laplacian by:
\beq
\label{cooling}
 \square=\p_t^2-\Delta\,\,\longrightarrow\,\,\p_t^2+\frac{d}{t}\p_t-\Delta
\eeq
The assumption of quenched initial fluctuations is relaxed and the initial 
field configurations are sampled from a thermal initial state at a given 
temperature $T$. Employing a Hartree approximation at $t=0$, the initial 
ensemble can be described by a thermal bath of non-interacting quasi-particles 
with self-consistently determined masses and the corresponding probability 
distribution ${\mathcal P}[\varphi_0,\dot\varphi_0]$, see \Eqn{statistical}, 
is simply given by a Gaussian \cite{Randrup:1996hk}.  Assuming a high-temperature 
initial state, the latter distribution is characterized by almost vanishing average values for 
the chiral field and its time derivative and by thermal two-point functions. Initial
field configurations sampled in this Gaussian ensemble are then evolved in time
according to the exact classical equations of motion with cooling, Eqs.~\eqn{eom} 
and \eqn{cooling}.

\begin{figure}[t]
\begin{center}
\epsfig{file=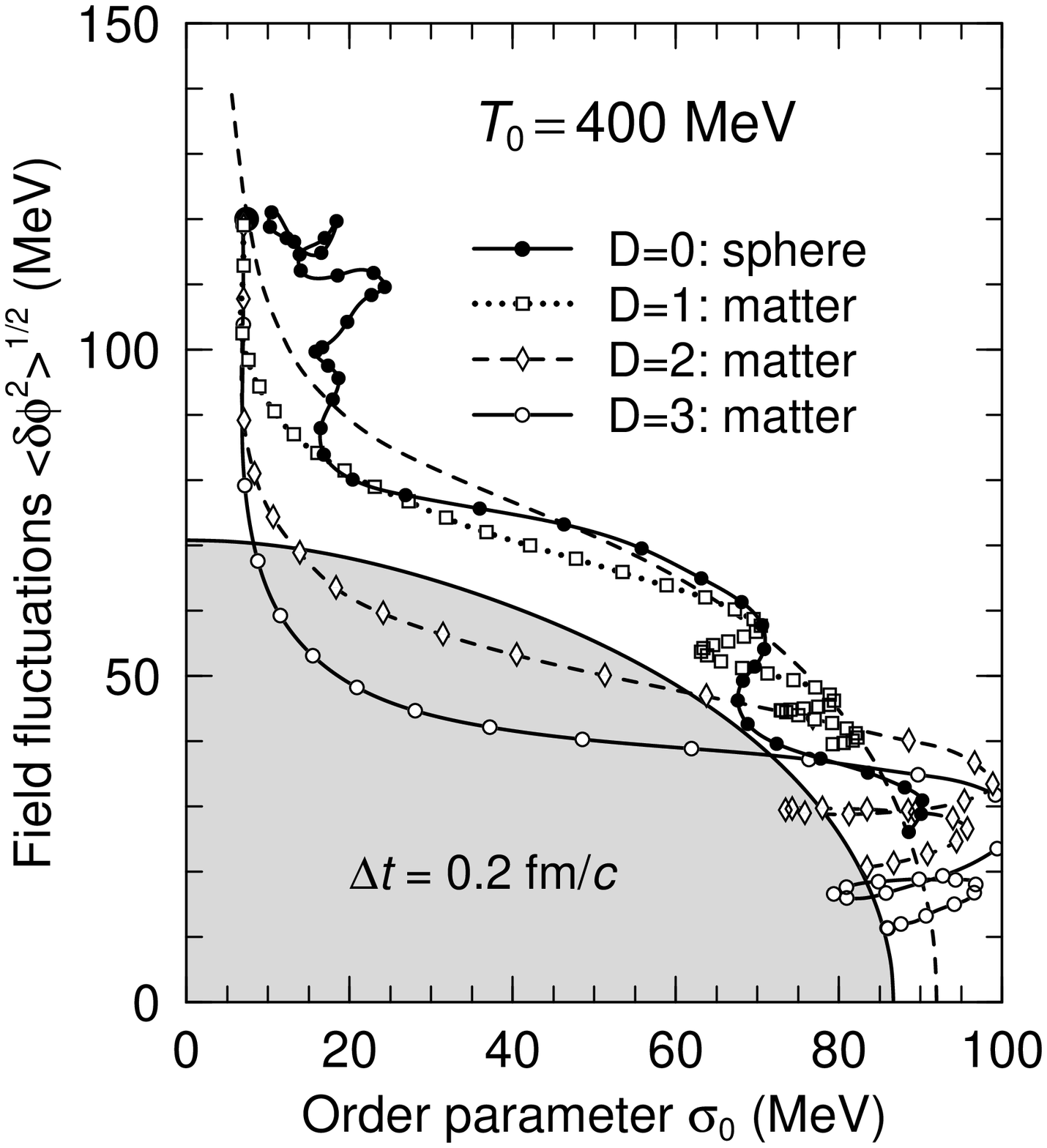,width=6.5cm}\hspace{.5cm}\epsfig{file=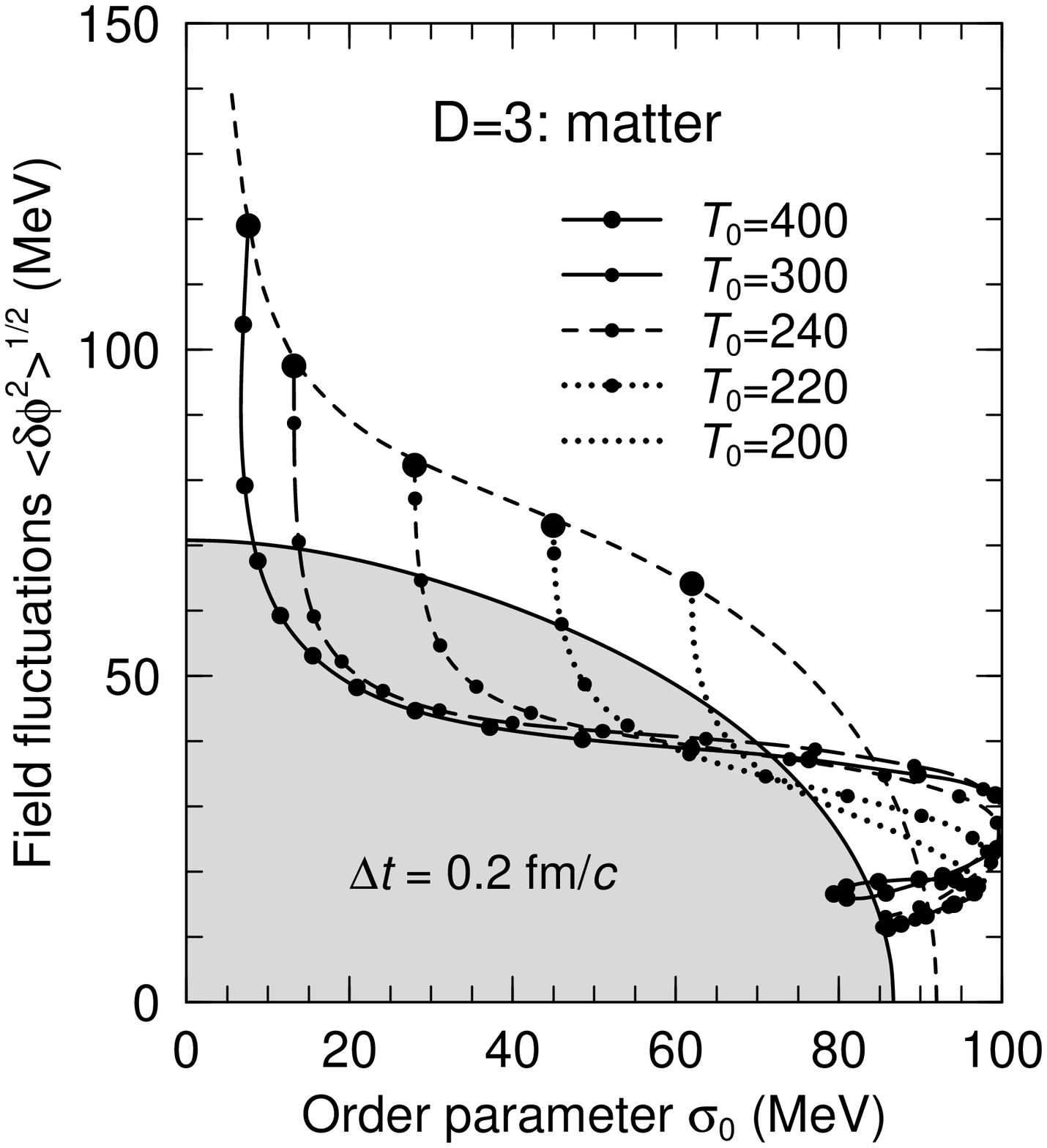,width=6.5cm}
\end{center}
\caption{\label{fig:cooling}
 \small Dynamical trajectories in the plane $\sigma_0$--$\bra\dvphi^2\ket^{1/2}$, where 
 $\sigma_0\equiv(\phi_a\phi_a)^{1/2}$, for various expansion
 scenarios (left) and various initial temperatures (right). The dashed line connects
 the equilibrium states from $T=0$ to above $500$~MeV and the shaded area corresponds
 to the instability region where the effective mass \eqn{effmass} is negative. Successive
 points on a given trajectory are separated by $\Delta t=0.2$~fm/c.
 From \cref{Randrup:1996ay}.}
\end{figure}

The result of the time evolution is most nicely illustrated by plotting the
time evolution of the order parameter $\phi^2(t)=\phi_a(t)\phi_a(t)$, where
$\phi_a(t)\equiv\bra\varphi_a(\bx,t)\ket$ is the ensemble averaged value
of the field at time $t$,\footnote{Assuming a spatially homogeneous ensemble,
the latter does not depend on $\bx$.} versus that of the square fluctuations 
$\bra\dvphi^2\ket(t)=\bra\varphi^2(\bx,t)\ket-\phi^2(t)$, where 
$\varphi^2\equiv\varphi_a\varphi_a$. The corresponding trajectories are shown
in \Fig{fig:cooling} for various initial temperatures and various types
of expansions, with $d\equiv D=1,2,3$. Due to the cooling term in the equations 
of motion, the initial thermal fluctuations are suppressed -- the higher the value
of $d$, the more efficient the cooling in \eqn{cooling} -- and might enter the 
instability region, where the effective mass squared \eqn{effmass} is negative,
represented by the shaded area in \Fig{fig:cooling}. 
When this happens, long-wavelength modes get amplified. In particular, the order
parameter -- which is nothing but the zero-mode -- grows rapidly and the system is 
pushed away from the instability region.
As expected, we see that the phenomenon is more pronounced for higher values of
$d$, corresponding to more rapid expansions. Also, it appears clearly that the 
time spent in the instability region, which eventually determines the amplitude 
of the amplification for low-momentum pion modes, strongly depends on the initial
temperature: For $T\lesssim 200$~MeV, the initial order parameter is too large 
and the system never enters the instability region. Similarly, one would expect 
that for very high initial temperatures $T\gg 500$~MeV,\footnote{Notice, however,
that the use of the sigma model cannot be justified at such high temperatures.} 
the initial fluctuations are too large for the system to ever become unstable
But one observes that there is a large range of temperatures, which roughly
correspond to those expected in high-energy nucleus-nucleus collisions, where
the expansion produces an efficient quenching of initial fluctuations.

However, a more detailed analysis of the results shown in \Fig{fig:cooling} reveals 
that, even for the most favorable case $d=3$, the time spent in the instability 
region is rather short and barely exceeds $1-1.5$~fm/c. This indicates that 
the typical trajectories do not get dramatically amplified, in contrast to what 
happens when the initial fluctuations are artificially quenched deep in the
instability region as discussed before. In fact, the ensemble average amplification 
factor for the most amplified mode $k=0$ was estimated in \cref{Randrup:1996ay} to 
be roughly of order $10$ in the most favorable case. This is very different from 
the large amplifications, typically $\sim 100$, obtained in the quench scenario. 
Thus, we can already conclude that, in a more realistic scenario with 
cooling, large amplifications seem to be a rather rare phenomenon. This conclusion 
has been corroborated by other studies (see \eg \cite{Biro:1997va}). A crucial question 
is, therefore, to know how likely these events are. We shall return to this later in 
this section (see subsection \ref{sec:likelihood}). We first discuss the issue of 
describing the quantum corrections to the dynamics in the next subsection.

\subsubsection{Quantum corrections to the dynamics}

A considerable amount of work has been devoted to include quantum fluctuations 
in the dynamical description of DCC formation. Initial-value problems in quantum 
field theory are, however, known to be of notorious difficulty.\footnote{For reviews 
concerning recent developments in this field, see \eg \cite{Berges:2003pc}.} 
Existing studies in the context of DCC physics have been mainly limited to 
the use of mean-field approximations, such as the so-called Hartree approximation, 
or the large-$N$ approximation. We describe these developments here, for 
the case of a spherically expanding geometry. We comment on recent progress in 
non-equilibrium quantum field theory beyond mean-field approximations at the end 
of this subsection.

\paragraph{Mean-field approximation}

In the present context, mean-field approximations consist in replacing the fully 
non-linear problem by an effective dynamics in a self-adjusting quadratic potential:
The system of interacting quantum fields is replaced by an ensemble of independent, 
self-consistently dressed quasi-particles. There are various ways one can formulate 
such approximations, such as Gaussian ans\"atze for the density matrix in the 
Schr\"odinger picture \cite{Boyanovsky:1994yk}, Gaussian-like factorization of 
correlation functions in the Heisenberg equations of motion 
\cite{Boyanovsky:1994yk,Randrup:1996ay}, non-perturbative resummation of so-called 
daisy and super-daisy diagrams \cite{Cooper:1996ii}, $1/N$-expansion at leading
order, where $N=4$ is the number of component of the chiral field 
\cite{Coleman:1974jh,Cooper:1994hr}, etc. 
Here, we employ the latter approach: the so-called large-$N$ approximation. 
We shall not enter, however, in the details of the $1/N$-expansion and
instead, we give a simple derivation of the equations relevant for
our purposes. 

The exact Heisenberg equations for the chiral quantum field $\varphi_a$ read 
(see \eqn{action}):
\beq
 \left[\square+\lambda(\varphi^2(x)-v^2)\right]\varphi_a(x)=H_a\,.
\eeq
As we did previously for the classical field equations (see 
Eqs.~\eqn{MFeom}-\eqn{effmass}), we replace
the non-linear term in parenthesis by its average $\varphi^2\to\bra\varphi^2\ket$, 
where, here, the brackets denote an average over the initial 
density matrix $\rho$, which encodes both statistical and quantum
fluctuations in the initial state: $\bra\cdots\ket\equiv\Tr\{\rho\,\cdots\}/\Tr\{\rho\}$. 
Writing the field $\varphi_a$ as a sum of its average value 
$\phi_a\equiv\bra\varphi_a\ket$ and a fluctuation field $\dvphi_a$, 
one gets:
\beq
\label{NN}
 \left[\square+\lambda(\phi^2(x)+\bra\dvphi^2(x)\ket-v^2)\right]
 (\phi_a(x)+\dvphi_a(x))=H_a\,.
\eeq
Taking the average of this equation with respect to the initial density 
matrix $\rho$, one obtain the following non-linear equation for the 
condensate $\phi_a$:
\beq
\label{fieldeom}
 \left[\square+M^2(x)\right]\phi_a(x)=H_a\,,
\eeq
where 
\beq
\label{MFmass}
 M^2(x)=\lambda\left[\phi^2(x)+\bra\dvphi^2(x)\ket-v^2\right]\,.
\eeq
We see that $\phi_a(x)$ evolves in what is essentially the classical 
potential $V(\phi^2)=\lambda(\phi^2-v^2)^2/4-H\sigma$, modified by quantum 
and statistical fluctuations through a term $\delta V\propto\bra\dvphi^2\ket\,\phi^2$.
Finally, subtracting Eqs.~\eqn{NN} and \eqn{fieldeom}, one obtains
the following Klein-Gordon--like equation for the fluctuation field:
\beq
\label{flucteom}
 \left[\square+M^2(x)\right]\dvphi_a(x)=0\,.
\eeq
The excitations of the fluctuation field $\dvphi_a$ describe 
self-consistently dressed quasi-particles evolving in a $x$-dependent 
quadratic potential, whose curvature depends on the local value of the 
condensate and on the mean effect of the fluctuations themselves.

Of course, $\bra\dvphi^2(x)\ket$ diverges and the above equations require
regularization, \eg through an ultra-violet cut-off $\Lambda$, and 
renormalization. The quadratic divergence in \Eqn{MFmass} can be argued 
to be $x$-independent by general power-counting arguments and is easily 
eliminated through the following vacuum subtraction \cite{Cooper:1987pt}:
\beq
\label{renmass}
 M^2(x)-\mpi^2=\lambda\left[\phi^2(x)-\fpi^2
 +\bra\dvphi^2\ket(x)-\bra0|\dvphi^2|0\ket\right]
\eeq
The remaining logarithmic divergence can be eliminated by introducing the 
renormalized coupling constant \cite{Coleman:1974jh,Cooper:1987pt}:
\beq
\label{rencoupling}
 \frac{1}{\lambda_R}=\frac{1}{\lambda}
 +\frac{N}{8\pi^2}\int_0^\Lambda\,\frac{k^2dk}{(k^2+\mpi^2)^{3\over2}}
\eeq
It is important to notice that the cancellation of the logarithmic 
divergence at any time requires a suitable choice of initial 
conditions (see Eqs.~\eqn{adiab0}-\eqn{adiab1} below): Not all initial conditions
are physically acceptable. Finally, it should be kept in mind that 
the present model is at best an effective theory for low-momentum
scales. The ultra-violet cut-off has therefore a physical meaning
and should be kept finite.\footnote{Moreover, it is well-known that 
the theory becomes trivial as $\Lambda\to\infty$ \cite{Luscher:1987ay}. 
The physical cut-off should be chosen smaller than the Landau pole 
$\Lambda_L$, which, in the present approximation, is roughly given by: 
$\Lambda_L/m_\pi\approx\exp(8\pi^2/N\lambda_R)$.} 
The aim of the renormalization procedure is to reduce the cut-off 
dependence of physical results.

\paragraph{\label{sec:expansion} Expanding geometry}

We describe the mean-field equations in more details in the case of a 
spherically expanding system, for which we essentially follow the treatment 
of \cref{Lampert:1996qw}. This requires one to quantify the theory on 
equal--proper-time hyper-surfaces $\tau={\rm const}$, see \Eqn{commutation} 
below,\footnote{For details concerning techniques of quantization in curved 
geometries, see \cite{Birell:1994}.} where the proper-time 
$\tau=\sqrt{t^2-r^2}$, with $r$ the radial distance to the origin, is 
the time measured by a co-moving observer sitting on an expanding shell.
Exploiting the spherical symmetry of the problem, the relevant 
coordinates are:
\beq
\label{hypercoord}
 \tau=\sqrt{t^2-r^2}\,\,,\,\,
 \eta=\frac{1}{2}\ln\left(\frac{t+r}{t-r}\right)\,\,,\,\,
 \hat\theta\,\,,\,\,\hat\varphi\,\,,
\eeq
where $\eta$ is the spatial radial rapidity, and $\hat\theta$ and $\hat\varphi$ the 
usual azimuthal and polar angles on the sphere.
With this choice of coordinates, the relativistic metric is of the
Friedrich-Robertson-Walker type and reads, explicitly:
\beq
\label{metrique}
 \dd s^2=g_{\mu \nu}\,\dd x^\mu\dd x^\nu=
 \dd\tau^2-\tau^2\left(h_{ij}\,\dd x^i\dd x^j\right)\,,
\eeq
where $h_{ij}$ is the metric on the $3$--hyperbolo\"{\i}d,
or pseudo-sphere, $\tau=1$: $h_{ij}={\rm diag}(1,\sinh^2\eta,
 \sinh^2\eta\sin^2\hat\theta)$. The four-dimensional Laplacian reads:
\beq
 \square\equiv\Delta^{(4)}
 =\p_\tau^2+\frac{3}{\tau}\p_\tau-\frac{1}{\tau^2} \Delta^{(3)}\,,
\eeq
where $\Delta^{(3)}$ 
is the corresponding three-dimensional Laplacian on the
unit pseu\-do-\-sphere:
\beq
\label{3dlap}
 \Delta^{(3)}=\frac{1}{\sqrt{h}}\,\p_i\,\sqrt{h}\,h^{ij}\,\p_j\,,
\eeq
with $\sqrt{h}=\sinh^2\eta\sin\hat\theta$ the determinant and  $h^{ij}$ the 
inverse of the three-dimensional metric $h_{ij}$.

Spherical symmetry and radial-boost invariance ensures that both the 
condensate $\phi_a$ and the effective mass $M^2$ only depend on the 
proper-time $\tau$ and \Eqn{fieldeom} reads:
\beq
 \ddot\phi_a(\tau)+\frac{3}{\tau}\,\dot\phi_a(\tau)+
 M^2(\tau)\,\phi_a(\tau)=H_a\,,
\eeq
The canonical quantization of the fluctuation field is more conveniently 
formulated in terms of the dimensionless rescaled fields (the various 
components of the fluctuation field $\dvphi_a$ being all equivalent in 
the present approximation, \Eqn{flucteom}, we drop the chiral indices for 
simplicity):\footnote{The rescaled fluctuation field introduced here should
not be confused with the original field $\varphi_a$. No confusion being
possible in the present section, we use the same notation to avoid a 
proliferation of symbols.}
\bea
 \varphi(\tau,\bx)&=&\tau\dvphi(\tau,\bx)\\
 \pi(\tau,\bx)&=&\tau\dot\varphi(\tau,\bx)=\varphi'(\tau,\bx)
\eea
where the dot represents the derivation with respect to $\tau$ and
the prime represents the derivation with respect to the conformal-time
variable $u=\ln\tau/\tau_0$. These operators satisfy the equal-time commutation 
relations:
\beq
\label{commutation}
 \left[\varphi(\tau,\bx),\pi(\tau,\bx')\right]=i\,\delta_{\bx,\bx'}\,,
\eeq
where we introduced the short-hand notations $\bx\equiv(\eta,\hat\theta,\hat\varphi)$ 
and $\delta_{\bx,\bx\,'}\equiv\delta(\eta-\eta')\delta(\hat\theta-\hat\theta')
\delta(\hat\varphi-\hat\varphi')/\sqrt{h}$.
Exploiting the symmetry of the problem, one projects the fluctuation field
$\varphi$ on eigenfunctions of the three-dimensional Laplacian 
\eqn{3dlap}:
\bea
 \varphi_\bs(\tau)&=&\int d^3x \sqrt{h}\,\cY_\bs^*(\bx)\,\varphi(\tau,\bx)\\
 \pi_\bs(\tau)&=&\int d^3x \sqrt{h}\,\cY_\bs^*(\bx)\,\pi(\tau,\bx)\,,
\eea
where the so-called hyperbolic harmonics are defined as \cite{Bander:1965im,Parker:1974qw}:
\beq
\label{eigenfunc}
 \Delta^{(3)}\,\cY_{\bs}(\bx)=-(s^2+1)\,\cY_{\bs}(\bx)\,.
\eeq
Here we introduced the notation $\bs\equiv(s,l,m)$, where $s$ is a positive 
real variable and $l$ and $m$ take integer values: $0\leq l<+\infty$ 
and $-l \leq m \leq l$. The hyperbolic harmonics form an orthonormal 
basis, spanning the space of functions on the unit pseudo-sphere. One 
has, in particular:
\beq
\label{ortho1}
 \int d^3x\,\sqrt{h}\,\,\cY_{\bs'}^*(\bx)\,
 \cY_{\bs}(\bx)=\delta_{\bs,\bs'}\,,
\eeq
where $\delta_{\bs,\bs'}\equiv\delta(s-s')\,\delta_{ll'}\,\delta_{mm'}$.

In the mean-field approximation, the equations of motion
\eqn{flucteom} for the fluctuation field are essentially linear
and the most general solution can, therefore, be written 
as:\footnote{Here, we have used the fact that the field 
$\varphi(\tau,\bx)$ is real, which implies, using $\cY_{\bs}^*(\bx)
=(-1)^m\,\cY_{-\bs}(\bx)$, with $-\bs\equiv(s,l,-m)$, that 
$\varphi_\bs^\dagger(\tau)=(-1)^m\varphi_{-\bs}(\tau)$.}
\bea
\label{phia}
 \varphi_\bs(\tau)&=&\psi_s(\tau)\,a_\bs
 +\psi^*_s(\tau)\,(-1)^m\,a^\dagger_{-\bs}\\
\label{pia}
 \pi_\bs(\tau)&=&\psi_s'(\tau)\,a_\bs
 +\psi^{*'}_s(\tau)\,(-1)^m\,a^\dagger_{-\bs}\,,
\eea
where, as before, the prime denotes differentiation with respect to 
$u=\ln\tau/\tau_0$. The annihilation and creation operators $a_\bs$ and 
$a^\dagger_{\bs}$ for quasi-particle excitations satisfy the following 
commutation relations:
\beq
 [a_{\bs},a^\dagger_{\bs'}]=\delta_{\bs,\bs'}\,,
\eeq
all other commutators being zero. The corresponding mode functions 
$\psi_s(\tau)$ satisfy the following differential equation:
\beq
\label{mode}
 \ddot\psi_s(\tau)+\frac{1}{\tau}\dot\psi_s(\tau)+
 \omega_s^2(\tau)\psi_s(\tau)=0\,,
\eeq
which describes a set of damped oscillators with 
time-dependent frequencies:
\beq
\label{frequency}
 \omega_s(\tau)=\sqrt{\frac{s^2}{\tau^2}+M^2(\tau)}\,.
\eeq
The physical momentum of the corresponding quasi-particle, $k=s/\tau$,
is red-shifted as time increases, as a simple consequence of expansion.
In terms of the variable $u=\ln\tau/\tau_0$, one has:
\beq
\label{OH}
 \psi_s''(\tau)+\tomega_s^2(\tau)\,\psi_s(\tau)=0\,,
\eeq 
where $\tomega_s(\tau)=\tau\omega_s(\tau)$ is a dimensionless frequency.
The problem therefore reduces to the familiar one of a set of parametrically 
excited oscillators.
The non-linear character of the mean-field dynamics enters through the 
calculation of the self consistent mass, \Eqn{MFmass}. For instance, 
assuming that in the initial state:
\beq
\label{initcorrel}
 \bra a_\bs^\dagger a_{\bs'}\ket=n_s^{(a)}\,\delta_{\bs,\bs'}
 \quad,\quad
 \bra a_\bs a_{\bs'}\ket=0\,,
\eeq
one has, using the property $\sum_{lm}\,|\cY_{\bs}(\bx)|^2=s^2/2\pi^2$
\cite{Bander:1965im}:
\beq
\label{MFmass2}
 \frac{M^2(\tau)}{\lambda}=\phi^2(\tau)-v^2+\frac{N}{\tau^2}
 \int_0^{\Lambda\tau}\frac{s^2ds}{2\pi^2}\,|\psi_s(\tau)|^2\,(2n_s+1)\,,
\eeq
where $N=4$ is the number of components of the chiral field. The integral 
on the RHS is limited to physical momenta $k=s/\tau\leq\Lambda$.
Finally, the quantization is completed by choosing appropriate
initial conditions for the mode functions. The cancellation
of divergences required by the renormalization of the mass
gap is ensured if one adopts the following adiabatic
condition at $\tau=\tau_0$:\footnote{With this choice, the 
solution of \Eqn{OH} for high momentum modes, 
$s^2/\tau^2\gg |M^2(\tau)|$ is approximately given by the 
second-order adiabatic function:
$$
 \psi_s(\tau)\sim\frac{1}{\sqrt{2\tomega_s(u)}}
 \exp\left\{-i\int_{0}^u du'\,\tomega_s(u')\right\}\,.
$$
where $u=\ln\tau/\tau_0$.
Using this solution, it is easy to check that the divergent part of 
the momentum integral in \Eqn{MFmass2} is exactly canceled at all 
times by the renormalization procedure described previously, 
Eqs.~\eqn{renmass}-\eqn{rencoupling}.}
\bea
\label{adiab0}
 \psi_s(\tau_0)&=&\frac{1}{\sqrt{2\tomega_s(\tau_0)}}\\
\label{adiab1}
 \psi_s'(\tau_0)&=&-i\sqrt{\frac{\tomega_s(\tau_0)}{2}}
 \left[1-i\frac{\tomega_s'(\tau_0)}{2\tomega_s^2(\tau_0)}\right]\,.
\eea

\paragraph{Physical particles, interpolating field and 
amplification factor}

Assume that Heisenberg and Schr\"odinger representations coincide
at $\tau=\tau_0$. With the choice \eqn{adiab0}-\eqn{adiab1} of initial 
mode functions, one can regard $a_\bs^\dagger$ as the
Schr\"odinger representation operator creating a particle with
frequency $\omega_s(\tau_0)$. These are the physical excitations 
appropriate for the description of the initial state. Similarly, 
one can introduce the particles with frequencies $\omega_s(\tau_f)$
appropriate to the final state at $\tau=\tau_f$. Let $b^{\dagger}_\bs$ 
denote the corresponding creation operator in the Schr\"odinger 
representation. The Heisenberg representation field operators, 
given by \eqn{phia}-\eqn{pia}, can then also be written as:
\bea
\label{phib}
 \varphi_\bs(\tau)&=&q_s\,b_\bs(\tau)
 +q^*_s\,(-1)^m\,b^\dagger_{-\bs}(\tau)\\
\label{pibb}
 \pi_\bs(\tau)&=&p_s\,b_\bs(\tau)
 +p^*_s\,(-1)^m\,b^{\dagger}_{-\bs}(\tau)\,,
\eea
with 
\bea
\label{adiabf0}
 q_s&=&\frac{1}{\sqrt{2\tomega_s(\tau_f)}}\\
\label{adiabf1}
 p_s&=&-i\sqrt{\frac{\tomega_s(\tau_f)}{2}}
 \left[1-i\frac{\tomega_s'(\tau_f)}{2\tomega_s^2(\tau_f)}\right]\,,
\eea
and
\beq
 b_\bs(\tau)=U(\tau,\tau_0)\,b_\bs\,U^\dagger(\tau,\tau_0)\,,
\eeq
where $U(\tau,\tau_0)$ denotes the unitary time-evolution operator,
which connects the Heisenberg and Schr\"odinger representations. 
Using \eqn{phia}-\eqn{pia} and \eqn{phib}-\eqn{pibb}, one easily 
obtains the Bogolyubov transformation connecting the operators 
$a_\bs$, $a^\dagger_\bs$ and $b_\bs(\tau)$, $b^\dagger_\bs(\tau)$:
\bea
 b_\bs(\tau)&=&\alpha_s(\tau)\,a_\bs
 +\beta_s(\tau)\,(-1)^m\,a^{\dagger}_{-\bs}\\
 (-1)^m\,b^\dagger_{-\bs}(\tau)&=&\beta^*_s(\tau)\,a_\bs
 +\alpha^*_s(\tau)\,(-1)^m\,a^{\dagger}_{-\bs}\,,
\eea
where
\bea
 i\alpha^*_s(\tau)&=&q_s\,\psi^{*'}_s(\tau)-p_s\,\psi^*_s(\tau)\\
 i\beta^*_s(\tau)&=&q_s\,\psi_s'(\tau)-p_s\,\psi_s(\tau)\,.
\eea
\begin{figure}[t]
\begin{center}
 \epsfig{file=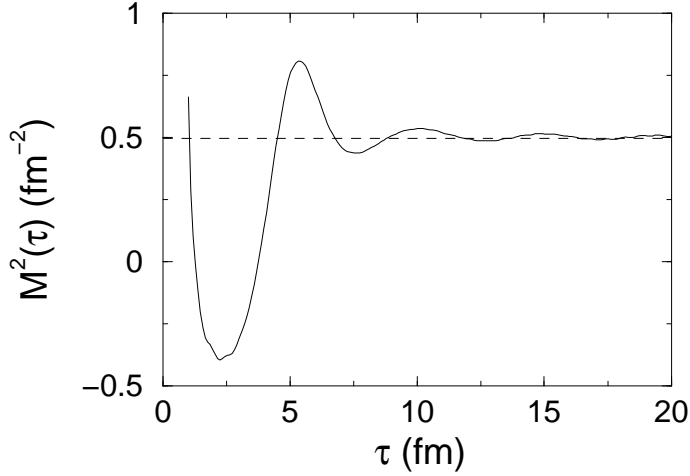,width=9.cm}
\end{center}
\caption{\label{fig:chi}
 \small Proper-time evolution of the effective mass squared 
 $M^2(\tau)$ for the initial condition of \cref{Lampert:1996qw}, 
 namely: $\phi_a(\tau_0) = (0.3,{\vec 0})$ [fm$^{-1}$] and 
 $\dot\phi_a(\tau_0) = (-1,{\vec 0})$ [fm$^{-2}$], at $\tau_0=1$~fm.
 The initial temperature is $T=200$~MeV.
 The employed parameters are: $\Lambda=800$~MeV, $m_\pi=139.5$~MeV,
 $f_\pi=92.5$~MeV and $\lambda_R=7.3$. The dashed line shows the value
 of the pion mass squared $m_\pi^2=0.5$~fm$^{-2}$.}
\end{figure}
\begin{figure}[t]
\begin{center}
 \epsfig{file=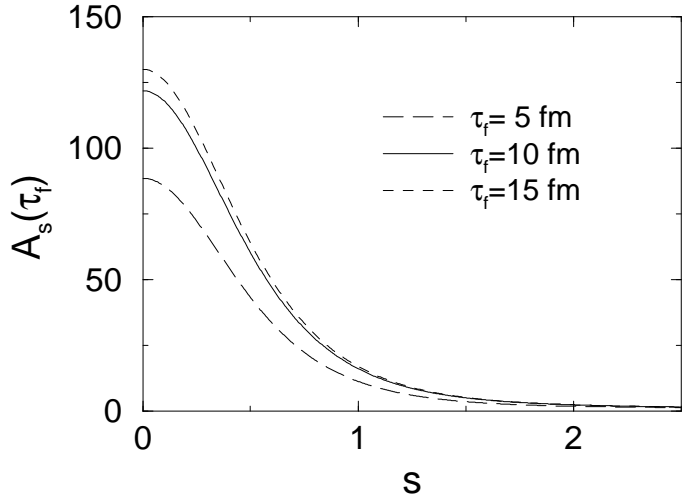,width=9.cm}
\end{center}
\caption{\label{fig:amplif}
 \small The amplification factor ${\mathcal A}_s (\tau_f)$
 corresponding to the event of \Fig{fig:chi}, for various
 values of the final time $\tau_f$.}
\end{figure}
Assuming that the initial state is characterized by the correlations
\eqn{initcorrel}, one finally obtains the number of $b$-particles at 
the final time $\tau_f$ for a given number of $a$-particles at the 
initial time $\tau_0$ as:\footnote{The derivation given here differs from 
that of \cref{Lampert:1996qw}, where the so-called adiabatic basis is 
unnecessarily used \cite{Krzywicki:1998sc}.}
\beq
\label{finalmult}
 n^{(b)}_s(\tau_f)+\frac{1}{2}
 ={\mathcal A}_s(\tau_f)\left(n^{(a)}_s+\frac{1}{2}\right)\,,
\eeq
where $n^{(b)}_s(\tau)\equiv\bra b_\bs^\dagger(\tau)b_\bs(\tau)\ket$. The 
amplification factor is given by:
\beq
\label{Amplif}
 {\mathcal A}_s(\tau_f)=1+2\,|\beta_s(\tau_f)|^2\,.
\eeq
Figure \ref{fig:chi} shows the time evolution of the effective mass
squared $M^2(\tau)$ for a particular set of initial conditions, taken
from \cref{Lampert:1996qw}. Here, one assumes a local thermal 
equilibrium at temperature $T$ in the initial state and sets, accordingly:
\beq
 n^{(a)}_s=\frac{1}{\e^{\omega_s(\tau_0)/T}-1}\,.
\eeq
One sees that, for the considered initial conditions (see the caption of
\Fig{fig:chi}), the system is driven into the unstable region and the 
effective mass squared becomes negative for some period of time. This 
leads to a dramatic amplification of low-momentum modes, as shown in
\Fig{fig:amplif}. Notice that due to the freeze-out of interactions at 
large enough $\tau_f$, one has $M^2(\tau_f)\simeq\mpi^2$ and the particles 
in the final state indeed correspond to physical pions. These results 
are the quantum analog of those obtained previously using classical 
statistical field theory techniques and demonstrates that significant
amplifications can be obtained for some appropriate choice of initial 
conditions $(\phi_a(\tau_0),\dot\phi_a(\tau_0))$.\footnote{\label{fn:initcond}
In the present set-up, the initial conditions are completely determined
by the choice of the eight real numbers $(\phi_a(\tau_0),\dot\phi_a(\tau_0))$:
From these, one computes the initial mass squared $M^2(\tau_0)$ and its
time derivative $(dM^2/d\tau)(\tau_0)$, which are then used to initialize 
the mode functions $(\psi_s(\tau_0),\dot\psi_s(\tau_0))$ through Eqs. 
\eqn{adiab0} and \eqn{adiab1}.} 

\paragraph{Non-equilibrium dynamics beyond the mean-field approximation}

Mean-field approximations, such as the large-$N$ approximation employed above, 
neglect scatterings between quasi-particles. The latter tend to redistribute 
the energy stored in the amplified modes among other modes and to thermalize the 
system, hence washing out any signal in the final state.\footnote{This effect is 
visible on \Fig{fig:quench} at late times. In 
contrast to mean-field approximations, the classical field approximation employed 
there treats the full non-linear character of the problem exactly. Notice, however, 
the processes of classical versus quantum thermalization are very different and, 
in particular, they are characterized by different time-scales (see \eg 
\cref{Aarts:2001yn}).} It is a crucial issue for the quench scenario to be 
successful, that freeze-out occurs before the early-time amplification get 
washed out.

The description of the far-from-equilibrium dynamics of quantum fields beyond 
mean-field--like approximations has long been a major difficulty in practice:
Similar to perturbation theory, standard approximation schemes, such as \eg a
coupling or a $1/N$-expansion of the one-particle-irreducible (1PI) effective 
action, are plagued by the problem that a secular (unbounded) time evolution 
prevents the study of the late-time behavior. Important progress in this field 
has been made in recent years with the use of efficient functional integration 
techniques, so-called $n$-particle-irreducible (nPI) effective actions 
\cite{Cornwall:1974vz,Norton:1974bm,DeDominicis:1964}, for which 
practicable non-perturbative expansion schemes are available. In particular, it 
has been demonstrated that approximations based on a systematic coupling or 
$1/N$-expansion of the 2PI effective action allow for practical, first principle 
calculations of the far-from-equilibrium dynamics as well as late-time quantum 
thermalization \cite{Berges:2000ur}.\footnote{Recent reviews on this topics as 
well as relevant literature can be found in \cite{Berges:2003pc,Berges:2004yj}.}

Clearly, it would be of definite interest to apply these methods to the problem 
of DCC formation. For instance, the large-$N$ calculation described in the previous 
subsection can be systematically improved using the 2PI $1/N$-expansion 
\cite{Berges:2001fi,Aarts:2002dj}. In particular, this provides a valid description 
of the dynamics at non-perturbatively high occupation numbers \cite{Berges:2002cz}, 
as is the case in the context of the out-of-equilibrium chiral phase transition.

\subsubsection{\label{sec:likelihood} The likelihood of DCC formation}

The quench scenario provides a plausible mechanism for DCC formation in high-energy 
heavy-ion collisions. A proper description of the cooling due to expansion reveals,
however, that significantly large amplifications are not very likely.
It is therefore crucial for phenomenology to estimate the probability 
that this happens. It is true that the present models are not realistic 
enough to be trusted at the quantitative level. Nevertheless, even within 
the existing framework it is legitimate to seek for a crude estimate of 
the probability in question. This problem has been addressed in 
\cref{Krzywicki:1998sc} (see also \cite{Biro:1997va}), which we review in 
the present subsection.

\paragraph{Physical picture and sampling strategy}

Consider the evolution of a spherical droplet of DCC in its rest frame. 
One starts with a small ball of radius $R_0$, filled with hot matter in local 
thermal equilibrium at a temperature $T$ and assumes that the ball undergoes 
a radial expansion at the speed of light. Due to the time dilation the 
temperature stays approximately constant within a layer near the boundary 
of the ball. The equations of motion of the sigma model are supposed to 
describe what happens in the interior of the ball, the cooling observed 
as one moves away from the surface towards the center. As a first approximation,
it is simpler to assume that the expanding bubble is connected forever to 
a heat bath kept at constant temperature $T$, so that the process never 
stops. Of course, the hot shell becomes thinner with increasing time and 
eventually disappears. We shall discuss the effect of switching the heat 
bath off at the end of this subsection.
The relevant boundary conditions for the pion field inside the bubble are 
specified on the hyper-surface $\tau=\tau_0=R_0$, where the proper-time 
$\tau=\sqrt{t^2-r^2}$, 
with $r$ the radial coordinate. A given field configuration completely 
determines the amplitude of the resulting pionic amplification. The
point is that the probability of occurrence of a particular initial field 
configuration can be completely determined once one has assumed that the 
initial droplet is in local thermal equilibrium. 

The quantum evolution of the system is treated in the large-$N$ approximation 
discussed previously. In that case, one can convince 
oneself (see also footnote \ref{fn:initcond} above) that the only relevant 
initial conditions are those concerning the average-value of the field 
$\phi_a(\tau_0)$ and its proper-time derivative $\dot\phi_a(\tau_0)$. 
For the sake of the argument, consider first the case of a classical 
field. The key observation is to realize that an observer living within the 
ball cannot distinguish the initial values $\phi_a(\tau_0)$ and $\dot\phi_a(\tau_0)$ 
from the spatial averages $\bar{\phi}_a$ and $\bar{\pi}_a$ of the field and of 
its time derivative respectively, calculated over the volume $V_0$ of the initial 
ball. The point is that the latter variables fluctuate 
in a predictable manner: Assuming that the initial ball, of fixed volume $V_0$, 
is in local thermal equilibrium at temperature $T$ means that the field 
fluctuates as if the ball were part of a larger system, with volume $V$ much 
larger than $V_0$, in equilibrium at the same temperature $T$. In this large 
system the variances of the spatial averages of the field and of its time 
derivative are very small, of order $1/V$ (since long range correlations 
are absent). The corresponding variances for spatial averages over the volume 
of the ball are just the same, enhanced by a factor of order $V/V_0$ (assuming 
that the radius of the ball is at least of the order of the correlation length 
at temperature $T$). 

In the quantum theory, the $c$-numbers $\bar{\phi}_a$ and $\bar{\pi}_a$ 
correspond to the possible measured values of the observables:
\bea
 \bar\Phi_a&=&\frac{1}{V_0}\int_{V_0}d^3x\,\varphi_a(\bx)\\
 \bar{\Pi}_a&=&\frac{1}{V_0}\int_{V_0}d^3x\,\dot\varphi_a(\bx)\,,
\eea
They can be sampled from the probability distribution characterized 
by the means:
\bea
\label{moyphi}
 {\rm E}[\bar\phi_a]&=&
 \bra\bar\Phi_a\ket_\T=\frac{H}{M^2_T}\,\delta_{a,\sigma}\,,\\
\label{moyphidot}
 {\rm E}[\bar{\pi}_a]&=&\bra\bar{\Pi}_a\ket_\T=0\,,
\eea
and by the variances (no sum over $a$):
\beq
\label{varphi}
 {\rm Var}\left[\bar\phi_a\right]=
 \bra\bar\Phi_a^2\ket_\T-\bra\bar\Phi_a\ket_\T^2
 =\frac{1}{V_0^2}\,\int_{V_0}d^3xd^3y\,
 \bra\dvphi_a(\bx)\dvphi_a(\by)\ket_\T
\eeq
and 
\beq
\label{varphidot}
 {\rm Var}\left[\bar{\pi}_a\right]=
 \bra\bar{\Pi}_a^2\ket_\T-\bra\bar{\Pi}_a\ket_\T^2
 =\frac{1}{V_0^2}\,\int_{V_0}d^3xd^3y\,
 \bra\delta\dot\varphi_a(\bx)\delta\dot\varphi_a(\by)\ket_\T\,,
\eeq
where thermal averages $\bra\dots\ket_\T$ correspond to the equilibrium density matrix 
$\rho\propto\exp(-H/T)$,
with $H$ the Hamiltonian. The thermal correlators in Eqs.~\eqn{varphi} and 
\eqn{varphidot} can be written as:
\beq
 \bra\dvphi_a(\bx)\dvphi_a(\by)\ket_\T=
 \int\frac{d^3 k}{(2\pi)^3}\,G^a_T(k)\,\e^{i\,\bk\cdot(\bx-\by)}\,,\nn
\eeq
and similarly for $\bra\delta\dot\varphi_a(\bx)\delta\dot\varphi_a(\by)\ket_\T$ 
with $G^a_T(k)\to K^a_T(k)$. In the large-$N$ approximation, one has: 
$G^a_T(k)=\coth(E_k/2T)/2E_k$ and $K^a_T(k)=E_k^2G^a_T(k)$, where $E_k=\sqrt{k^2+M_T^2}$.
Here, the equilibrium mass-gap at temperature $T$, $M_T$, is the solution of the 
gap-equation \eqn{MFmass} written in thermal equilibrium:
\beq
\label{gapthermik}
 \frac{M^2_T}{\lambda}=\phi_T^2-v^2+N\int\dk\,\frac{\coth(E_k/2T)}{2E_k}\,,
\eeq
where $\phi_T=H/M^2_T$ is the thermal condensate (cf. \Eqn{moyphi}).

In the large-$N$ approximation the interacting system is effectively 
replaced by an ensemble of independent excitations with mass $M_T$ and the 
probability distribution we are interested in is, therefore, simply given by a 
Gaussian. The latter is completely characterized by the parameters given in
\eqn{moyphi}-\eqn{varphidot}. The variances \eqn{varphi}-\eqn{varphidot} 
can be estimated analytically when the radius $R_0$ of the ball
is much larger than the correlation length $\lambda_T = 1/M_T$. 
In that case, it is sufficient to calculate the variances for the quasi-infinite 
volume $V$ and to multiply the result by $V/V_0$, since the $V/V_0$ small 
cells fluctuate independently. Obviously, only the modes $k\lesssim1/R_0$ 
contribute to the integrals in Eqs.~\eqn{varphi}-\eqn{varphidot}. Therefore,
for large volumes, there is only the zero mode contribution and one immediately 
gets:
\bea
\label{varphiapprox}
 {\rm Var}\left[\bar\phi_a\right]&\approx&\frac{G^a_T(0)}{V_0}=
 \frac{1}{2V_0 M_T}\,\coth\left(\frac{M_T}{2T}\right)\,,\\
\label{varphidotapprox}
 {\rm Var}\left[\bar{\pi}_a\right]&\approx&\frac{K^a_T(0)}{V_0}=
 \frac{M_T}{2V_0}\,\coth\left(\frac{M_T}{2T}\right)\,.
\eea
The dispersion of $\bar{\phi}_a$ calculated exactly
\cite{Serreau:2001bu} is smaller by a factor of $2$ ($1.5$) 
for $R_0=\lambda_T$ ($R_0=2\lambda_T$) and $T=200$ to $400$~MeV. 
The dispersion of $\bar\pi_a$ obtained from \eqn{varphidot} 
differs from the estimation \eqn{varphidotapprox} by 20\% (8\%), 
respectively. For $R_0 < \lambda_T$ the discrepancy between the 
analytical formulas and the exact results increases rapidly \cite{Serreau:2001bu}. 
Finally, as expected, the fluctuations within the ball depend rather 
weakly on the environment provided $R_0\gtrsim\lambda_T$.

The formalism of \cref{Lampert:1996qw}, reviewed in subsection 
\ref{sec:expansion} above, together with the sampling method proposed in
\cref{Krzywicki:1998sc}, and described above, enable one to estimate the 
likelihood of a coherent amplification of the pion field. More precisely, 
one can calculate the probability 
that the amplification factor $A_s$ given by \Eqn{Amplif} takes a 
given value. In such a calculation the size $R_0$ of the initial ball
appears as a free parameter. Remember, however, that one has to set 
$\tau_0=R_0$ and that the friction force responsible for the quench is 
proportional to $1/\tau$.  Thus the likelihood of DCC formation decreases
with increasing $R_0$. This parameter should be assigned the smallest 
possible value in order to get an upper bound for the probability we 
are looking for.

At this point, it is important to realize that the model described 
here only makes sense for $R_0\gtrsim\lambda_T$. Indeed, the concept 
of local thermal equilibrium is meaningful for a cell whose degrees 
of freedom fluctuate more or less independently from what happens in 
the neighbouring cells. Also, the validity of the mean-field approximation 
requires the size of the cell to be larger than the Compton wavelength of 
a typical excitation. With these arguments in mind, one focuses 
on the values of $R_0$ in the range of one or two correlation length.

\paragraph{Results}

\begin{figure}[t]
\begin{center}
 \epsfig{file=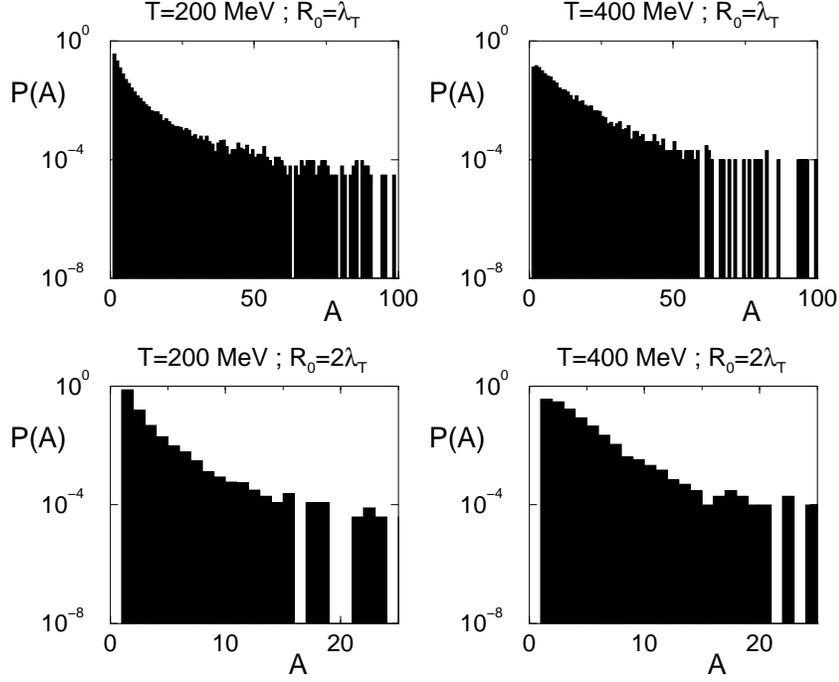,width=11.cm}
\end{center}
\caption{\label{fig:proba}
 \small Probability distribution of the amplification factor $A$ 
 of the softest mode for various choices of the initial temperature 
 $T$ and size of the ball $R_0$ (in units of the correlation length). 
 All histogram are made from at least $10^4$ events. 
 From Ref.~\cite{Krzywicki:1998sc}.}
\end{figure}
\begin{figure}[t]
\begin{center}
 \epsfig{file=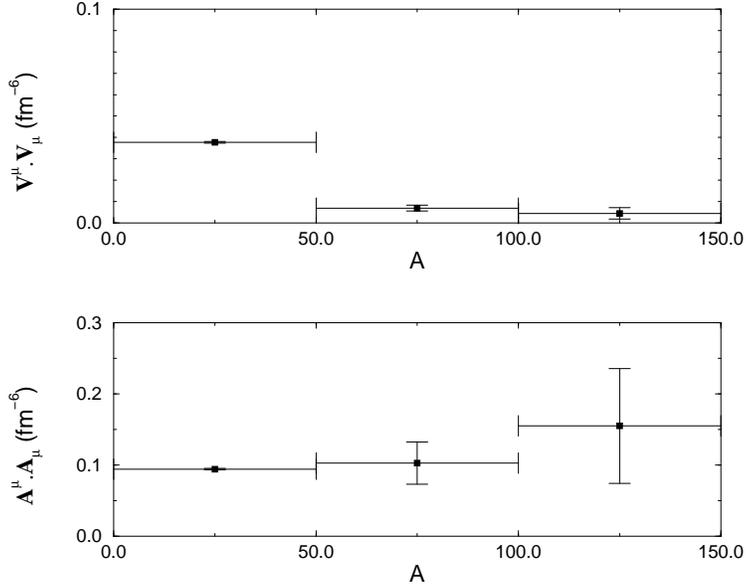,width=8.cm}
\end{center}
\caption{\label{fig:currents}
 \small The strengths of the initial classical iso-vector and
 iso-axial-vector currents, $\vV_\mu \cdot \vV^\mu$ and 
 $\vA_\mu \cdot \vA^\mu$ respectively (in fm$^{-6}$),  
 versus the amplification factor $A$. The parameters are 
 $R_0=\lambda_T$ and $T=200$ MeV. From Ref.~\cite{Krzywicki:1998sc}.}
\end{figure}

The main result of \cref{Krzywicki:1998sc} is reproduced in \Fig{fig:proba},
which shows the histograms of the probability $P(A)$ that the amplification 
factor of the $s=0^+$ mode takes the value $A_0=A$. The relevant parameters 
are the same as in \cref{Lampert:1996qw}, see Figs.~\ref{fig:chi} and \ref{fig:amplif}. 
The amplification is calculated at $\tau_f=10$~fm/c, where the system is 
in the stationary regime. Clearly, large amplifications occurs in a small 
fraction of events only. It is interesting to notice that one condition for
a large amplification appears to be the smallness of the absolute strength 
of the initial classical iso-vector current $\vV_{\mu}=\vpi_0\times\p_{\mu}\vpi_0$ 
as compared to that of the corresponding axial current $\vA_{\mu}=\vpi_0\p_{\mu}\sigma_0
-\sigma_0\p_{\mu}\vpi_0$, where $(\vpi_0,\sigma_0)\equiv\phi_a(\tau_0)$ and 
$(\dot\vpi_0,\dot\sigma_0)\equiv\dot\phi_a(\tau_0)$.
This is illustrated by the histograms of \Fig{fig:currents}, which 
shows the the respective currents versus the amplification factor 
$A$. Recall that a similar condition was required in order that the classical 
solution of Blaizot and Krzywicki, described in \Sec{sec:DCC}, actually corresponds 
to a DCC, that is to a linearly polarized in isospin space.

In order to judge what amplification should be considered as large
for phenomenological purposes, one can estimate the multiplicity of 
produced pions. The invariant one-pion inclusive spectrum can be obtained 
as \cite{Cooper:1974mv}:
\beq
\label{cf}
 E\frac{dn}{d^3p}=\int d^4x\sqrt{-g}\,\delta(\tau-\tau_f)\,
 f(x,p)\,p^\mu u_\mu
\eeq
where $g$ is the determinant of the metric tensor, see \eqn{metrique}, $f(x,p)$ is 
the invariant phase-space density of produced particles and $u^\mu=dx^\mu/d\tau=
x^\mu/\tau$ is the $4$-velocity of the co-moving volume centered at $x^\mu$. 
It is a unit $4$-vector orthogonal to the hyper-surface $\tau=\tau_f$, 
where the particles are counted. In the present case, one has 
\cite{Krzywicki:1998sc}: $f(x,p)=N(2\pi)^{-3}n^{(b)}_s(\tau_f)$, 
where $n^{(b)}_s(\tau_f)$ is given by \Eqn{finalmult} above. The 
momentum $s/\tau_f$ in the expanding frame is related to the $4$-momentum 
$p^\mu$ in the rest frame by the obvious relation \cite{Lampert:1996qw} 
(see \Eqn{frequency}):
\beq
 p^\mu u_\mu=\omega_s(\tau_f)=\sqrt{\frac{s^2}{\tau_f^2}+M^2_f}\,,
\eeq
where $M^2_f=M^2(\tau_f)\approx \mpi^2$ is the mass gap at time $\tau_f$.
In the present case, the integrand in \eqn{cf} depends on a single external 
$4$-vector, \ie 
$p^\mu$, and therefore the integral, being a Lorentz scalar, only depends 
on $p^2=M_f^2$: The invariant spectrum is flat. Using hyperbolic
coordinates, see \Eqn{hypercoord}, one finally obtains \cite{Serreau:2001bu}:
\beq
\label{spectrum2}
 E\frac{dn}{d^3p}\,(\bp)=E\frac{dn}{d^3p}\,(\bp=0)=
 \frac{N}{M^2_f}\int_0^{\Lambda\tau_f}\frac{s^2ds}{2\pi^2}\,
 n^{(b)}_s(\tau_f)\,.
\eeq
Notice that the RHS of \Eqn{spectrum2} does not depend on 
the choice of $\tau_f$, provided the latter is large enough (see also 
\Fig{fig:amplif}).

The flatness of the spectrum is, of course, an artifact of the 
unrealistic assumption that the boost invariant expansion continues 
forever. In a real collision process the expansion would last a finite
time and the resulting spectrum would be cut, the value of the cut-off  
reflecting the behavior of the environment. Obviously, the predicted 
total multiplicity depends strongly on this cut-off and cannot be 
estimated in a reliable manner within the present model. However, the 
RHS of \eqn{spectrum2} is presumably a reasonable estimate of the invariant 
momentum-space density of soft pion radiation (see, however, 
\cite{Amelino-Camelia:1997in}). The latter can be compared to the
corresponding density of incoherently produced pions.
A simple example is instructive: The one-particle spectrum in the central 
rapidity region of a heavy-ion collision can be parametrized as follows:
\beq
 \left.E\frac{dn}{d^3p}\right|_{inc}=
 \left.\frac{dn}{dyd^2p_t}\right|_{inc}=
 \frac{h}{\pi\langle p_t^2\rangle} 
 \exp\left(-\frac{p_t^2}{\langle p_t^2\rangle}\right)\,,
\eeq
where $h$ denotes the height of the rapidity plateau.
In a central Pb-Pb collision at the CERN SPS one observes \cite{Appelshauser:1998vn} 
about 200 $\pi^-$ per unit rapidity, \ie for all pions $h\approx 600$.
Thus the invariant momentum space density at very small transverse
momentum is roughly $1900$~GeV$^{-2}$, using $\langle p_t^2\rangle=0.1$~GeV$^2$.
The corresponding density fluctuation is expected to be 
$\sim \sqrt h/\pi\bra p_t^2\ket \approx 75$~GeV$^{-2}$. For the DCC 
signal to be detectable, the RHS of \eqn{spectrum2} should be significantly 
larger than the expected fluctuation. One finds \cite{Krzywicki:1998sc} that 
the signal is more than three times the fluctuation for events where 
the amplification factor for the zero mode is $A\gtrsim 45$. Setting 
$R_0=\lambda_T$, the corresponding probability is roughly $4\times 10^{-3}$ 
both for $T=200$~MeV and $T=400$~MeV.
Of course, this is a conditional probability, as it has been assumed 
that the initial plasma droplet was formed in the collision. Moreover,
one should keep in mind that the present estimate has been obtained in the 
most favorable scenario and should, therefore, be considered as an upper-bound. 

Thus, the probability of a potentially observable DCC signal appears 
small. The present crude estimate indicates that in a central Pb-Pb 
collision at CERN SPS this probability is at best:
\beq
 {\rm Probability}(\mbox{``observable'' DCC})\lesssim 10^{-3}\,.
\eeq 
This should be taken into account by experimenters designing DCC hunt 
strategies as well as theorists studying possible DCC signatures. It is
worth emphasizing that the above prediction agrees with the present experimental 
upper limits for DCC formation obtained at CERN by both the WA98 and the NA49 
collaborations \cite{wa98_global,na49,wa98_local_cen}.\footnote{We stress that 
the present estimate is pertinent to the CERN SPS experimental conditions and would
certainly lead to a different result for RHIC or LHC energies. Naively, one would 
expect that the likelihood of a potentially observable DCC fluctuation decreases 
at higher collision energies, where both the initial temperature and the average 
multiplicity of normal pions are increased compared to the present case.}

\subsubsection{\label{sec:unpolarized} The unpolarized DCC}

The quench scenario has been widely accepted as a microscopic description of DCC 
formation in heavy-ion collisions and, as such, has been extensively used to study 
DCC phenomenology~\cite{Randrup:1996es,Rajagopal:1997au,Petersen:1999jc}, see \Sec{sec:pheno}. 
However, if the rapid suppression of initial fluctuations indeed provides a 
robust mechanism for the formation of a strong coherent pion field, it is not 
clear whether it can explain the hypothetical polarization in iso-space 
originally proposed (see \eg \Eqn{dccpol}). In fact, the expected large 
event-by-event fluctuations of the neutral ratio, see \Eqn{DCCsign}, have 
never been observed in actual simulations of the out-of-equilibrium phase 
transition~\cite{Gavin:1993bs,Randrup:1996es,Rajagopal:1997au}. 
Of course, deviations from the ideal law \eqn{DCCsign} are to be expected 
in a realistic situation, but the complete absence of large fluctuations 
in the quench scenario suggests that the original picture of linearly polarized 
waves coherently aligned in isospin space might actually not be realized in the 
context of the far-from-equilibrium chiral phase transition.

\begin{figure}[t]
\begin{center}
 \epsfig{file=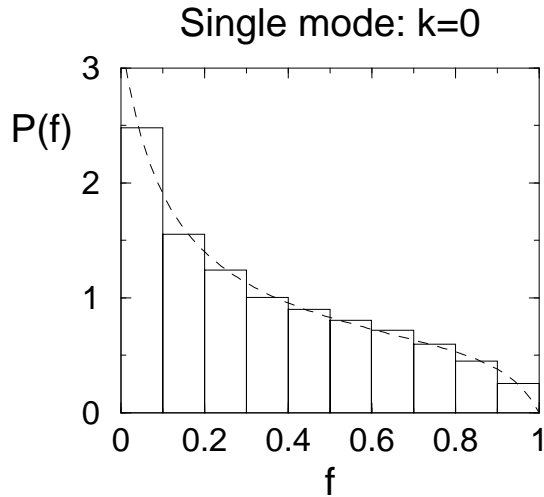,width=8.cm}
\end{center}
\caption{\label{fig:mode}\small Neutral fraction distribution in
 a single mode, here $\bk=0$, at initial time (dashed line) and 
 at final time, here $t_f=10$~fm (histogram). From \cref{Serreau:2000tb}.}
\end{figure}

It is a question of great phenomenological relevance to inquire to what extent 
is the original picture realized in a realistic microscopic model.
The point is that the usual assumption of a (locally) thermalized 
initial state implies that the field modes are completely uncorrelated at 
initial time. In order to actually generate a DCC configuration, the 
microscopic mechanism at work needs not only to be efficient in amplifying 
the modes amplitudes, but it should also build correlations between amplified 
modes. This question has been investigated in \cref{Serreau:2000tb}, where a 
detailed statistical analysis of the pion field configurations produced after 
a quench has been performed, with particular emphasis on their isospin structure. 
For this purpose, it is sufficient to consider the original model of Rajagopal 
and Wilczek \cite{Rajagopal:1993ah}, see subsection \ref{sec:quench}, which 
includes all the relevant physical features. In particular, the classical field 
approximation allows one to take into account the full non-linear character 
of the problem, which is crucial to study correlations. Moreover, in the quench 
approximation, the system is artificially prepared in an unstable situation, all 
configurations undergo a dramatic amplification, as in \Fig{fig:quench}. 
This automatically selects the events which are relevant for the present
analysis.

\begin{figure}[t]
\begin{center}
 \epsfig{file=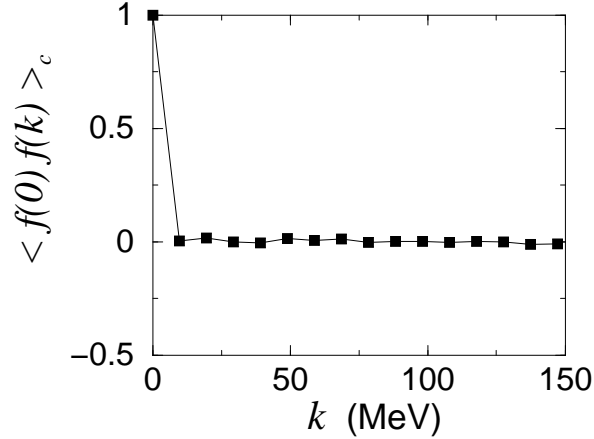,width=8.cm}
\end{center}
\caption{\label{fig:correl}\small Normalized correlation 
 between the fluctuations of the neutral fraction in the zero 
 mode and in mode k as a function of momentum. From \cref{Serreau:2000tb}.}
\end{figure}

\begin{figure}[t]
\begin{center}
 \epsfig{file=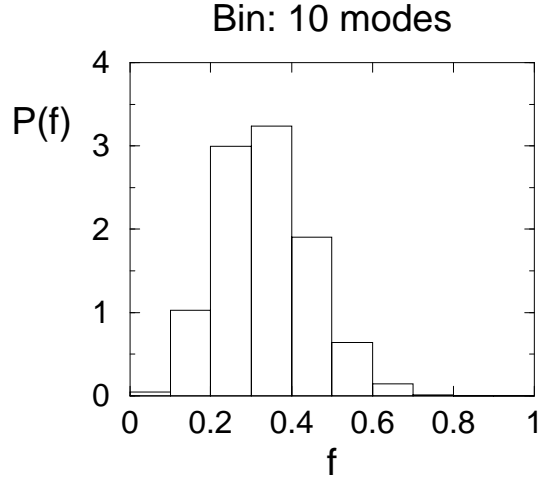,width=8.cm}
\end{center}
\caption{\label{fig:bin}\small Event-by-event distribution of the 
 neutral fraction in a bin in momentum space containing 10 amplified
 modes. The expected DCC signal is already considerably reduced, due 
 to the absence of correlation between modes. From \cref{Serreau:2000tb}.}
\end{figure}

To analyze the isospin orientations of distinct field modes, a sensitive 
observable is the neutral fraction of pions in each mode $\bk$,  
defined as:\footnote{For a more detailed analysis, see \cref{Serreau:2000tb}.}
\beq
\label{ratio}
 f ({\bf k}) = \frac{\bar n_3({\bf k})}
 {\bar n_1({\bf k})+\bar n_2({\bf k})+\bar n_3({\bf k})} \, ,
\eeq
where $\bar n_a(\bk)\equiv\bar n_a(\bk,t_f)$ represent the 
averaged occupation numbers corresponding to the classical field configuration 
at the final time $t_f$, see \Eqn{number1}. The event-by-event distribution of 
the neutral fraction \eqn{ratio} for the most amplified mode, $\bk=0$, is shown 
in \Fig{fig:mode} for $t_f=10$~fm/c, corresponding roughly to the end of the 
spinodal instability period, cf. \Fig{fig:amplifquench}. The corresponding distribution
at time $t_f=56$~fm/c, characteristic of the parametric amplification mechanism,
looks exactly the same \cite{Serreau:2000tb}. In both cases, all the amplified
modes exhibit similar distributions. In fact, the neutral fraction distributions 
for amplified modes are found to be essentially the same as the corresponding 
ones in the initial state. The latter can be computed exactly
\cite{Serreau:2000tb,Serreau:2001bu} and is represented by the dashed line 
in \Fig{fig:mode}. For a given mode $\bk$, it reads:\footnote{The following 
formula is obtained for Neumann boundary conditions (where the Fourier components
of the field are real numbers) which are convenient for discussing the question 
of polarization.}
\beq
\label{indist}
 P_k(f)=\frac{1}{2}\Big[F_{\Omega_k}(f)+F_{-\Omega_k}(f)\Big]\,,
\eeq
where
\beq
 F_\Omega(f)=[\Omega-(1-f)]\left(\frac{\Omega+1}{\Omega-(1-2f)}\right)^{3/2}\,,
\eeq
and
\beq
 \Omega_k=\frac{\sigma_2^2+\omega_k^2\sigma_1^2}{\sigma_2^2-\omega_k^2\sigma_1^2}\,.
\eeq
with $\sigma_1^2$ and $\sigma_2^2$ the variances of the field and its time derivative 
respectively (see subsection \ref{sec:quench}). Although not exactly~$1/\sqrt f$, 
the distribution is very broad, exhibiting large fluctuations around the mean 
value $\bar f=1/3$, which is the relevant point for phenomenology. 
As emphasized previously, these large fluctuations are a direct manifestation 
of the classical nature of the field modes. The deviations from the 
ideal law \eqn{DCCsign} come from the fact that the latter are not strictly 
linearly polarized waves \cite{Serreau:2000tb}.

However, one finds that distinct modes have completely independent 
polarizations in isospin space, as can be seen on \Fig{fig:correl},
which shows the statistical correlation between the neutral fractions 
in different amplified modes: Their directions of oscillation in 
isospin space are completely random. In other words, different modes 
act as independent DCCs.
This has the important phenomenological consequence that the large 
event-by-event fluctuations of the neutral fraction are rapidly washed 
out when the contributions of several modes are added in a momentum bin,
even when one limits one's attention to soft modes only. This is 
illustrated in \Fig{fig:bin}, which shows the neutral fraction 
distribution in a bin containing $10$ modes. The expected signal is 
considerably reduced, already for such a small bin. This explains the 
absence of large fluctuations reported in previous studies 
\cite{Gavin:1993bs,Randrup:1996es,Rajagopal:1997au}, where the authors
typically considered the contribution of a large number of 
modes.\footnote{Similar conclusions 
have been reached in a slightly different context in \cref{Holzwarth:2002wv}.} 

\begin{figure}[t]
\begin{center}
 \epsfig{file=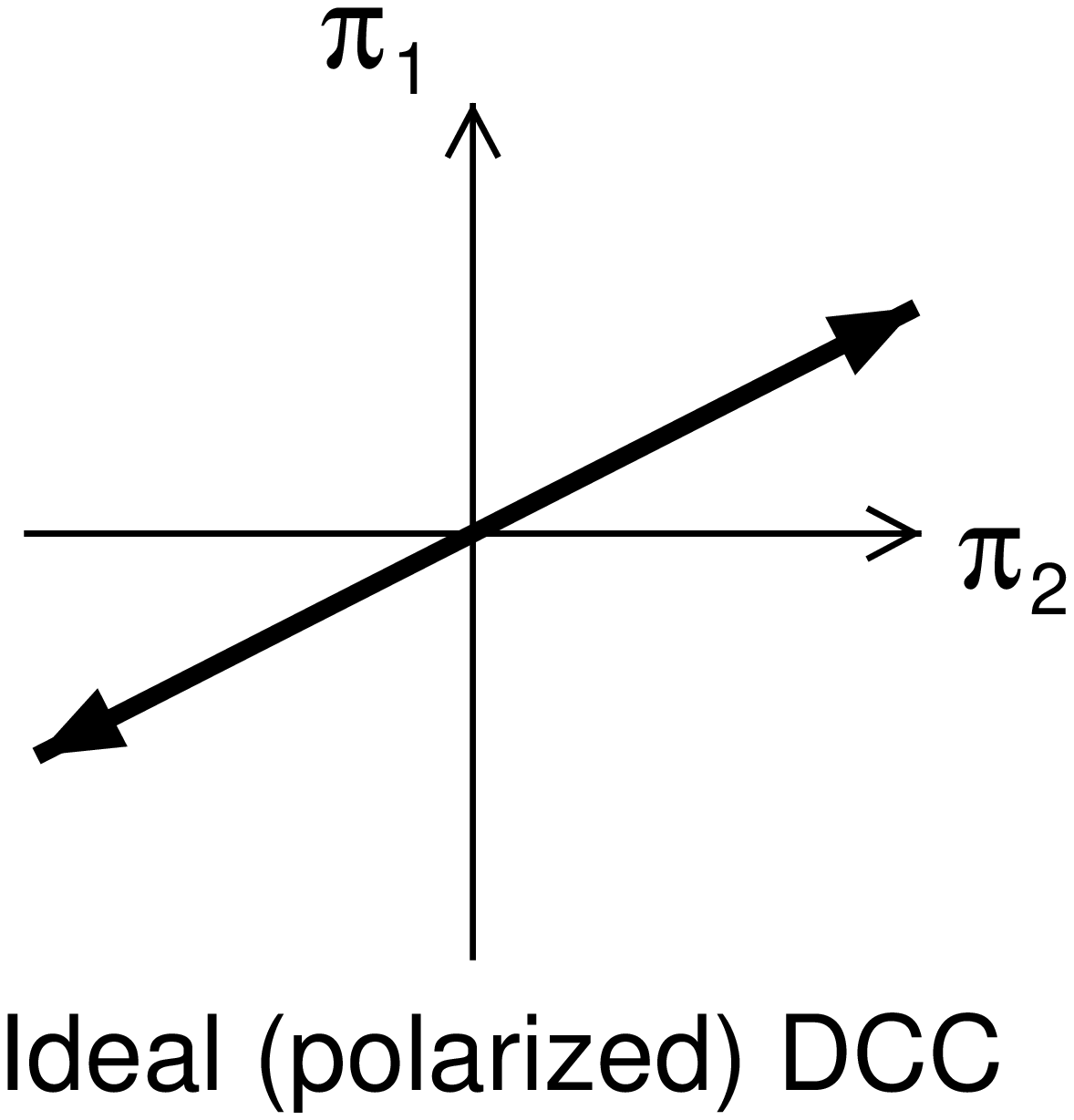,width=3.cm}\hspace{3.cm}\epsfig{file=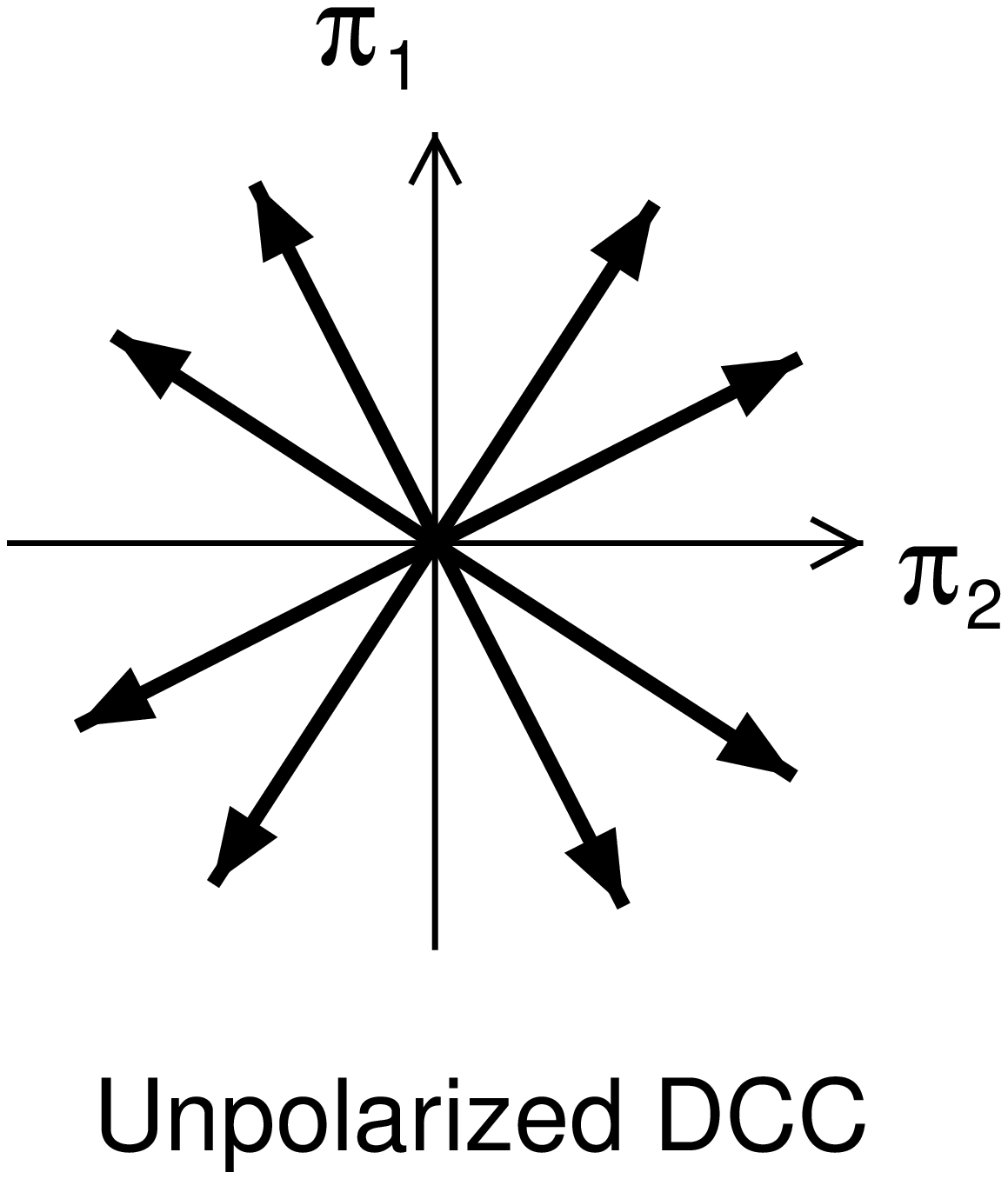,width=3.cm}
\end{center}
\caption{\label{fig:cartoon}\small Schematic representation of the 
ideal DCC configuration (left), where all field modes oscillate in the 
same direction in isospin space (see \Eqn{DCCsign}), and of the 
unpolarized DCC formed after a quench (right), where the modes have 
independent directions of oscillation.}
\end{figure}

In conclusion, the non-linear dynamics of the linear sigma model does not 
build the required correlation between modes:
The state produced in the simplest form of the quench scenario, where no 
correlations are present initially, is not identical to the originally 
proposed DCC. Instead, a more realistic picture is that of a superposition 
of waves having independent orientations in isospin space: an ``unpolarized'' 
DCC configuration, as depicted on \Fig{fig:cartoon}. 
Of course, one cannot exclude the possibility that the required correlations 
between modes be indeed (partially) formed in actual nuclear or hadronic 
collisions by means of some other mechanism.\footnote{See, for instance, 
\cref{Asakawa:1998st}.} Reality most probably lies between the two extreme 
pictures represented on \Fig{fig:cartoon}. The point is that the only presently 
existing microscopic scenario for DCC formation predicts an unpolarized 
state. This information should be taken into account in further theoretical
as well as experimental investigations. 

\subsection{\label{sec:lifetime} Lifetime of the DCC}

So far we have been mainly concerned with the description of the 
intrinsic DCC dynamics, that is the possible mechanism responsible
for its formation. We now discuss extrinsic aspects of DCC dynamics,
namely those concerning the interactions of the DCC bubble with its 
hadronic environment. If at all, the DCC is produced in a high-multiplicity 
environment and must be considered as an open system interacting with the 
surrounding degrees of freedom \cite{Krzywicki:1993vz}. In the context of 
ultra-relativistic heavy-ion collisions, one typically imagines 
a thermal bath of pions and nucleons. The interactions with the latter 
tend to destroy the coherence of the DCC excitation and, if it were not
for expansion -- which eventually causes the interactions to freeze out, 
the produced DCC state would simply melt in the external
bath: The DCC pions would thermalize before being emitted. 
The possibility to detect any hadronic signal from a DCC therefore 
crucially depends on the lifetime of the latter in the environmental bath.

\subsubsection{Interactions with the hadronic environment: effective dissipation}

The interactions of the DCC with the debris of the collisions have been studied
by many authors in various contexts 
\cite{Krzywicki:1993vz,Biro:1997va,Abada:1997mb,Steele:1998ye,Dumitru:2000qy}. 
Although a quantitative 
description is clearly out of reach within the present theoretical understanding, a 
qualitative discussion is already useful. A widely used approach in the literature
consists in integrating out the degrees of freedom of the bath to obtain an effective 
dynamics for the DCC state 
\cite{Biro:1997va,Rischke:1998qy,Xu:1999aq,Bettencourt:2001xd,Mocsy:2002hv}.
As a simple picture, one imagines a DCC excitation, characterized by a classical pion 
field $\pi_a$, in contact with a thermal bath of pions and nucleons at a given temperature 
$T$. The DCC being essentially a soft excitation, it is reasonable to assume 
that the dynamics of the bath is characterized by comparatively short time scales. Tracing
over the rapid degrees of freedom in analogy with the standard description of Brownian 
motion \cite{Feynman:1963fq,FeynmanHibbs}, one essentially obtains an effectively dissipative dynamics for 
the DCC pion field. In general, the latter is characterized by a non-local memory integral 
and an associated colored random noise, reflecting the presence of the bath \cite{Greiner:1996dx}. 
For the present argument, however, it is sufficient to assume the so-called Markov limit 
\cite{Greiner:1996dx}, where the dissipation kernel becomes local, thus leading to a simple 
friction term $\sim\eta_k\dot\pi_a(\bk,t)$ in the equations of motion for the DCC pion 
field excitation in momentum space, and where the associated fluctuating 
noise field $\xi^a_\bk(t)$ is a white noise (see \Eqn{corrnoise} below). The relevant 
equations of motion for a given realization of the noise field $\xi^a_\bk(t)$ can be 
written as:
\beq
\label{noise}
 \ddot\pi_a(\bk,t)+\eta_k\,\dot\pi_a(\bk,t)+\omega_k^2\pi_a(\bk,t)
 +\cdots=\xi^a_\bk(t)\,,
\eeq
where $\omega_k=\sqrt{k^2+\mpi^2}$ and were, the dots on the LHS represent 
non-linear terms in the field, which we neglect for the present discussion.\footnote{In 
general, the presence of the bath modifies the effective mass and couplings of 
the soft DCC modes and induces higher-order couplings as well 
\cite{Greiner:1996dx,Rischke:1998qy}. However, these effects do not play a major role 
in the present qualitative discussion and we neglect them for simplicity.}
Physical results are obtained after averaging over all possible realizations of the noise, 
with an appropriate weight. Denoting by $\bra\dots\ket_\B$ the corresponding average over 
the degrees of freedom of the bath, one has, for instance, $\bra\xi^a_\bk(t)\ket_\B=0$.
Moreover, for a bath in thermal equilibrium the correlation function of the random noise 
and the corresponding damping coefficient $\eta_k$ are related by:
\beq
\label{corrnoise}
 \bra\xi^a_\bk(t)\xi^b_{\bk'}(t')\ket_\B
 =\eta_k\,\omega_k\coth\left(\frac{\omega_k}{2T}\right)\,
 \frac{\delta_{ab}\delta_{\bk,\bk'}\delta(t-t')}{V}\,,
\eeq
as a consequence of the fluctuation-dissipation theorem. Here, $V$ is the volume 
of the system and $V\delta_{\bk,\bk'}\equiv(2\pi)^3\delta^{(3)}(\bk-\bk')$. 
Equation \eqn{corrnoise} ensures that the DCC eventually thermalizes with the 
heat bath at temperature $T$. For instance, one finds that, for large times: 
\beq
 \bra\pi^a_\bk(t)\pi^b_{\bk'}(t)\ket_\B\to\delta^{ab}\delta_{\bk,\bk'}
 \frac{\coth(\omega_k/2T)}{2\omega_kV}\,,
\eeq which indeed corresponds to the required thermal 
correlator. Finally, the friction coefficient $\eta_k$, which determines the rate
at which equilibrium is approached, is simply given by the on-shell damping rate
at temperature $T$ \cite{Greiner:1996dx}: $\eta_k\equiv\gamma_k=\sigma_k^>-\sigma_k^<$, 
where $\sigma^>_k$ and $\sigma^<_k$ denote the in-medium absorption and production 
rates for on-shell pionic excitations of momentum $\bk$ respectively.

\subsubsection{The lifetime of a DCC excitation}

Averaging \Eqn{noise} over the noise, one obtains that the amplitude of the DCC field 
excitation decays as\footnote{This assumes that $\omega_k > \eta_k/2$, which is 
reasonable if the interactions with the heat bath are weak enough.} 
$\bra\pi_a(\bk,t)\ket_\B\sim\e^{-\eta_k t/2}$. The corresponding number of pions 
therefore decays with a rate $\eta_k$:
\beq 
\label{numberB}
 \bar n_a(\bk,t)\equiv\frac{1}{2\omega_k}\left|i\bra\dot\pi_a(\bk,t)\ket_\B
 +\omega_k\bra\pi_a(\bk,t)\ket_\B\right|^2\sim\e^{-\eta_k t}\,.
\eeq
Assuming a thermal bath of pions -- which are the most abundantly
produced particles in heavy-ion collisions, one can compute the
relevant friction term $\eta_k$. As a first estimate, the authors of
\cref{Biro:1997va} have computed the latter at lowest order in perturbation 
theory in the context of the linear sigma model in the high temperature, 
chirally symmetric phase, resulting in a rather short lifetime. For instance,
they obtain, for the zero mode: $\eta_{k=0}^{-1}\sim 1$~fm/c, which is a
characteristic scale of typical hadronic processes. However, as pointed out
in \cite{Rischke:1998qy}, this estimate is too crude as it neglects the fact 
that, after formation, the DCC is supposed to evolve in the phase of spontaneously 
broken symmetry, where the dynamics is strongly affected by the presence 
long-wavelength Goldstone modes.
In particular, the interactions of the latter are suppressed at low energies 
and the associated time-scales are expected to be correspondingly longer.
Various possible contributions to the DCC decay rate have been studied
in the literature \cite{Rischke:1998qy,Steele:1998ye,Mocsy:2002hv}. We review 
the main results below.

In \cref{Rischke:1998qy}, Rischke has investigated the damping of a DCC
excitation in a thermal bath of pions in the context of the linear sigma model.
In the broken phase, damping of DCC pions may arise from the 
absorption of a thermal pion, $\pi_{\rm DCC}\pi_{\rm bath}\to\sigma_{\rm bath}$, 
together with the reverse process, $\sigma_{\rm bath}\to\pi_{\rm DCC}
\pi_{\rm bath}$, namely the decay of a thermal $\sigma$. These processes, however,
are strongly suppressed due to the restricted available phase space \cite{Rischke:1998qy}. 
The corresponding contribution to the damping rate reads:
\bea
\label{pipisigma}
 \eta_k^{\pi\pi\leftrightarrow\sigma}&=&
 \frac{1}{2\omega^\pi_k}\int\tdp_\pi\tdq_\sigma\,
 (2\pi)^4\delta^{(4)}(k_\pi+p_\pi-q_\sigma)\,
 |{\mathcal M}_{\pi\pi\to\sigma}|^2\nn
 &&\qquad\times\Big\{n^\pi_p(1+n^\sigma_q)
 -n^\sigma_q(1+n^\pi_p)\Big\}\,,
\eea
where we introduced four-momenta as \eg $k_i^\mu\equiv(\omega^i_k,\bk)$ 
and we used the notation $\tdp_i\equiv \frac{d^3p}{(2\pi)^32\omega^i_p}$, 
with $i=\pi,\sigma$. Here $\omega^i_k=\sqrt{k^2+m_i^2}$ and 
$n^i_k=[\exp\omega^i_k/T-1]^{-1}$ is the Bose-Einstein distribution
for the respective species. The terms $1+n$ in brackets correspond
to Bose-enhancement factors in the final state. Finally, 
$|{\mathcal M}_{\pi\pi\to\sigma}|^2$ denotes the relevant matrix element. 
The rate $\eta_{k=0}^{\pi\pi\leftrightarrow\sigma}$ has been estimated in 
\cite{Rischke:1998qy} using perturbation theory as a qualitative guide. At 
lowest order, one has $|{\mathcal M}_{\pi\pi\to\sigma}|^2=(2gm_\sigma)^2$, 
where $gm_\sigma=\lambda f_\pi$ is the dimensionful $\pi\pi\sigma$ coupling, 
and the momentum integrals in \eqn{pipisigma}
can be easily computed \cite{Rischke:1998qy}. Neglecting terms of relative 
order $m_\pi^2/m_\sigma^2$, one obtains the following approximate 
expression:\footnote{An exact expression for $\eta_{k=0}^{\pi\pi
\leftrightarrow\sigma}$ at this order of perturbation theory can be found 
in \cite{Rischke:1998qy}. We mention that a more general expression, including 
non-vanishing momentum, can be found in \cref{Mocsy:2002hv}.}
\beq
\label{rate11}
 \frac{\eta_{k=0}^{\pi\pi\leftrightarrow\sigma}}{m_\pi}\approx
 \frac{g^2}{4\pi}\,\frac{m_\sigma^4}{m_\pi^4}\,
 \Big(\e^{m_\pi/T}-1\Big)\,\e^{-m_\sigma^2/2m_\pi T}\,.
\eeq
As expected, this results in rather long lifetimes for temperature of interest
in heavy-ion collisions. For instance, one obtains 
$1/\eta_{k=0}^{\pi\pi\leftrightarrow\sigma}\approx20~(9)$~fm/c for 
$T=150~(170)$~MeV. Of course, the result \eqn{rate11} should be regarded only as 
a rough qualitative estimate, being based on a perturbative expansion (recall 
that $g\approx m_\sigma/2f_\pi\approx 3$). Still, this indicates that the DCC 
lifetime in the broken phase is considerably longer than typical hadronic 
time-scales. 

For a more quantitative estimate, it is important to include other processes, 
which are formally of higher-order in perturbation theory, but which can give 
significant contribution to the damping rates. The first such contribution comes 
from two-body elastic scatterings $\pi\pi\to\pi\pi$. A non-perturbative 
estimate\footnote{For a perturbative calculation, see \cref{Mocsy:2002hv}.} of the 
latter has been performed in \cref{Steele:1998ye}, where the authors used experimental 
data to constrain the relevant matrix element.\footnote{We mention that the authors 
of \cref{Steele:1998ye} employ a slightly different approach to describe the damping 
of the DCC excitation. Indeed, they derive an expression for the DCC decay rate by 
generalizing the LSZ reduction formula to include the DCC coherent state in the
initial and final states. The expression they obtain for the damping rate of 
a given mode $\bk$ is identical to the one used here, that is the usual on-shell 
damping rate (see \Eqn{pipipipi} below).} The corresponding contribution to the 
damping rate reads:
\bea
\label{pipipipi}
 \eta_k^{\pi\pi\leftrightarrow\pi\pi}&=&
 \frac{1}{2\omega^\pi_k}\int\tdp\tdq\tdr\,
 (2\pi)^4\delta^{(4)}(k+p-q-r)\,|{\mathcal M}_{\pi\pi\to\pi\pi}|^2\nn
 &&\qquad\times\frac{1}{2}
 \Big\{n^\pi_p(1+n^\pi_q)(1+n^\pi_r)-
 n^\pi_q n^\pi_r (1+n^\pi_p)\Big\}\,,
\eea
where the factor $1/2$ in the second line accounts for identical particles in the 
initial (final) state. The isospin averaged matrix element can be written as
$|{\mathcal M}_{\pi\pi\to\pi\pi}|^2=\frac{1}{3}\sum_I(2I+1)|M^I_{\pi\pi}|^2$,
with the standard partial-wave decomposition of the amplitude for scattering
in a state of total isospin $I=0,1,2$ \cite{Steele:1998ye}:
\beq
 M^I_{\pi\pi}=32\pi\sum_\ell(2\ell+1)P_\ell(\cos\theta)\frac{\sqrt s}{2q}
 \e^{i\delta^I_\ell}\sin\delta^I_\ell\,,
\eeq
where $\sqrt s$, $q$ and $\theta$ denote the center of mass energy, three-momentum 
transfer and scattering angle of the binary collision respectively.
Using a parametrization of the data for the three dominant phase shifts 
$\delta^0_0$, $\delta^1_1$ and $\delta^2_0$ \cite{Bertsch:1987ux},
and assuming, as above, that the DCC pions have essentially zero-momentum, 
the authors of Ref.\ \cite{Steele:1998ye} obtain the estimate 
$1/\eta_{k=0}^{\pi\pi\leftrightarrow\pi\pi}\approx8.9~(5.6)$~fm/c for
$T=150~(170)$~MeV. Again, as expected from low-energy theorems, this is much
longer than typical hadronic time scales and is instead characteristic
of Goldstone dynamics. Note also that the contribution to the damping
rate from two-body scatterings is larger than the previous one from 
pion absorption and sigma decay. Taking both contributions into account
one obtains $(\eta_{k=0}^{\pi\pi\leftrightarrow\sigma}+
\eta_{k=0}^{\pi\pi\leftrightarrow\pi\pi})^{-1}\approx 6.1~(3.5)$~fm/c at
the respective temperatures.

\begin{figure}[t]
\begin{center}
 \epsfig{file=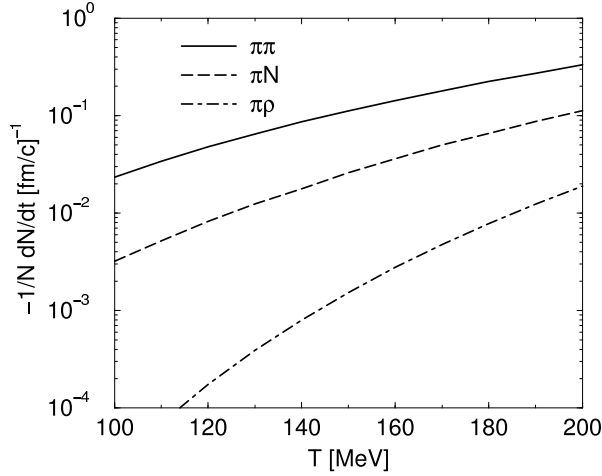,width=8.cm}
\end{center}
\caption{\label{fig:lifetime_decay}\small Various contributions to the decay rate
  $\eta_{k=0}=-\dot N/N$ of an homogeneous ($k=0$) DCC excitation as a function of 
  the temperature $T$ of the surrounding heat bath. From \cref{Steele:1998ye}}
\end{figure}

Similarly, the authors of \cite{Steele:1998ye} have estimated the 
contributions from $\pi\rho$ scattering, assuming that the latter is 
dominated by the formation of a $a_1$(1240) meson in the $s$-channel, 
as well as from scatterings of DCC pions with baryons. While the former 
is found to give a negligible contribution due to kinematics, it was argued
that the latter can play a substantial role since a number of baryons are 
produced even in the central rapidity region. The contribution of pion-nucleon 
scattering to the damping rate has a similar expression as \Eqn{pipipipi}, 
but replacing the combination of statistical factor on the second line (including
the overall $1/2$) by:
\beq
 \bar n^N_p(1+n^\pi_q)(1-\bar n^N_r)
 -n^\pi_q\bar n^N_r(1-\bar n^N_p)\,,
\eeq
where $\bar n^N_p=[\exp(E_p-\mu)/T+1]^{-1}$ is the Fermi-Dirac distribution for 
nucleons, with $E_p=\sqrt{p^2+m_N^2}$. The baryon chemical potential $\mu$ 
is introduced to account for the finite baryon density in the central rapidity 
region and is chosen to reproduce the observed ratio between pions and nucleons. 
The authors of \cite{Steele:1998ye} use the value $\mu=260$~MeV at $T=150$~MeV
in order to enforce a $5:1$ pion-to-nucleon ratio as observed at SPS energies 
\cite{Braun-Munzinger:1998cg}.\footnote{The pion-to-nucleon ratio quoted in
\cref{Braun-Munzinger:1998cg} is actually of $6$ to $1$, corresponding to 
$\mu=230$~MeV at $T=150$~MeV. However, this small difference has little impact
on the results described below \cite{Steele:1998ye}.}
Using, again, a parametrization of experimental data to describe the relevant 
scattering matrix elements, they find that the effect of nucleons is about a 
$20$\% reduction of the DCC lifetime at the SPS.\footnote{At RHIC energies,
the pion-to-nucleon ratio $\pi/N$ is increased by about a factor $2$, but this 
is partly compensated by the increase of anti-nucleon production (see \eg 
\cite{Braun-Munzinger:2001ip,Braun-Munzinger:2003zd}). One has, roughly, 
$\pi/(N+\bar N)\approx 7$.}
This is shown in \Fig{fig:lifetime_decay}, where the 
various contributions to the DCC decay rate are shown as a function of temperature. 
The chemical potential is adjusted in order to maintain a fixed pion-to-nucleon ratio 
at all temperatures. The figure also shows the contribution from $\pi\rho$ scattering.

\begin{figure}[t]
\begin{center}
 \epsfig{file=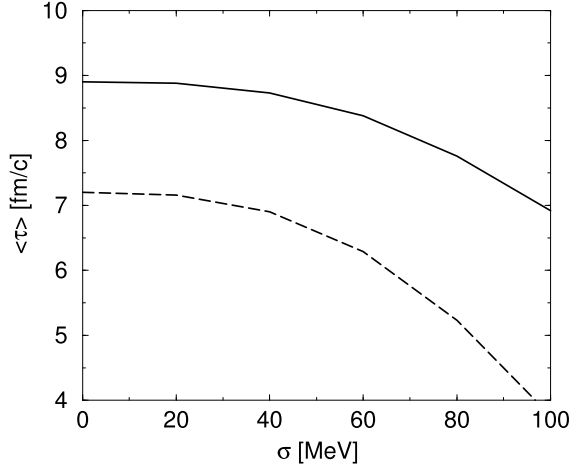,width=8.cm}
\end{center}
\caption{\label{fig:lifetime_width}\small The momentum-averaged DCC lifetime 
  \eqn{rateav}-\eqn{rateav2} as a function of the momentum spread $\Delta k\equiv\sigma$ 
  of the DCC excitation, assuming a Gaussian shape (see text). The solid line shows the
  contribution from $\pi\pi$ scattering alone, whereas the dashed curve is obtained by including
  $\pi\rho$ and $\pi N$ scatterings as well. From \cref{Steele:1998ye}}
\end{figure}

Finally, it is interesting to estimate the influence of the finite size of the DCC, 
that is to take into account the spread of the momentum distribution of the DCC 
pions. Adapting the treatment of \cref{Steele:1998ye}, we define a momentum averaged 
lifetime $\bar\tau$ as:
\beq
\label{rateav} 
\frac{1}{\bar\tau}=-\frac{\dot N(t)}{N(t)}\approx
 -\frac{\int_k\,\eta_k\,\bar n(\bk,t)}
   {\int_k\,\bar n(\bk,t)}\,,
\eeq
where we used the notation $\int_k\equiv\int\frac{d^3k}{(2\pi)^3}$. Here, 
$N(t)\equiv\int_k\,\bar n(\bk,t)$ is the total number of DCC pions, with 
$\bar n(\bk,t)\equiv\sum_a \bar n_a(\bk,t)$ and $\bar n_a(\bk,t)$ the number 
of DCC pions of isospin $a$ in mode $\bk$, as defined in \eqn{numberB}. 
We used the fact that $\dot{\bar n}(\bk,t)\approx-\eta_k\,\bar n(\bk,t)$ in 
the limit $\eta_k\ll\omega_k$ (see \Eqn{noise}) to approximate the RHS of 
\Eqn{rateav}. For a rough estimate, we write $\bar n(\bk,t)\simeq N(t)\,
{\mathcal F}(\bk)$, where ${\mathcal F}(\bk)$ describes the momentum shape of the 
DCC spectrum. Inserting this ansatz in \eqn{rateav}, we get:
\beq
\label{rateav2} 
 \frac{1}{\bar\tau}\simeq-\int_k\,\eta_k\,{\mathcal F}(\bk)\,.
\eeq
This expression is similar to the one used by the authors of \cref{Steele:1998ye}. 
Their result is reproduced on \Fig{fig:lifetime_width}, which shows the momentum
averaged lifetime \eqn{rateav2} as a function of the momentum spread $\sigma\equiv\Delta k$ 
of the DCC. This assumes a Gaussian shape, ${\mathcal F}(\bk)\propto\exp[-k^2/2\Delta k^2]$, 
and includes the contributions from $\pi\pi$, $\pi\rho$ as well as $\pi N$ binary 
scatterings (with baryonic chemical potential adjusted to maintain a fixed pion-to-nucleon 
ratio). The corresponding matrix elements are parametrized as above. Obviously, increasing 
the width $\Delta k$ decreases the lifetime: Smaller DCC domains decay faster. The typical 
momentum spread predicted by the quench scenario discussed in previous sections is given 
by the symmetry breaking scale $\Delta k\lesssim f_\pi\simeq 90$~MeV, corresponding to 
an estimated average lifetime $\bar\tau\gtrsim 5$~fm/c at SPS energies.
Although not very large, this is reasonably long for a hadronic DCC signal
to be observable. Notice that larger DCCs, which have a better chance 
to survive until freeze-out, are also the rarest.

\subsection{\label{sec:pheno} Phenomenology}

The possible detection of a DCC in high-energy hadronic or nuclear
collisions\footnote{We mention that the possibility of DCC production at
a photon collider has been advocated in \cref{Kuraev:2003tr}.} would 
bring valuable information concerning the chiral 
structure of the QCD vacuum and/or concerning the chiral phase transition.
Although, as we have seen in the previous sections, the present theoretical 
understanding of DCC formation is far from being accurate enough to produce 
reliable numbers, it is important to characterize even qualitatively 
possible experimental signatures of the phenomenon.
The DCC pions are expected to be mostly concentrated at low momentum in the 
DCC rest frame and the DCC emission should, therefore, be characterized by a 
cluster of pions with low relative momenta. Ideally, the latter would present 
characteristic isospin fluctuations. In the present section, we describe various 
hadronic signatures which have been discussed in the literature as well as 
experimental tools used in actual DCC searches (see \Sec{sec:exppart}). 
We also discuss possible electromagnetic signatures.

As already emphasized, the formation of a DCC is expected to be a 
rare phenomenon and the main difficulty is to isolate any DCC signal from 
the enormous background of ``incoherent'' pions, produced by standard 
mechanisms. In particular, the presence of DCC pions would hardly be detectable 
in global observables, such as \eg the single-pion inclusive 
spectrum\footnote{Multi-pion correlations have also been 
proposed as possible probes of the formation of a DCC 
\cite{Greiner:1993jn,Biyajima:1998yh,Nakamura:1999ai,Hiro-Oka:1997xi}. 
It is indeed well-known that the presence of a coherent source  
reduces the so-called Hanburry-Brown-Twiss (HBT) effect on identical
pion correlations \cite{Gyulassy:1979yi}. However, the interpretation of HBT 
measurement is highly non-trivial \cite{Weiner:1999th,Gyulassy:1979yi} and 
such effects are very difficult to observe in practice.} and event-by-event 
analysis seem preferable. We focus on the latter in the following.

\subsubsection{Multiplicity fluctuations}

A simple hadronic observable is the pion multiplicity in a given bin
in mo\-men\-tum-\-spa\-ce. In particular, multiplicity fluctuations can reveal
the dynamics underlying pion production. They can be characterized 
by means of the so-called factorial moments, defined as:
\beq
 f_i=\bra N(N-1)\cdots(N-i+1)\ket=\sum_{N\ge0} P(N)\frac{N!}{(N-i)!}\,,
\eeq
where $P(N)$ denotes the multiplicity distribution in the considered bin of
momentum-space. Clearly, the $i$-th factorial moment $f_i$ probes events 
where the multiplicity $N\ge i$. Purely statistical fluctuations are
characterized by a Poisson distribution, for which $f_i=\bra N\ket^i$. 
It is therefore useful to introduce reduced factorial moments:
\beq
\label{facmom}
 F_i=\frac{f_i}{\bra N\ket^i}\,.
\eeq
Deviations from $F_i=1$ reflect the presence of dynamical (non-random) 
fluctuations.

\begin{figure}[t]
\begin{center}
 \epsfig{file=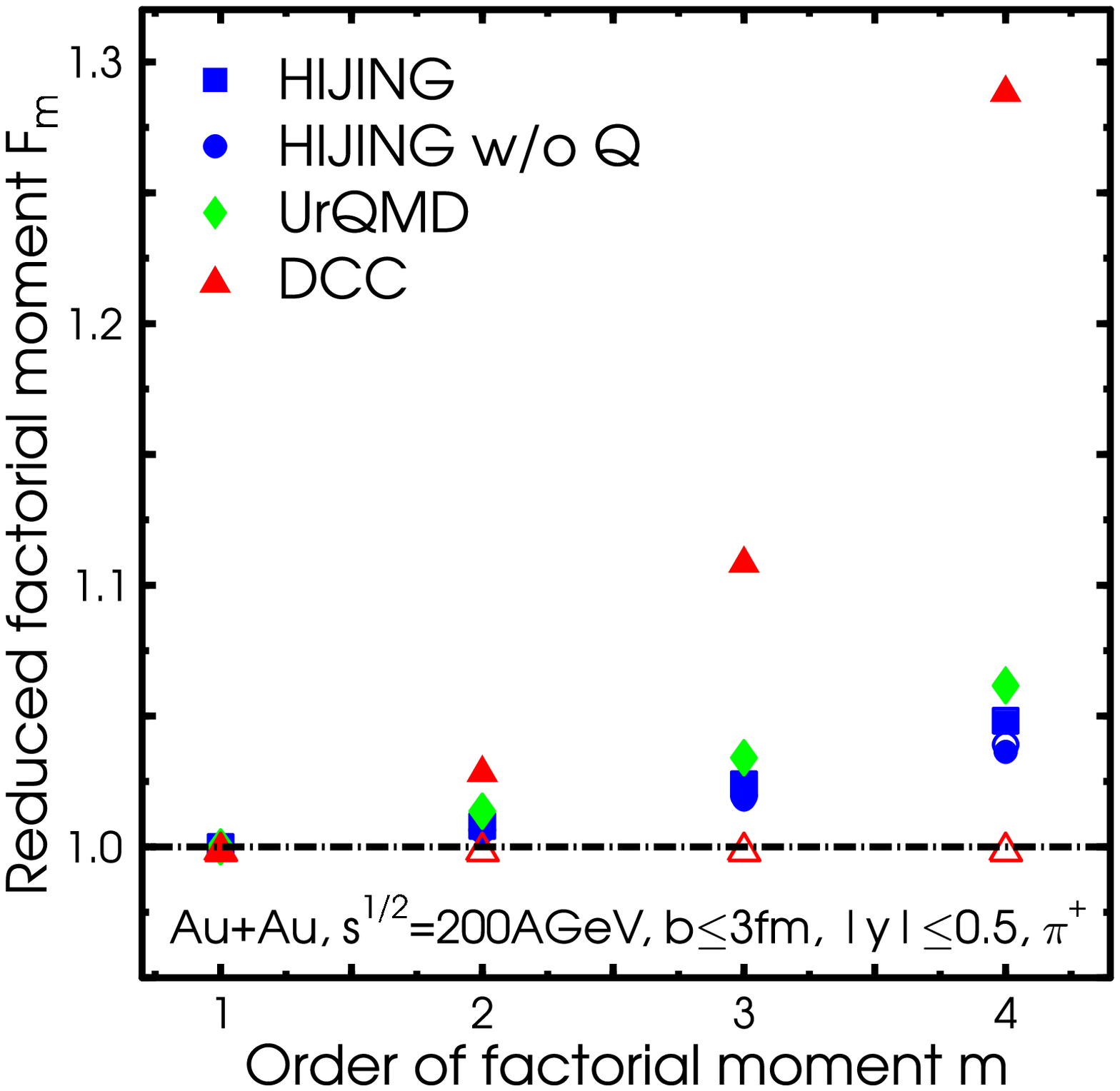,width=6.cm}\hspace{.5cm}\epsfig{file=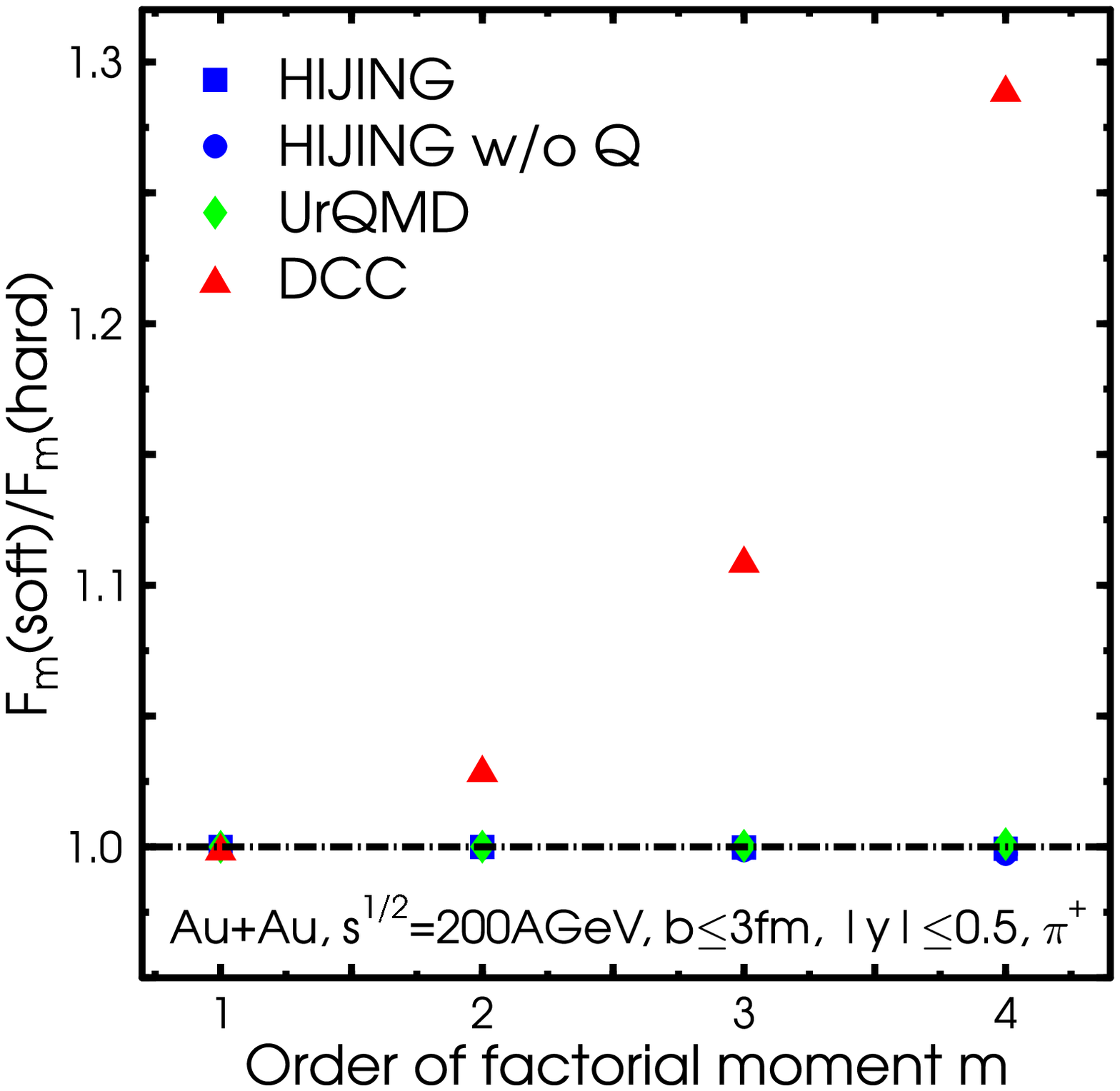,width=6.cm}
\end{center}
\caption{\label{fig:fluct_Randrup}\small Left: Reduced factorial moments for 
  the multiplicity of positively charged pions emitted at mid rapidity in 
  central $Au+Au$ collisions at the full RHIC energy, as obtained from 
  semi-classical solutions of the linear sigma model in an expanding cylindrical 
  geometry (triangles) \cite{Petersen:1999jc}. Also shown are the results generated 
  by HIJING with (squares) or without (circles) jet quenching and by UrQMD (diamonds). 
  Solid and open symbols correspond to soft ($p_T\le200$~MeV) and hard 
  ($p_T\ge200$~MeV) pions respectively. Right: The corresponding ratios of 
  reduced factorial moments calculated for soft and hard pions. 
  From \cref{Bleicher:2000tr}.}
\end{figure}

It is interesting to compute the reduced factorial moments in a specific
model of DCC production. In \cref{Bleicher:2000tr}, Bleicher {\it et al.}
have performed such an analysis in the context of a semi-classical treatment
of the linear sigma model dynamics, in an expanding geometry with cylindrical symmetry
\cite{Petersen:1999jc}:\footnote{This is mainly relevant to the case of nuclear 
collisions. For analysis of multiplicity fluctuations in models based on 
coherent and/or squeezed states descriptions of the DCC, see \eg
Refs.~\cite{Amado:1994fu,Dremin:1995nd,Hiro-Oka:1999xk,Bambah:2004tq}.} 
Using the classical field theory methods described
in \Sec{sec:chiralPT}, one computes the average occupation number $\bar n_a(\bk)$ 
in a given momentum mode $\bk$ from an expression analogous to \Eqn{number1},
suitably adapted to the expanding geometry. Identifying each classical field 
configuration with a quantum-mechanical coherent state, the actual multiplicity 
is then sampled from the associated Poisson distribution with average given
by $\bar n_a(\bk)$. The latter fluctuates from event to event and one finally 
performs a statistical average over initial field configurations in a thermal 
ensemble at a given initial temperature $T_0$. The corresponding reduced factorial 
moments are shown in the left panel of \Fig{fig:fluct_Randrup}, for both soft 
($p_T\le200$~MeV) and hard ($p_T\ge200$~MeV) pions. Also shown are the 
corresponding results obtained from the event generators HIJING \cite{Gyulassy:1994ew} and 
UrQMD \cite{Bass:1998ca,Bleicher:1999xi}, based on standard production mechanisms. The latter clearly
exhibit non-Poissonian fluctuations, with a gentle increase of the reduced
factorial moments $F_i$ with the order $i$, both for soft and hard pions.
In contrast, the dynamics of the linear sigma model leads to reduced factorial
moments that remain close to unity for the hard pions while increasing rapidly
for the soft pions, indicating enhanced multiplicity fluctuations.\footnote{
Similar results have been obtained in a slightly different context in \cref{Xu:1999aq}.
Such large multiplicity fluctuations are also already visible from the broad 
distributions of the amplification coefficient reported in \cref{Krzywicki:1998sc} 
(see \Fig{fig:proba} of the present report).}

The qualitative difference with standard production mechanisms is better 
emphasized by considering the ratios between the reduced moments for soft and 
hard pions, as illustrated in the right panel of \Fig{fig:fluct_Randrup}.
These ratios may provide a useful observable to search for dynamical fluctuations 
beyond what has been included in standard event generators.

\subsubsection{Charged-neutral fluctuations and robust observables}

The most striking feature of the coherent (semi-classical) nature of DCC
emission is the anomalously large event-by-event fluctuations of 
the neutral fraction of produced pions, see \Eqn{DCCsign}. This is 
at the basis of most existing DCC hunt strategies. Similar to the previous
analysis of multiplicity, such charged-neutral multiplicity fluctuations 
can be conveniently studied by means of the bivariate factorial moments
\beq
\label{bivariate}
  \bra \nc(\nc-1)\cdots(\nc-i+1)\,n_0(n_0-1)\cdots(n_0-j+1)\ket\,,
\eeq
where $\nc$ and $n_0$ denote the charged and neutral-pion multiplicity 
respectively. Here we review the discussion of charged-neutral fluctuations
presented in \cref{Brooks:1996nu} for the analysis of the MiniMAX experiment 
at the Fermilab Tevatron. In particular, these authors have introduced specific 
observables (see \Eqn{ri1} below) which are rather insensitive to the fluctuations 
of the total multiplicity and, therefore, essentially probe the charged-to-neutral 
ratio fluctuations for a wide class of production models. These so-called robust
observables are also largely insensitive to detector efficiencies.

\paragraph{Generating functionals for bivariate factorial moments}

Let $p(\nc,n_0)$ denote the probability for the occurrence of $\nc$ charged and 
$n_0$ neutral pions in a multi-particle event, 
in a given region of phase-space. Equivalently, one introduces the generating
function:
\beq
\label{generating}
 G(\zc,z_0)=\sum_{\nc,n_0\ge0}p(\nc,n_0)\,z_{\rm ch}^\nc\,z_0^{n_0}\,, 
\eeq
which encodes the bivariate factorial moments \eqn{bivariate}:
\beq
\label{bivgen}
 \Big<\frac{\nc!}{(\nc-i)!}\frac{n_0!}{(n_0-j)!}\Big>
 =\left.\frac{\p^{i+j}G(\zc,z_0)}
 {\p z_{\rm ch}^i\p z_0^j}\right|_{\zc=z_0=1}\,.
\eeq
Denoting by $P(N)$ the total multiplicity distribution in
the considered region of phase-space, one writes
\beq
\label{conditional}
 p(\nc,n_0)=P(N)p_N(\nc,n_0)\,,
\eeq
where $p_N(\nc,n_0)$ denotes the conditional probability that $\nc$
of the $N$ produced pions be charged and $n_0=N-\nc$ be neutral.
A wide class of models can be described by the following ``binomial
transform'' \cite{Brooks:1996nu}: 
\beq
\label{models}
 p_N(\nc,n_0)=\frac{N!}{\nc!n_0!}\int_0^1df\,p(f)\,f^{n_0}(1-f)^\nc\,.
\eeq
For instance, the case of generic pion production can be described by 
a standard binomial distribution, which corresponds to $p(f)=\delta(f-\bar f)$, 
where $\bar f=1/3$ is the mean fraction of neutral pions. Equation \eqn{models}
also includes the case of ideal DCC production, which corresponds to 
$p(f)=1/2\sqrt f$, see \Eqn{DCCmodel}.
Using Eqs.~\eqn{generating}, \eqn{conditional} and \eqn{models}, one can write:
\beq
\label{genfunc}
 G(\zc,z_0)=\int_0^1df\,p(f)\,G(fz_0+(1-f)\zc)
\eeq
where
\beq
\label{genfuncN}
 G(z)=\sum_{N\ge0}P(N)\,z^N
\eeq
is the generating function for the factorial moments of the total
multiplicity fluctuations.

\paragraph{Charged-pions--photons fluctuations}

Neutral pions are de\-tec\-ted through their decay in a pair of photons 
$\pi^0\to\gamma\gamma$. It is, therefore, useful to measure directly
the bivariate factorial moments for charged-pions--photons fluctuations.
The latter can be defined as in \Eqn{bivariate} with the replacement $n_0\to n_\gamma$,
the number of detected photons in the region of momentum-space under consideration, and 
$\p/\p z_0\to\p/\p z_\gamma$.
For ideal detection efficiency, the corresponding generating functional can 
be obtained from \Eqn{genfunc} by replacing $z_0\to z_\gamma^2$. More 
realistically, there are probabilities $\epsilon_{0,1,2}$ that $0$, $1$, 
or $2$ photons from the $\pi^0$ decay be actually detected, with $\epsilon_0
+\epsilon_1+\epsilon_2=1$. These three possibilities can be taken into account 
by introducing the following generating function:
\beq
 g_0(\zg)=\epsilon_0+\epsilon_1\zg+\epsilon_2\zc^2\,.
\eeq
Similarly, taking into account the detection efficiency for charged pions, there 
is a probability $\epsilon_{\rm ch}$ that a charged pion be actually observed in 
the detector and a probability $1-\epsilon_{\rm ch}$ for not observing it. As before,
these two possibilities can be described by the following generating function:
\beq
 g_{\rm ch}(\zc)=(1-\epsilon_{\rm ch})+\epsilon_{\rm ch}\zc\,.
\eeq
Thus the relevant generating functional for observed charged-pions--photons 
fluctuations, taking into account finite detection efficiencies, can be obtained 
from \Eqn{genfunc} with the replacement $\zc\to g_{\rm ch}(\zc)$ and $z_0\to g_0(\zg)$ 
\cite{Brooks:1996nu}:
\beq
\label{genfuncobs}
 G_\obs(\zc,\zg)=G(g_{\rm ch}(\zc),g_0(\zg))\,.
\eeq
One has, for the observed bivariate moments:
\beq
 \Big<\frac{\nc!}{(\nc-i)!}\frac{\ng!}{(\ng-j)!}\Big>_\obs
 =\left.\frac{\p^{i+j}G_\obs(\zc,\zg)}
 {\p z_{\rm ch}^i\p \zg^j}\right|_{\zc=\zg=1}\,.
\eeq
For instance, one easily obtains, for the first moments:
\bea
 \bra\ng\ket_\obs&=&(\epsilon_1+2\epsilon_2)\,\bra f\ket\,\bra N\ket\\
 \bra\nc\ket_\obs&=&\epsilon_{\rm ch}\,\bra1-f\ket\,\bra N\ket\\
 \bra\ng\nc\ket_\obs&=&\epsilon_{\rm ch}(\epsilon_1+2\epsilon_2)\,
 \bra f(1-f)\ket\,\bra N(N-1)\ket\\
 \bra\ng(\ng-1)\ket_\obs&=&(\epsilon_1+2\epsilon_2)^2
 \bra f^2\ket\bra N(N-1)\ket+2\epsilon_2\bra f\ket\bra N\ket\\
 \bra\nc(\nc-1)\ket_\obs&=&\epsilon_{\rm ch}^2\,\bra(1-f)^2\ket\,\bra N(N-1)\ket\,,
\eea 
where the brackets on the RHS denote an average with respect either to the
multiplicity distribution $P(N)$ or to the distribution $p(f)$ 
(cf. \Eqn{models}). 

From the above expressions, one observes that the ratio\footnote{Here and
in the following, we assume exact isospin symmetry, that is 
$\bra f\ket\equiv\int_0^1 df\,p(f)f=1/3$, for simplicity.}
\beq
 r_{1,1}=\frac{\bra\nc\ng\ket_\obs\bra\nc\ket_\obs}{\bra\nc(\nc-1)\ket_\obs\,\bra\ng\ket_\obs}
 =2\frac{\bra f(1-f)\ket}{\bra(1-f)^2\ket}
\eeq
is independent of the efficiencies introduced above as well as of the 
multiplicity distribution and only depends on the distribution $p(f)$. 
More generally, it is easy to check that the following ratios:
\beq
\label{ri1}
 r_{i,1}=\frac{F_{i,1}}{F_{i+1,0}}\,,
\eeq
of reduced bivariate factorial moments:
\beq
\label{Fij}
 F_{i,j}=\frac{\bra \nc(\nc-1)\cdots(\nc-i+1)\ng(\ng-1)\cdots(\ng-j+1)\ket_\obs}
 {\bra\nc\ket_\obs^i\bra\ng\ket_\obs^j}\,.
\eeq
are robust observables in the above sense:
\beq
 r_{i,1}=2\frac{\bra f(1-f)^i\ket}{\bra(1-f)^{i+1}\ket}\,.
\eeq
Moreover, these ratios are sensitive to the difference between generic 
and DCC production. For generic production, $\bra f^i\ket=\bra f\ket^i=(1/3)^i$ and one 
immediately gets:
\beq
\label{ri1gen}
 r_{i,1}({\rm generic})=1
\eeq
for all $i$'s. In contrast, for ideal DCC production, one has: 
\beq
 \bra f^i(1-f)^j\ket_{\rm DCC}=\frac{\Gamma(i+\frac{1}{2})\Gamma(j+1)}
      {2\Gamma(i+j+\frac{3}{2})}\,,
\eeq
from which one obtains \cite{Brooks:1996nu}:
\beq
\label{ri1DCC}
 r_{i,1}({\rm DCC})=\frac{1}{i+1}\,.
\eeq
The increasing difference between the generic and ideal DCC production 
mechanisms with increasing order $i$ of the ratios $r_{i,1}$ clearly reflects
the broadness of the charged-to-neutral distribution in the latter case.

\paragraph{Mixed events}

The above discussion concerns the cases of pure generic versus pure DCC 
production. More realistic descriptions should include intermediate situations,
where both generic and DCC production mechanisms contribute. Various types of 
such mixed events have been studied in \cite{Brooks:1996nu}. Here, we consider 
a model where a DCC is formed with probability $\alpha$ and, when this happens, 
the detected pions are either of generic or of DCC origin. The corresponding 
generating function is given by the product:
\beq
 G_{\rm mix}(\zc,z_0)=G_{\rm DCC}(\zc,z_0)\,G_{\rm gen}(\zc,z_0)\,.
\eeq
where $G_{\rm DCC}(\zc,z_0)$ is given by the generic form 
Eqs.~\eqn{genfunc}-\eqn{genfuncN} with multiplicity distribution 
$P_{\rm DCC}(N)$ and with $p_{\rm DCC}(f)=1/2\sqrt f$. Similarly,
$G_{\rm gen}(\zc,z_0)$ corresponds to a given multiplicity distribution
$P_{\rm gen}(N)$ and $p_{\rm gen}(f)=\delta(f-1/3)$. In the cases where no 
DCC is formed, which occur with probability $1-\alpha$, the production is 
purely of generic type. The total generating function takes the 
form:\footnote{Taking into account the detection efficiencies as in the 
previous discussion, the actual generating function is still given by 
\Eqn{genfuncobs}, with the RHS corresponding to \Eqn{genmix}. It is easy
to check that the ratios \eqn{ri1} remain independent of the efficiencies.}
\beq
\label{genmix}
 G(\zc,z_0)=\alpha G_{\rm mix}(\zc,z_0)+(1-\alpha)G_{\rm gen}(\zc,z_0)\,.
\eeq

From the definitions~\eqn{ri1} and \eqn{Fij}, and using \eqn{bivgen} and 
\eqn{genmix} above, one can compute the robust ratios $r_{i,1}$ in the 
present model. For instance, one finds:
\beq
 r_{1,1}=\frac{1+\alpha(A+B)}{1+\alpha(A+2B)}\,,
\eeq
where
\bea
 A&=&2\frac{\bra N\ket_{\rm DCC}\bra N\ket_{\rm gen}}
 {\bra N(N-1)\ket_{\rm gen}}\\
 B&=&\frac{3}{5}\frac{\bra N(N-1)\ket_{\rm DCC}}{\bra N(N-1)\ket_{\rm gen}}\,.
\eea
Here, the brackets denote averages with respect to the multiplicity distributions
corresponding to either the DCC or the generic production mechanisms, with obvious
notations. The case of pure generic production is obviously recovered for 
$\alpha=0$. Assuming that the probability of DCC formation is small, $\alpha\ll1$,
one obtains the approximate expression:
\beq
 r_{1,1}\approx 1-\frac{3\alpha}{5}\frac{\bra N(N-1)\ket_{\rm DCC}}
 {\bra N(N-1)\ket_{\rm gen}}\,.
\eeq
\begin{figure}[t]
\begin{center}
 \epsfig{file=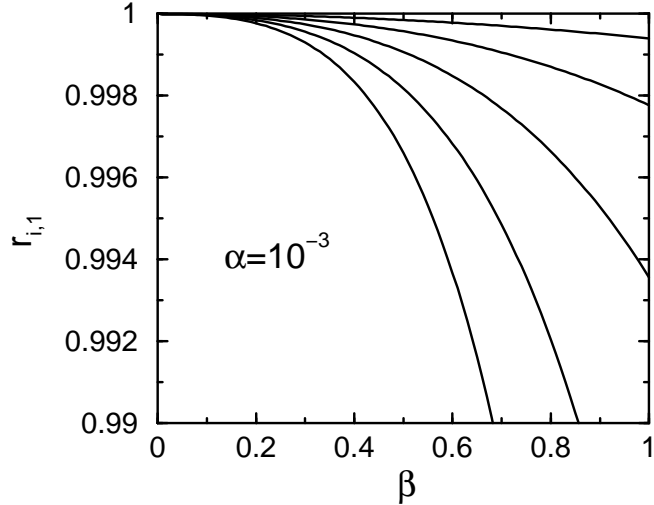,width=7.cm,angle=-90}
\end{center}
\caption{\label{fig:robust}\small The robust ratios $r_{i,1}$ for the model
  described in the text (cf. \Eqn{ri1approx}) with, from top to bottom: $i=1,\ldots,5$, 
  as a function of the fraction $\beta$ of DCC pions in the considered region 
  of phase space. The probability that a DCC forms is $\alpha=10^{-3}$.
  We have assumed that the total multiplicity distributions $P_{\rm gen}(N)$
  and $P_{\rm DCC}(N)$ are both Poissonian for simplicity.}
\end{figure}
Similarly, one can compute the ${\mathcal O}(\alpha)$ deviation from pure generic 
production due to the presence of a DCC for higher-order ratios. One obtains, 
after some calculations:
\beq
\label{ri1approx}
 r_{i,1}\approx1-\alpha\sum_{k=1}^i\frac{i!}{(i-k)!}
 \,\frac{k\,(3\beta)^{k+1}}{(2k+3)!!}
 \,\frac{F_{k+1}^{\rm DCC}F_{i-k}^{\rm gen}}{F_{i+1}^{\rm gen}}
\eeq
where $\beta=\bra N\ket_{\rm DCC}/\bra N\ket_{\rm gen}$ is the fraction 
of DCC pions in the region of phase-space considered and the $F_j$'s denote
the reduced factorial moments, cf. \Eqn{facmom}, corresponding either 
to the DCC or to the generic multiplicity distributions. 
As expected, the deviation from $r_{i,1}=1$ is larger for higher-order ratios
and increases as the probability of DCC formation and/or as the fraction 
$\beta$ of DCC pions increase. This is illustrated in \Fig{fig:robust},
where we show the ratios $r_{i,1}$ for $i=1,\ldots,5$ as a function of
$\beta$ for a given value of the probability $\alpha$. For illustration, 
we have taken both the multiplicity distributions $P_{\rm gen}(N)$ and 
$P_{\rm DCC}(N)$ to be Poissonian, that is all the reduced factorial 
moments $F_j^{\rm gen}=F_j^{\rm DCC}=1$ in \eqn{ri1approx}. This
constitute a lower bound on the expected deviation from $1$ for the
robust ratios $r_{i,1}$.

The results described here assume the general form \eqn{models} for the multiplicity 
distribution. This includes ideal DCC production as well as simple modelizations
of generic production. It would be interesting to extend the analysis of robust
ratios to dynamical scenarios of DCC formation, such as \eg semi-classical simulations
of the out-of-equilibrium chiral phase transition, as described in previous sections.

\subsubsection{Multi-resolution discrete wavelet analysis}

As described above, a possible DCC signal would appear as a localized structure 
in momentum space. To search for such structure in the lego plot, the authors 
of \cref{Huang:1995tv} have proposed to use the so-called multi-resolution wavelet 
analysis.\footnote{For a review see \eg \cite{Fang:1996ju}.} This provides a useful 
tool to identify possible structures 
simultaneously in term of their size and location in momentum space. This 
generalizes the standard Fourier analysis. The technique
has been applied to classical simulations of the linear sigma model
\cite{Huang:1995tv,Randrup:1997kt}. It has also been used by various 
experimental collaborations (see \Sec{sec:exppart}).
Here, we review the basics of the discrete wavelet analysis. 

To illustrate the method, we consider a one-dimensional phase-space, 
described by the dimensionless variable $x$ in the interval $[0,1]$. 
Any observable can be represented by a function $f(x)$ on this interval. 
Let us cut the accessible phase-space in $2^j$ bins of size $\Delta x=1/2^j$, 
where $j$ is a positive integer $j<j_{max}$, with $j_{max}$ corresponding to 
the finest accessible resolution. To each collision event corresponds a given 
sample of the function $f(x)$. Call $f^j_k=f(x=k\Delta x)$ the value of the 
sampled function in the $k$th bin, the complete sample can be represented by 
the following function:
\beq
\label{mother}
 f^{(j)}(x)=\sum_{k=0}^{2^j-1}f_k^j\,\phi_k^j(x)
\eeq
where $\phi_k^j(x)$ is zero everywhere but in the $k$th bin:
\bea
 \phi^j_k(x)=\left\{ 
 \begin{array}{ll}
   1 & k/2^j\leq x < (k+1)/2^j \\
   0 & {\rm otherwise}
 \end{array} \right.\,.
\eea
Notice that the ``bin functions'' $\phi^j_k(x)$ can be obtained as translations 
and dilation of a single function $\phi(x)$, called the mother function:
\beq
\label{motherfunc}
 \phi^j_k(x)=\phi(2^jx-k)\,,
\eeq
where, in the present case, $\phi(x)=\theta(1-x)\theta(x)$.

Equation \eqn{mother} describes the sampled observable at the resolution scale $j$.
To detect structures at lower resolution scales, one can apply the following
coarse-graining procedure: Replace two adjacent bins $2k$ and $2k+1$ by a 
single bin of size $2\Delta x=1/2^{j-1}$, with corresponding bin 
function given by: 
\beq
 \phi^{j-1}_k(x)=\phi^j_{2k}(x)+\phi^j_{2k+1}(x)\,,
\eeq
and define the value $f^{j-1}_k$ of the function in the new bin $k$ as
the average of the values in the previous smaller bins:
\beq
 f^{j-1}_k=\frac{1}{2}(f^j_{2k}+f^j_{2k+1})\,.
\eeq
The resulting coarse-grained sample at scale $j-1$ is simply given by:
\beq
\label{mother2}
 f^{(j-1)}(x)=\sum_{k=0}^{2^{j-1}-1}f^{j-1}_k\,\phi^{j-1}_k(x)\,.
\eeq
Equation \eqn{mother2} defines the mother function representation of 
the distribution $f(x)$ at scale $j-1$. The $f^{j-1}_k$'s are accordingly 
called the mother function coefficients (MFCs). Clearly the procedure described 
above eliminates all fluctuations at shorter scales than the resolution scale
$j-1$. These steps can be repeated to search for possible structures from the
finest experimental resolution scale $j_{max}$ to the lowest one $j=0$.

The information lost at each step in the coarse-graining procedure is 
encoded in the difference $\tilde f^{(j-1)}(x)\equiv f^{(j)}(x)-f^{(j-1)}(x)$.
It is easy to show that the latter can be given a similar representation:
\beq
\label{FFrep}
 \tilde f^{(j-1)}(x)=\sum_{k=0}^{2^{j-1}-1}\tilde f^{j-1}_k\,
 \psi^{j-1}_k(x)\,,
\eeq
where the functions $\psi^{j-1}_k$ are given by:
\beq
 \psi^{j-1}_k(x)=\phi^j_{2k}(x)-\phi^j_{2k+1}(x)\,.
\eeq
They can be obtained as translations and dilations of the so-called 
father function: $\psi^j_k(x)=\psi(2^jx-k)$, where, in the present case, 
$\psi(x)=\phi(2x)-\phi(2x-1)$. The corresponding father function 
coefficients (FFCs) in \eqn{FFrep} are related to the MFCs
at the previous resolution scale $j$ through: 
\beq
\label{FFC}
 \tilde f^{j-1}_k=\frac{1}{2}(f^j_{2k}-f^j_{2k+1})\,.
\eeq
Equation \eqn{FFrep} defines the father function representation 
of the sampled distribution $f(x)$ at the scale $j-1$.\footnote{The sample 
at scale $j$ can be fully reconstructed from the mother function representation 
at lower resolution scale $j-k$ with the help of all father function representations 
at the intermediate scales $j-k,\ldots,j$:
$$
 f^{(j)}(x)=f^{(j-k)}(x)+\sum_{m=0}^k\tilde f^{(j-k+m)}(x)\,.
$$}
Clearly, the FFCs \eqn{FFC} at a given scale $j$ measure the variation
of the sampled distribution $f$ between two adjacent bins. The information 
about the size of the fluctuations at each scale can be conveniently encoded 
in the following power spectrum:
\beq
 P_j=\frac{1}{2^j}\sum_{k=0}^{2^j-1}|\tilde f^j_k|^2\,.
\eeq
Similar to the Parseval theorem for the Fourier transform, one has: 
\beq
\label{pspec}
 \sum_{j\ge0}P_j=\int_0^1dx\,|f(x)|^2\,.
\eeq

\begin{figure}[t]
\begin{center}
 \epsfig{file=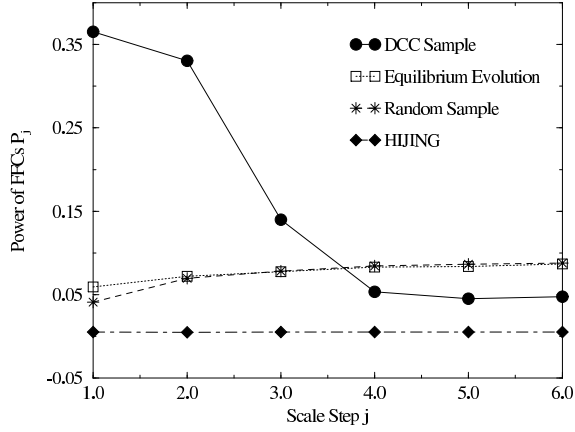,width=8.cm}
\end{center}
\caption{\label{fig:waveletPS}\small Scale dependence of the discrete wavelet 
power spectrum for various dynamically generated data. From \cref{Huang:1995tv}.}
\end{figure}

To demonstrate the method, the authors of \cite{Huang:1995tv} have applied 
the above analysis to various dynamically generated data. They have considered
the evolution of the pion field as described by the linear sigma model in the 
Hartree approximation, for a boost invariant $1+1$ expanding geometry. 
The initial field configurations are sampled from the quenched ensemble 
described in \Sec{sec:quench}. The discrete wavelet analysis is 
applied to the ratio:
\beq
\label{ratioeta}
 h(\eta)=\frac{\pi_3^2(\eta)}{\pi_1^2(\eta)+\pi_2^2(\eta)+\pi_3^2(\eta)}\,,
\eeq
where $\pi_{1,2,3}(\eta)$ are the isospin Cartesian components of the pion 
field as a function of the spatial rapidity $\eta=\frac{1}{2}\ln(t+z)/(t-z)$, 
where $z$ is the collision axis. The ratio \eqn{ratioeta} is related to the 
neutral pion fraction discussed in previous sections. The rapidity-interval 
$-\eta_{max}<\eta<\eta_{max}$, with $\eta_{max}=5$, is bined with a bin size 
$\Delta\eta=0.08$, corresponding to a finest resolution scale $j_{max}=7$, 
such that $2\eta_{\max}/\Delta\eta=2^{j_{max}}$. Figure~\ref{fig:waveletPS} 
shows the corresponding power spectrum \eqn{pspec} at various scales, for both
the initial quenched ensemble (random sample) and the final time evolved ensemble
(DCC sample). Also shown are the results corresponding to a thermal ensemble
as well as the power spectrum obtained from HIJING Monte Carlo data.\footnote{In that
case the sampled observable is the fraction of neutral pions as a function of rapidity.} 
The DCC sample clearly exhibit a non-trivial structure, reflecting clustering
in spatial rapidity. Similar results have been obtained in \cref{Randrup:1997kt} 
for three-dimensional classical simulations of the linear sigma model in a static 
geometry with a cooling term, as described in subsection~\ref{sec:exp} above.

It is important to mention that the analysis of Refs.~\cite{Huang:1995tv,Randrup:1997kt} 
concerns data sampled in coordinate space. Although this is sufficient to demonstrate 
the wavelet analysis technique, it would be of interest, for phenomenological purposes,
to perform similar analysis in momentum space. 

\subsubsection{Electromagnetic signatures}

Because they interact strongly with the background of other hadrons,
hadronic signals may rapidly loose the information they carry before
they leave the interaction region. In contrast, electromagnetic signals do not 
undergo multiple rescatterings and, therefore, provide a sensitive probe of the 
various stages of the collision. If the DCC lives long enough, they might probe 
its formation even if the latter is diluted in the hadronic environment before 
freeze-out.

It has been proposed that anomalous pion production from a DCC would
result in anomalous radiation of low-momentum photon and/or low-mass dileptons 
\cite{Huang:1994xu,Huang:1996kq,Kluger:1997cm,Boyanovsky:1997an,Charng:2002ak}. 
Here we illustrate these ideas in the case of dilepton production. 
At lowest order in the electromagnetic coupling constant $\alpha_{\rm em}$, 
the dilepton production rate reads:
\beq
\label{emrate}
 \frac{dN_{\ell^+\ell^-}}{d^4 q}=\frac{\alpha_{\rm em}^2}{6\pi^3}
 \frac{B(q^2)}{q^4}(q^\mu q^\nu-q^2g^{\mu\nu})\,W_{\mu\nu}(q)\,,
\eeq
where $q^\mu$ is the four-momentum of the lepton pair, such that $q^2\ge4m_\ell^2$,
with $m_\ell$ the lepton mass, $B(q^2)=(1+2m_\ell^2/q^2)(1-4m_\ell^2/q^2)^{1/2}$ 
and $W^{\mu\nu}(q)$ is the $4$-dimensional Fourier transform of the electromagnetic 
current correlator:
\beq
\label{emcor}
 W^{\mu\nu}(q)=\int d^4xd^4y\,\e^{-iq\cdot(x-y)}\bra J^\mu(x)J^\nu(y) \ket\,,
\eeq
where the brackets represent the average in the state under consideration.
and $J^\mu(x)$ is the electromagnetic current operator:
\beq
\label{emcurrent}
 J^\mu(x)=\frac{i}{2}[\pi^\dagger(x)
 \stackrel{\leftrightarrow}{\p^\mu} \pi(x)
 -\pi(x)\stackrel{\leftrightarrow}{\p^\mu} \pi^\dagger(x)]\,, 
\eeq
where $f\!\!\stackrel{\leftrightarrow}{\p_\mu}\!\!g=
f\p_\mu g-(\p_\mu f)g$ and where $\pi(x)$ and $\pi^\dagger(x)$ denote 
the charged pion field operators.

To illustrate the anomalous electromagnetic emission due to a DCC, we essentially 
follow the schematic model presented in \cite{Huang:1996kq,Kluger:1997cm}. The 
presence of a DCC may be modeled by splitting the pion field operators into a 
classical part $\pi_\cl(x)$, which represent the coherent DCC excitation, and the 
rest $\tpi(x)$, which describes the background pions:
\beq
 \pi(x)=\pi_\cl(x)+\tpi(x)
\eeq
The electromagnetic current \eqn{emcurrent} can be written as a sum of three
contributions:
\beq
 J^\mu(x)=J^\mu_\cl(x)+J^\mu_1(x)+J^\mu_2(x)\,,
\eeq
where $J^\mu_\cl(x)$ is the purely classical contribution, obtained from 
\Eqn{emcurrent} with the replacement $\pi\to\pi_\cl$, and $J^\mu_{n=1,2}(x)$
contains $n=1,2$ powers of the field $\tpi$ respectively.
Inserting this decomposition in \Eqn{emcor} and assuming that the system of 
background pions is electrically neutral, one obtains:
\beq
\label{terms}
 W^{\mu\nu}(q)=W^{\mu\nu}_\cl(q)+W^{\mu\nu}_1(q)+W^{\mu\nu}_2(q)\,,
\eeq
where the first term on the RHS,
\beq
 W^{\mu\nu}_\cl(q)=J^\mu_\cl(q)\,J^{\nu *}_\cl(q)
\eeq
describes the purely classical contribution and
\beq
 W^{\mu\nu}_n(q)=\bra J^\mu_n(q)\,J^{\nu\dagger}_n(q)\ket\,,
\eeq
with $J^\mu_\cl(q)$ and $J^\mu_n(q)$ the four-dimensional Fourier transform of 
$J^\mu_\cl(x)$ and $J^\mu_n(x)$ respectively.

The contribution to the rate \eqn{emrate} from the first term on the RHS
on \Eqn{terms} corresponds to the anomalous bremsstrahlung from the DCC itself.
As an illustrative example \cite{Huang:1996kq}, it can be estimated using the 
Blaizot-Krzywicki classical solution \cite{Blaizot:1992at}, described in 
\Sec{sec:DCC}. For this purpose, it is useful to notice that the electromagnetic
current \eqn{emcurrent} actually coincides with the third component of the
iso-vector current $\vV^\mu=\vpi\times\p^\mu\vpi$: 
\beq
 J^\mu(x)=\pi_1(x)\p^\mu\pi_2(x)-\pi_2(x)\p^\mu\pi_1(x)=V^\mu_3(x)\,.
\eeq
For the one-dimensional boost-invariant solution of Refs.~\cite{Blaizot:1992at,Blaizot:1994ih}, 
one has $\vpi(x)\equiv\vpi(\tau)$, with proper-time $\tau=\sqrt{t^2-z^2}$, and the 
classical iso-vector current can be written as (cf. \eqn{isovec}):
\beq
\label{Vcurrent}
 \vV^\mu_{\rm cl}(x)=u^\mu\frac{\va}{\tau}\,\theta(\tau-\tau_0)\,g(\bx_\perp)
\eeq
where $u^\mu=(\cosh\eta,\sinh\eta,0,0)$, with spatial rapidity $\eta$
defined as $\tanh \eta=z/t$. Here, following the authors of \cref{Huang:1996kq},
we have multiplied the boost-invariant solution of Blaizot and Krzywicki by a 
slowly varying function of the transverse coordinates $g(\bx_\perp)$ to account 
for the finite transverse extent of the system. The Fourier transform of the 
current \eqn{Vcurrent} is easily evaluated and one obtains, after averaging 
over the possible orientations of the iso-vector $\va$ \cite{Huang:1996kq}:\footnote{A 
similar expression can be obtained for the spectrum of real photon produced by the DCC 
classical field, see \cref{Huang:1996kq}.}
\beq
\label{ratecl}
 \frac{dN_{\ell^+\ell^-}^{\rm cl}}{dydMd^2q_\perp}=
 \frac{\alpha_{\rm em}^2}{24\pi}\bra a_3^2\ket\,
 \frac{q_\perp^2B(M^2)}{M_\perp^2\,M^3}
 \,[J_0^2(M_\perp\tau_0)+N_0^2(M_\perp\tau_0)]\,|\tilde g(\bq_\perp)|^2\,,
\eeq
where $M=\sqrt{q^2}$ is the invariant mass of the lepton pair,
$y$ and $\bq_\perp$ its longitudinal rapidity and transverse momentum
respectively, and $M_\perp\equiv\sqrt{q_\perp^2+M^2}$. 
Here, $J_0$ and $N_0$ are usual Bessel functions and $\tilde g(\bq_\perp)$ 
is the Fourier transform in transverse space of the shape function
$g(\bx_\perp)$. High transverse momentum dilepton are suppressed due
to the finite transverse extent of the source. For instance, for a 
Gaussian shape $g(\bx_\perp)=\exp(-x_\perp^2/R_\perp^2)$, one has a 
strong exponential suppression: 
$|\tilde g(\bq_\perp)|^2\propto\exp(-q_\perp^2R_\perp^2/2)$.
Notice that, as a consequence of boost-invariance in the present model, 
the rate \eqn{ratecl} does not depend on the dilepton rapidity $y$.
Finally, $\bra a_3^2\ket$ measures the fluctuations of the strength 
of the iso-vector current in the initial state. Neglecting the pion mass,
The latter can be related to the mean initial energy density $\bra \epsilon_0\ket$ 
using \Eqn{endens}.\footnote{The energy density seems to be over-estimated by a factor $4$ in 
\cite{Huang:1996kq}.}
Assuming a high temperature symmetric initial state where the 
values of the fields $(\vpi,\sigma)$ and their proper-time 
derivatives at $\tau=\tau_0$ are Gaussian random numbers of zero 
mean and of variance $\sigma_1$ and $\sigma_2$ respectively,
as in \eqn{probab}, one obtains:
\beq
 \bra a_3^2\ket=2\kappa_0^2=\frac{\tau_0^2}{3}\bra \epsilon_0\ket\,.
\eeq
with $\kappa_0=\sigma_1\sigma_2\tau_0$.

\begin{figure}[t]
\begin{center}
 \epsfig{file=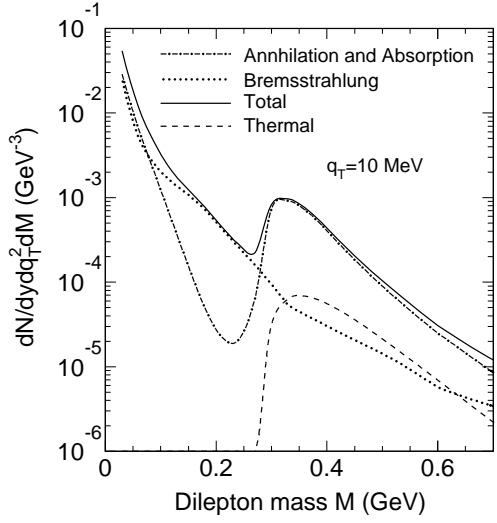,width=7.5cm}
\end{center}
\caption{\label{fig:dileptons}\small 
The $e^+e^-$ yield \eqn{emrate} as a function of the dilepton invariant mass. 
The dotted line corresponds to the anomalous bremsstrahlung from the DCC classical 
field configuration;
The dashed line represents the contribution from the surrounding thermal bath
at temperature $T$; Finally, the dot-dashed curve represents the contribution
arising from the interactions between DCC and thermal pions. The relevant 
parameters (see text) are: $T=145$~MeV, $\bra\epsilon_0\ket=58$~MeV/fm$^3$,
$\tau_0=1$~fm. The absolute normalization of the DCC contribution depends on 
the details of the shape function $g(\bx_\perp)$ for the DCC and is given here 
for a Gaussian shape $g(\bx_\perp)=\exp(-x_\perp^2/R_\perp^2)$ with $R_\perp=2$~fm. 
The relative normalization of the various contributions are not realistic (see text).
From \cref{Kluger:1997cm}.}
\end{figure}

The dependence of the yield \eqn{ratecl} on the dilepton mass at fixed 
transverse momentum is shown by the dotted curve in \Fig{fig:dileptons}. 
Also shown are the contributions from the annihilations of on-shell background 
pions, assuming that the latter form 
a thermal bath (dashed curve), as well as the contribution arising from 
the interactions between thermal and DCC pions as described below (dot-dashed curve). 
One observes that the bremsstrahlung contribution \eqn{ratecl} from off-shell 
DCC pions have no threshold and is actually mainly concentrated at low 
invariant mass $M\lesssim 2\mpi$. In this region, however, the yield
is extremely sensitive to the initial time $\tau_0$ due to the singular 
behavior of Bessel Neumann function $N_0(z)$. Moreover, this signal would probably
be very difficult to discriminate in a heavy-ion collision due to the 
large background from Dalitz decays $\pi^0\to e^+e^-\gamma$ in this 
kinematical region (see e.g. \cite{Rapp:1999ej}). 

Another possibility, suggested in \cref{Kluger:1997cm}, is that
non-coherent pions from the surrounding heat bath may annihilate on the 
coherent DCC state, forming dilepton pairs of typical invariant masses 
in a rather narrow window around $M\simeq 2m_\pi$, thus avoiding the 
$\pi^0$ Dalitz-decay region. The corresponding contribution to the dilepton
rate \eqn{emrate} comes from the second term on the RHS side of \Eqn{terms}.
Assuming a thermal bath of non-interacting pions at temperature $T$, 
one obtains, after some calculations, the following contribution \cite{Kluger:1997cm}:
\bea
 \frac{dN^{(1)}_{\ell^+\ell^-}}{dydMd^2q_\perp}
 &=&\frac{2\alpha_{\rm em}^2}{3\pi^3}\frac{1}{M}\int\frac{d^3k}{2\omega_k (2\pi)^3}
 \left[\frac{(k\cdot q)^2}{M^2}-m_\pi^2\right] \nn
 &&\left\{ (1+n^-_k)\,|\pi_{\rm cl}(r)|^2+(1+n^+_k)\,|\pi_{\rm cl}(-r)|^2
 \right.\nn
\label{rate1}
 &&\,\,+\left. n^-_k\,|\pi_{\rm cl}(l)|^2+n^+_k\,|\pi_{\rm cl}(-l)|^2\right\}.
\eea
where $\omega_k=\sqrt{k^2+m_\pi^2}$, $k^\mu=(\omega_k,\bk)$, $r^\mu=
(k+q)^\mu$, $l^\mu=(k-q)^\mu$ and $n_k^\pm$ denotes the equilibrium
Bose-Einstein distribution at temperature $T$ for positive and negative
pions. The first two terms in curved brackets on the RHS of \eqn{rate1}
correspond to the decay of the classical DCC field into an on-shell
thermal pion and a dilepton pair: $\pi_\cl\to\tpi\ell^+\ell^-$, whereas
the last two terms represent the annihilation of a DCC pion against a 
thermal one: $\pi_\cl\tpi\to\ell^+\ell^-$. The latter contribution has
a threshold at $M=2\mpi$, and produces an enhancement of the dilepton
yield around $M\simeq2\mpi$, as is clearly visible on \Fig{fig:dileptons}.
A more detailed analysis has been performed in \cref{Kluger:1997cm}, using
classical field  dynamics, which indicates that this enhancement is restricted 
to dilpeton momenta $|\bq|\lesssim300-500$~MeV. 

We would like to stress that \Fig{fig:dileptons} is only meant here to illustrate 
the dependence of the various contributions to the dilepton yield with the dilepton mass. 
The relative normalization of the various contributions should not be taken too seriously
as it is most probably far from realistic. First, because the present model completely 
neglects the small probability that a DCC actually forms. In a more realistic description, 
the rates \eqn{ratecl} and \eqn{rate1} should be accordingly rescaled. Second, because 
the parameters used here correspond to an unexpectedly large contribution from the DCC. 
Indeed, as a rough estimate of the energy density of the background thermal pion gas, 
one may use the Stefan-Boltzmann law for a gas of non-interacting massless pions: 
$\epsilon_\pi=g(\pi^2/30)T^4$ with $g=3$ the pion degeneracy. For $T=145$~MeV used in
\Fig{fig:dileptons}, one obtains $\epsilon_\pi\approx55$~MeV/fm$^3$ which is similar 
in magnitude as the initial energy density of the DCC state in the above calculation: 
$\bra\epsilon_0\ket=58$~MeV/fm$^3$. This is certainly unrealistic, at least in the 
context of the non-equilibrium chiral phase transition scenario discussed in earlier 
sections. In a more realistic situations, one expects $\bra\epsilon_0\ket\ll\epsilon_\pi$
and the various contributions to the dilepton rate shown in \Fig{fig:dileptons} should 
be appropriately rescaled.

Finally, we mention that the generic dilepton spectrum in this region is dominated by 
$\eta$ Dalitz-decays, thus making the measurement of dileptons from a DCC a very difficult 
task. The possibility of such a measurement would certainly require making appropriate
cut on the data, probably in correlation with other hadronic observables. This certainly 
deserves further investigations.

\subsubsection{Further speculations}

It is usually believed that, in the context of high-energy nuclear collisions,
the best place for DCC searches is the central rapidity region of the most central 
collisions, because this is where one expects the deposited energy density to be the 
highest, thereby providing the most favorable conditions for chiral symmetry 
restoration. However, as we have seen previously, this is also a place where
DCC signatures might be difficult to observe. Here, we briefly mention alternative
suggestions.

Rajagopal has argued in \cref{Rajagopal:2000yt} that semi-peripheral collisions at 
RHIC energies might provide more favorable conditions for efficient quenching as 
well as sooner freeze-out, hence a better chance that DCC pions be visible in the 
detectors. Another interesting motivation for looking at such collision events has 
been proposed by the authors of \cref{Minakata:1995gq,Asakawa:1998st}, who argued 
that strong electromagnetic fields could result in a coherent excitation of the $\pi_0$ 
field through the chiral $U(1)$ anomaly. Such an excitation could be subsequently 
amplified by the quench mechanism, as discussed at length in the present report. 
More peripheral collisions might also be interesting for DCC searches because they 
essentially consists in the collision of the pion clouds of the respective nuclei. 
The relevant dynamics is, therefore, that of low-energy pion fields and the 
corresponding physical picture might be closer to the original baked-Alaska scenario
(see \Sec{sec:DCC}). One could further imagine to favor a particular isospin direction 
by colliding asymmetric (\eg neutron rich) nuclei.\footnote{This suggestion is due 
to D.~Vautherin (private communication).} 

Another place of special interest to look for DCC signatures is the projectile or
target fragmentation region of nuclear collisions, as advocated by Bjorken 
\cite{Bjorken:eRHIC,Bjorken:private}. The basic idea is 
that, viewing the collision in the rest frame of one of the ions, the projectile 
traverses the target at essentially the speed of light and ejects all the resting 
nuclear matter away almost instantaneously. If the smoking ruins left behind relax 
sufficiently slowly toward ordinary vacuum, the relevant dynamics should be that 
of soft pion fields and relaxation process should be characterized by low-momentum 
pion emission.

It is also important to stress that present theoretical calculations in the context 
of the quench scenario in heavy-ion collisions indicate that the DCC might actually 
be more difficult to observe than originally thought 
\cite{Gavin:1993bs,Randrup:1996es,Rajagopal:1997au,Serreau:2000tb,Serreau:2003gj} 
(see subsection \ref{sec:unpolarized}). 
In particular, the results of \cref{Serreau:2000tb} suggest that one should try to 
isolate as much as possible the various Fourier modes of the DCC pion field. 
Multi-resolution analysis techniques such as the discrete wavelets described above 
provide very efficient tools for such task. Moreover, in regards to the rather small 
expected probability that a DCC forms, it would be desirable to perform appropriate 
cuts on the data sample. For instance, one could restrict the analysis to events 
with a particularly large pion multiplicity, say larger than the average, in a 
given region of momentum space. We shall discuss such type of analysis in the 
experimental part of this review.

Finally, we mention that a number of other observables, which we have not
described here, have been discussed in the literature. This includes \eg
non-trivial charged-pion fluctuations \cite{Hwa:2001xn} -- which present the 
advantage that one does not have to reconstruct the $\pi_0$'s, or anomalous 
strangeness production \cite{Gavin:2001uk,Kapusta:2000ny}, etc.

\section{\label{sec:exppart} Experiment}     

\subsection{\label{sec:experiments} DCC-search experiments}

Can a DCC domain be produced in actual hadron or heavy-ion collisions and, if so, 
are we capable of detecting them? In the search for the answers to these 
questions, several experiments have been carried out and some have been planned for 
future. Some of them are based on the study of cosmic ray showers and others are 
accelerator-based. The list of all these experiments is given in Table~\ref{expt_list}. 
Below we discuss some of these experiments and their capability to look for DCC in 
terms of measurement of right observables. The results from these experiments will 
be discussed in detail in \Sec{sec:expt_results}.

\begin{table}[h]
\caption{\label{expt_list} \small Experiments related to search for disoriented chiral 
condensates.}
\vspace{0.5cm}
\begin{tabular}{|c|c|c|c|c|}
\hline
Year & Type & Experiment & $\sqrt{s}$ & Observable \\
\hline
1972-80  & Cosmic Ray, Balloons & JACEE \cite{jacee} & $\ge$ 1.7 TeV & No. of $\gamma$ \& $ch$ in 5 $\le$ $\eta$ $\le$ 9 \\
\hline
1980  & Cosmic Ray,  & Brazil-Japan & $\ge$ 1.7 TeV & No. of $\gamma$ \& $ch$ in  \\
  & Mt. Chacaltaya \& Fuji &  &  & forward $\eta$  \\
\hline
1977-91  & Cosmic Ray, Mt. Pamir & PAMIR \cite{pamir} & $\ge$ 1.7 TeV & No. of $\gamma$ \& $ch$ in forward $\eta$  \\
\hline
1982  & Accelerator based,  & UA5 \cite{ua5_540,ua5_900}&  540 GeV & No. of $\gamma$ \& $ch$ in  $\sim$ 4$\pi$ \\
and  & nucleon-nucleon  &  & and   & \& asymmetry in electro- \\
1986  & collision at CERN  &  & 900 GeV  & magnetic \& hadronic energy \\
\hline
1983  & Accelerator based,  & UA1 \cite{ua1} & 540 GeV & asymmetry in electromagnetic  \\
  & nucleon-nucleon  &  & &\& hadronic energy \\
  & collision at CERN  &  &  & in -3.0 $\le$ $\eta$ $\le$ 3.0 \\
\hline
1996  & Accelerator based,  & D0  & 1.8 TeV & asymmetry in electromagnetic  \\
  & nucleon-nucleon  & and & &\& hadronic energy \\
  & collision at FERMILAB  & CDF \cite{cdf}  &  & in -4.0 $\le$ $\eta$ $\le$ 4.0 \\
\hline
1997  & - do -  & MiniMAX~\cite{minimax}  & 1.8 TeV &No. of $\gamma$ \& $ch$ in   \\
  &   &  & & 3.2 $\le$ $\eta$ $\le$ 4.2 \\
\hline
1998  & Accelerator based,  & WA98~\cite{wa98_global,wa98_local,wa98_local_cen}  &  17.3 GeV/n &No. of $\gamma$ \& $ch$ in   \\
and  & nucleus-nucleus  &  & & 2.9 $\le$ $\eta$ $\le$ 3.75 \\
2001  & collision at CERN  &  &  &  \\
\hline
1999  & - do - & NA49~\cite{na49}  & 17.3 GeV/n &$p_{T}$ fluctuation  in charged\\
 &  &  & & particles in 4.0 $\le$ $\eta$ $\le$ 5.5 \\
\hline
2001  & Accelerator based,  & PHENIX~\cite{phenixqm}  & 200 GeV/n &  No. of $\gamma$ \& $ch$ in\\
and & nucleus-nucleus  &  & & -0.35 $\le$ $\eta$ $\le$ 0.35 \\
2002 & collision at RHIC  &  &  & \\
\hline
\end{tabular}
\end{table}

\subsubsection{\label{sec:cosmic_ray_expt} Cosmic ray experiments}

\begin{figure}[t]
\begin{center}
\epsfig{figure=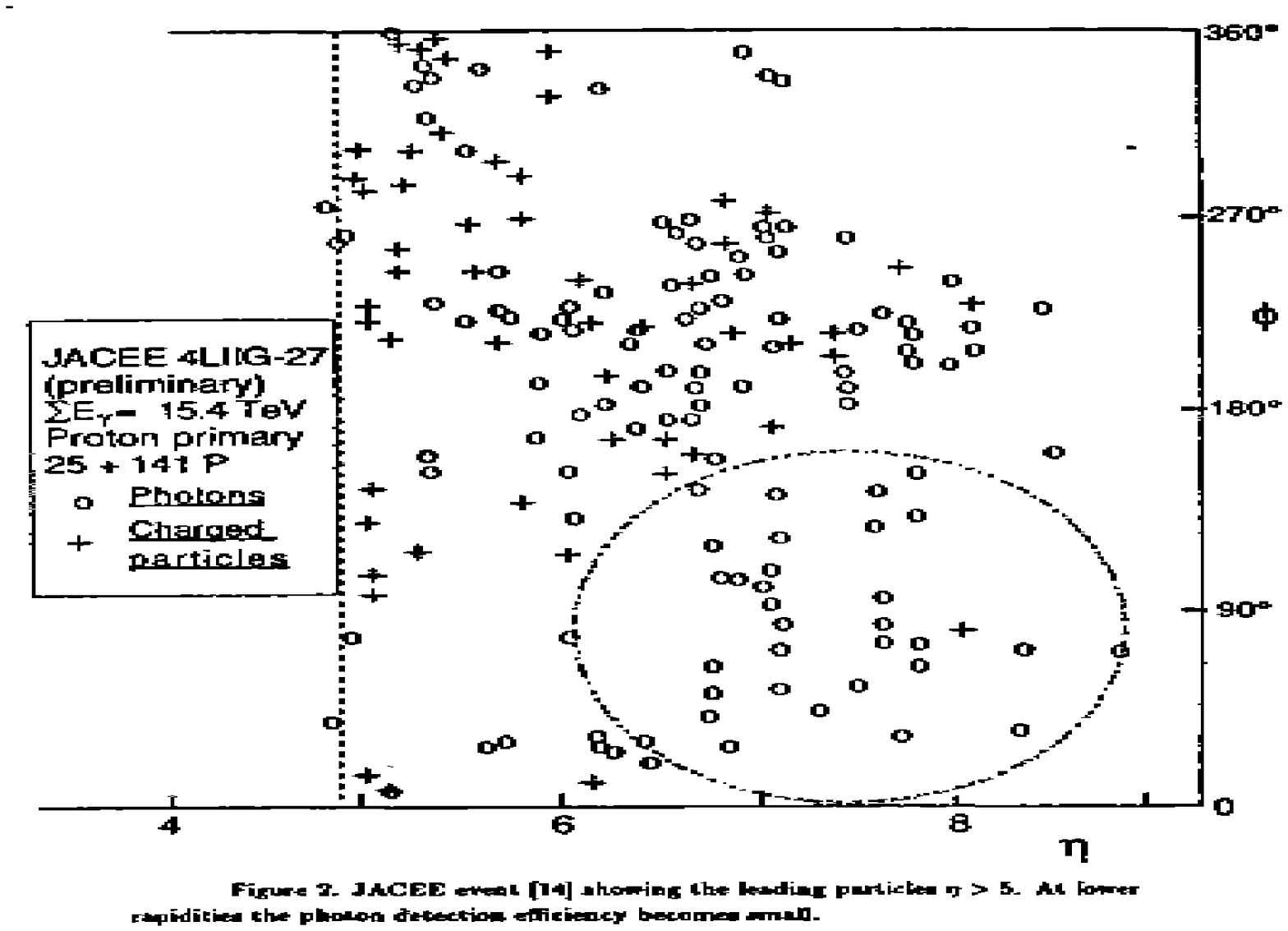,width=10.cm}
\caption{\label{jacee_expt}\small
The $\eta-\phi$ phase space distribution  of an event from the 
JACEE collaboration~\cite{jacee}. The region indicated by the circle shows 
a region with anomalously high ratio of gammas to charged secondaries,
called as anti-Centauro event.}
\end{center}
\end{figure}

Cosmic ray experiments were performed by exposing stacks of
nuclear emulsion chambers at high mountain tops or sending
them to high altitudes in balloons. They recorded secondary 
particles (mostly pions and gamma rays) originating in the 
successive interactions of a particle of primary cosmic 
radiation (such as proton) with atmospheric nuclei. 
The primary cosmic ray particle energies are in the range of
$100$~TeV to $10000$~TeV~\cite{jacee}. A cosmic ray
event can therefore be thought of as a result of
collision between a projectile nucleon of an
estimated energy of about 1500 TeV (equivalent
center of mass energy $\sqrt{s}\geq~1700$~GeV) and
an atmospheric nuclei~\cite{ua5_900}.
The mountain top experiments, 
carried out at Mt. Chacaltaya in Bolivia, at Pamir in Russia 
and Mt. Fuji in Japan, reported events having large number of 
charged hadrons (about 63 - 90) and very few or no photons. 
These events are known as ``Centauro-type'' events. The first 
Centauro-type event called ``Centauro-I''  was reported in 1972. 
The JACEE collaboration~\cite{jacee}, which had sent emulsion chambers in 
balloons to high altitudes also observed events having anomalously high 
number of gammas 
in comparison to charged secondaries. Such events are called
``anti-Centauro-type'' events. One such interesting event
reported by the JACEE collaboration~\cite{jacee}
is shown in \Fig{jacee_expt}, where the 
distributions of charged particles and gammas are presented in 
$\eta-\phi$ space. One of the possible explanations for these 
peculiar events with large isospin fluctuations in cosmic ray 
experiments is the possibile production of a DCC in high-energy collisions. 

There has been critical reviews of the
  exotic characteristics of Centauro-I. In Refs.
  ~\cite{kopenkin,ohsawa} the
  authors found that there is no original
  correspondence between the clusters in upper and
  lower chambers of the emulsion chamber experiment.
  The authors of Ref.~\cite{kopenkin} claim that such an
  event no longer required any new physics to
  describe the event structure. In Ref.~\cite{ohsawa}
  there is a suggestion that the event structure may have
  arisen due a bundle of target interactions. They
  however claim that the characteristics, like
  hadrons not accompained by any $\gamma$ rays,
  can still not be understood and are very different
  from commonly observed cosmic ray events.


Recent results from a systematic study of the neutral pion fraction 
distribution (using sensitive techniques developed for DCC search) 
in high-energy cosmic ray families detected at high altitudes at 
Pamir~\cite{pamir} do not exclude such a possibility. The Pamir experiment took
data by exposing thick lead chambers of homogeneous structure at Pamir in 
1977-1991. The chamber thickness was between 40 - 110 cm Pb and, hence, had 
a detection probability of $\gamma$-hadron families close to 1.
One of the limitations of cosmic ray experiments is that, 
because of the high energy detection threshold
in experiments with emulsion chambers, the investigation of high-energy
interaction is limited to the projectile particle fragmentation region. 
The situation is opposite to that in collider experiments where secondary 
particles emitted with small angles in far forward regions of the collision
cannot be detected because of the beam pipe.

\subsubsection{\label{sec:nncoll} Nucleon-Nucleon collisions}

The observation of Centauro-type events motivated experimentalists to
look for such events in accelerator-based experiments, even before the concept
of the DCC came into picture. The UA1 experiment~\cite{ua1} 
searched for events having characteristics of cosmic ray Centauros 
in $p-\bar{p}$ collisions at $\sqrt{s}=540$~GeV, using information on 
charged particle  multiplicities, transverse momenta and the criteria of 
large hadronic and low electromagnetic energy content in calorimeters. 
The central image chamber detector of the UA1 apparatus was used to get
charged particle tracks. Momenta were obtained from the curvature of
the tracks in uniform magnetic field of 0.56 Tesla. The energy information
was obtained from layers of finely divided sheets of lead and 
plastic scintillators which formed the electromagnetic calorimeter.
The calorimeter surrounded the central detector. The electromagnetic 
energy deposition was measured from the first 4 radiation length of material.
The hadronic energy contribution was estimated from the rest of the 
electromagnetic calorimeter and from energy deposition in the 
iron scintillator calorimeter behind it. No Centauro-like events 
were seen out of a total data sample of 48K events. Based on this the 
UA1 collaboration reported an upper limit for the 
production cross section of the order of a micro-barn for a Centauro events 
at collider energies equivalent to 155 TeV incident energy on stationary 
nuclei.

The UA5 experiment~\cite{ua5_540,ua5_900}, through an
analysis of 3600 minimum bias events for 
correlation between photon and charged-particle
multiplicities, reported no events having features corresponding to the 
Centauro events as observed in cosmic-ray emulsion chambers.
The detector used for charged particle measurement
was a streamer chamber, with a coverage of almost the entire 4$\pi$ solid angle.
The photon measurements were based on electromagnetic showers resulting
from photons converting through 
the walls of the vacuum chamber (0.05 to 2.6 radiation length of
material) situated between the two streamer chambers. The search were
carried out at $\sqrt{s}$ energies of 540 GeV and 900 GeV without any success.
The search was extended to high $\sqrt{s}$ energy of 1.8 TeV in the 
D0 and CDF~\cite{cdf} experiments at Fermilab.
These experiments used the technique of the asymmetry in hadronic to 
electromagnetic energies for DCC search. They also 
reported null results. 

Subsequently a dedicated experiment at Fermilab Tevatron, 
called MiniMAX \cite{minimax}, 
was designed by Bjorken, Kowalsky, Taylor and collaborators 
to look for DCC signals
in the far forward rapidity region of $p-\bar{p}$ collisions 
at $\sqrt{s}=1.8$~TeV. 
This experiment, located at the C0 area of the  Fermilab Tevatron was
designed to measure the ratio of charged-to-neutral pions. The
acceptance in the lego space of pseudo-rapidity $\eta$ and azimuthal
angle $\phi$ is roughly a circle of radius $0.65$ centered at
$\eta=4.1$. The MiniMAX apparatus included Multi-Wire Proportional
Chambers (MWPC) to measure charged particles, movable 1 radiation length 
($X_{0}$) Pb converter to improve the resolution of charged-particles, 
a trigger scintillator, an array of 28 lead-scintillators acting 
as electromagnetic calorimeter located behind the MWPC to measure 
photons and upstream tagging 
detectors. The analysis of the data with about $8 \times 10^6$ events 
was done using the technique of robust observables, discussed earlier in \Sec{sec:pheno}, 
to take care of the small acceptance of the detector and various efficiency 
factors.
The analysis yielded results consistent with no DCC production.

\subsubsection{\label{sec:AAcol} Nucleus-Nucleus collisions}

Various DCC searches have also been pursued in high-energy heavy-ion collision
experiments. This is motivated by the high energy density and temperature achieved 
in such collisions, which provide favorable conditions for the DCC production 
scenario based on the non-equilibrium chiral phase transition \cite{Rajagopal:1993ah}, 
as discussed in \Sec{sec:chiralPT}. 
Among the various heavy-ion experiments, WA98 was the first one
which had the capability of measuring photon and charged particle
multiplicities over a common region of phase-space, which is a key ingredient 
for DCC search through neutral fraction fluctuations~\cite{wa98_local,wa98_global}. 
The data analyzed by the WA98 collaboration were taken with the $158$~AGeV Pb beam 
of the CERN SPS on a Pb target of $213~\mu$m thickness during a period of WA98 
operation without magnetic field. The analysis makes use of a subset of detectors 
of the WA98 experiment which were used to measure the multiplicities of charged 
particles and photons. Charged-particle hits ($N_{\mathrm ch}$) were counted 
using a circular Silicon Pad Multiplicity Detector (SPMD) 
located $32.8$~cm downstream from the target. It provided uniform 
pseudo-rapidity coverage in the region $2.35<\eta<3.75$. 
The photon multiplicity was measured using a preshower Photon Multiplicity 
Detector (PMD) placed $21.5$~m downstream of the target 
and covering the pseudo-rapidity range $2.9<\eta<4.2$.  
It consisted of an array of $53,200$ plastic scintillator pads placed behind $3X_0$ 
thick lead converter plates. Clusters of hit pads having total energy 
deposit above a hadron rejection threshold are identified as
``photon-like''. 
For the DCC analysis the pseudo-rapidity region of common coverage of the
SPMD and PMD was selected ($2.9< \eta< 3.75$) by WA98 collaboration.  
The acceptance in
terms of transverse momentum extends down to $p_T=30$~MeV/c,
although no explicit $p_T$-selection was applied.
As far as analysis goes, there has been a
global search of DCC formation ~\cite{wa98_global} in the WA98 experiment. 
Asymmetry in $N_{\gamma}$ to $N_{ch}$ ratio over the full common phase space 
of the photon and charged particle detectors was used to get some idea
regarding DCC formation in Pb-Pb collisions at $158$~AGeV.
Following the theoretical expectations \cite{Bjorken:1991sg,Blaizot:1992at,Rajagopal:1993ah}
that the isospin fluctuations, caused by formation of DCC, may produce clusters of
coherent pions in phase space forming localized domains, the WA98
collaboration has also searched for localized domains of DCC 
\cite{wa98_local,wa98_local_cen}. 
This was done by trying to detect large and localized
fluctuations in the ratio of photons--to--charged-particles numbers.

An analysis using the momentum information of charged particles without
looking at photons was carried out by the NA49 experiment~\cite{na49} 
at the CERN SPS to look for DCC. There, particles were selected that had a 
measured track length of more than $2$~m in one of the two  
Main Time Projection Chambers (MTPC) outside the magnetic field and were 
also observed in at least one of the Vertex TPCs inside the superconducting 
magnets. The NA49 collaboration studied particles in the region: 
$0.005<p_T<1.5$~GeV/c in transverse momentum and $4<y_\pi<5.5$ in rapidity.
Both the WA98 and NA49 experiments reported results in terms of upper limit 
on the frequency of DCC formation. 

At the RHIC, the STAR experiment \cite{starnim} plans to look for DCC 
formation using the combination of photon-multiplicity detector (PMD) and the 
electromagnetic calorimeter (EMCAL) for photon detection along with the time 
projection chamber (TPC) and forward-time projection chamber (FTPC) for 
charged-particle detection \cite{starnim}. 
The PHENIX collaboration \cite{phenixqm} has also demonstrated its capability 
to study isospin fluctuations, essentially through 
charged-particle measurements 
using drift and pad chambers and photon measurements from the electromagnetic 
calorimeter of the PHENIX detector \cite{phenixqm}. 
Finally, we mention that the ALICE experiment \cite{alicetp} at the upcoming 
LHC at CERN will have the combination of a PMD and a charged-particle 
multiplicity detector (FMD) in the forward rapidity region as well as an 
electromagnetic calorimeter (PHOS) and a time projection chamber (TPC) near 
mid-rapidity, which allow for DCC search in different rapidity regions.

\subsection{\label{sec:dcc_model} Simulated DCC events}

It is important to model DCC pion production in realistic heavy ion collision 
event generators in
order to study the sensitivity of the various techniques developed to look for 
DCC-like fluctuations in experimental data. This is also useful to get a clear 
idea of the limitations of both the experiments and the analysis 
tools at hand. 
Finally this provides a basis to interpret actual data, as done \eg by the 
WA98~\cite{wa98_local,wa98_global} and NA49~\cite{na49} collaborations. 
Here, we present a simple model proposed in Refs.~\cite{wa98_global,strenght}, 
where the generic particle production is governed by the VENUS event generator \cite{venus} 
and the DCC pions are introduced by hand at the freeze-out stage. 
The goal is to observe the effect of DCC domains on measured 
quantities, assuming 
DCC pions survive till the freeze-out time. As discussed previously in 
\Sec{sec:lifetime}, this assumption is justified by theoretical calculations
\cite{Steele:1998ye}. The main features of the model are the following:
\begin{itemize}
\item {\it DCC-type fluctuations}:\\ For DCC simulation in a given domain, 
  the identity of charged pions taken pairwise ($\pi^+\pi^-$) is changed to 
  neutral ones ($\pi^0\pi^0$) and vice-versa~\cite{wa98_global,strenght}, according to 
  the ideal DCC inverse square-root law, \Eqn{DCCsign}.\\
\item {\it Domain size}:\\ The size of a domain is defined 
  in terms of its extent in pseudo-rapidity ($\eta$) and azimuthal angle 
  ($\phi$). In most analysis described in the following, one choses a domain-size 
  of one unit in rapidity: $\Delta\eta=1$ and a variable extent $\Delta\phi$ in 
  azimuthal angle. This is not a bad assumption as it is expected that
  DCC doamins will occur in small regions and will mostly modify the 
  production of low momentum pions. The influence of DCC on the charge-neutral
  pion ratio may then be limited to localized $\eta$-$\phi$ regions due to
  the motion of the DCC domain within the overall collective motion or 
  strong radial flow in heavy ion collisions.

\item {\it Number and $p_{T}$ spectrum of DCC pions}:\\
  For simplicity all the pions inside a chosen domain are considered to 
  be DCC-type. As already emphasized, theoretical calculations suggest that 
  DCC pions have low $p_{T}$. In order to study the usefulness of $p_{T}$ 
  information of particles, as discussed in \Sec{sec:poslim_expt} below, 
  one typically consider two cases \cite{poslim}: 
  (i) all pions with $p_{T}\le 150$~MeV are taken as DCC pions in a given 
  domain; (ii) a variable number (depending on the domain size) of DCC-pions
  with $p_{T}\le 150$~MeV are added by hand to the generic sample of pions 
  in a given domain. The number of these additional pions depends on the size 
  and energy density of the DCC domain. Domains of typical radius $R\sim 3-4$~fm
  are expected from numerical simulations of the non-equilibrium chiral phase 
  transition in the context of heavy-ion collisions \cite{Gavin:1993bs,Gavin:1993px,kapusta}. 
  Assuming the energy density $\epsilon_{\rm DCC}$ of the DCC domain to be 
  about $50$~MeV/fm$^3$, the number of DCC pions is estimated from the relation
  $\frac{4}{3} \pi R^3(\epsilon_{\rm DCC}/m_{\pi})$ to be between $40$ and $100$. This
  probably gives a reasonable upper bound on actual numbers to be expected.\\ 
\item {\it $\pi^{0}$ decay and photon fraction}:\\
  After introducing DCC-type fluctuations the $\pi^{0}$'s are allowed
  to decay. The neutral pion fraction $f$ then gets approximately modified to: 
  \begin{equation}
    \label{new_f}
    f\approx f^\prime=\frac{\Ng/2}{\Ng/2 + \Nc}\,,
  \end{equation}
  where $\Ng$ is the photon multiplicity and $\Nc$ is the total multiplicity 
  of charged pions. \\
\item {\it Percentage of events being DCC-type}:\\
  The sets of events with DCC will be referred to as ``nDCC'' events.
  The ensemble of nDCC events may or may not have all the events of DCC type and
  the percentage $\alpha$ of DCC-type events in a given ensemble can be varied 
  to study the sensitivity of various physical and detector related effect to 
  the observation of DCC.
\end{itemize}

\begin{figure}[t]
\begin{center}
\epsfig{figure=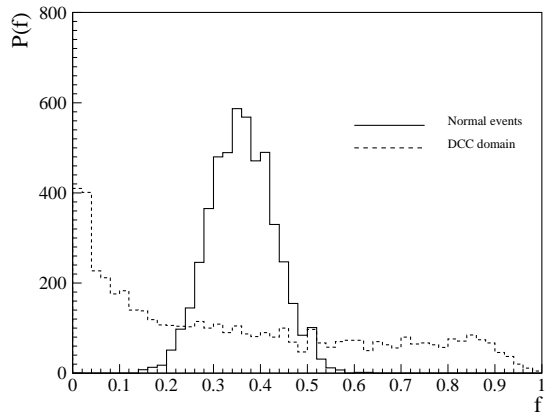,width=8cm}
\caption{\label{f_model_expt}\small
Neutral pion fraction ($f$) distribution from a DCC-domain as incorporated
in the model for simulated DCC events (see text). The corresponding distribution 
for normal (non-DCC) events is also shown for comparison. From \cref{bedanga_thesis}.}
\end{center}
\end{figure}

In order to understand the various techniques used for DCC search, one 
can apply them to ``realistic'' simulated DCC events as described above.
For this, one assumes \cite{bedanga_thesis} a single DCC domain lying within the 
common pseudo-rapidity coverage of two hypothetical multiplicity detectors for photon 
and charged particles. To simulate what happens in a true experimental situation, 
it is necessary to include detector related effects. For photon and charged-particle 
detection the detection efficiencies are taken to be about $70\pm 5\%$ and $95\pm2\%$ 
respectively. It is also known that charged particles sometimes lead to photon-like 
signals. Such a contamination can be as high as $25 \%$ \cite{pmdnim}. For simulations,
one therefore includes a $25 \%$ charged particle contamination in the photon signal. 
These parameters are taken considering the WA98 PMD \cite{pmdnim} and SPMD \cite{spmdnim} 
detectors in view. Finally, 
we have the size of DCC domain in $\phi$, i.e., $\Delta \phi$, varied. To study the 
sensitivity one also prepares several sets of data with different DCC fractions
$\alpha$ from $1\%$ to $100\%$, mixing normal events and DCC-type events in an 
appropriate manner. The $f$-distribution in a localized DCC region of $\Delta\eta=1$ 
and $\Delta\phi=90^\circ$ within the common acceptance of the detectors is shown in 
\Fig{f_model_expt}. Also shown is the $f$-distribution of normal events.
With the model for DCC event simulation in hand, we now discuss on analysis 
techniques developed specifically for DCC search. 

\subsection{\label{sec:expt_tech} Experimental techniques developed for DCC search}

In this section we discuss the various techniques developed for looking 
at charged-neutral fluctuations or DCC-type fluctuations. We shall apply 
some of these techniques to simulated DCC events as discussed above and study 
their sensitivity to look for anomalous fluctuations. We follow the  
several such analysis methods proposed in the literature, among
which:
\begin{enumerate}
\item $\Ng-\Nc$ correlation~\cite{wa98_global,wa98_local,wa98_local_cen};
\item discrete wavelet analysis~\cite{Huang:1995tv,strenght};
\item robust observables~\cite{minimax};
\item $\Phi$-measure~\cite{bedanga};
\item event-shape analysis~\cite{eventshape}.
\end{enumerate}
The techniques listed in the first three items have been used in actual experimental 
analysis, so we describe them in detail in what follows. 
The technique based on event shape analysis is discussed only briefly. In later 
section~\ref{sec:expt_results} we present the experimental results based on these 
analysis techniques.

\subsubsection{\label{sec:corr} Photon--to--charged-particle multiplicity correlation}

In order to look for any possible fluctuation
in photon and charged particle multiplicities, which may have
non-statistical origin, it is necessary to
look at the correlation between $\Ngl$ and 
$N_{\rm ch}$ as a function of resolution in $\eta-\phi$
phase-space~\cite{wa98_global,wa98_local,wa98_local_cen}. 
We first discuss the actual method of analysis and 
then show the sensitivity of the method.

\begin{figure}[t]
\begin{center}
\epsfig{file=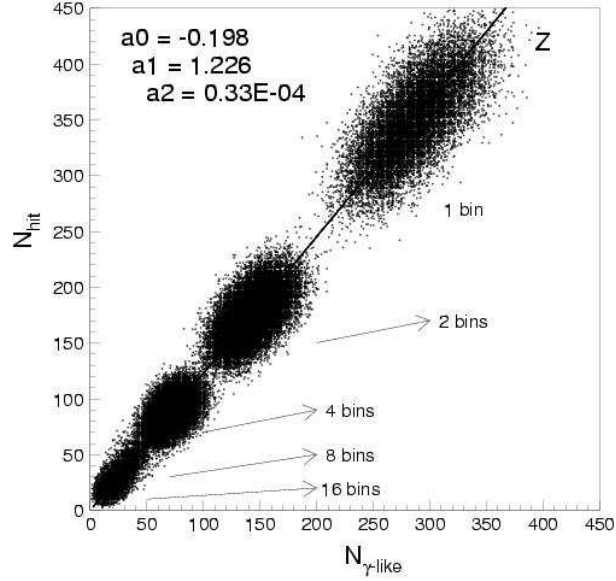,width=8.cm}
\caption{\label{ngam_nch_cor1}
\small $\Ngl$ vs. $\Nc$ correlation plot
for various bins in $\phi$, obtained from normal events. Also shown
are the correlation line and its parameters. From \cref{bedanga_thesis}.}
\end{center}
\end{figure}

In this analysis method event-by-event correlation between 
$\Ngl$ and $\Nc$ can be studied in different $\phi$-segments, 
by dividing the entire  $\phi$-space into $2$, $4$, $8$ and $16$ bins of equal
size. The correlation plot for each of the $\phi$-segments, starting with the 
case of one single bin -- no segmentation -- are shown in \Fig{ngam_nch_cor1} 
for a sample of normal events. A common correlation axis ($Z$) is then 
obtained for the full distribution by fitting the $\Ngl$ 
and $\Nc$ correlation with a second order polynomial. 
The correlation axis with fit parameters are shown in the figure.
The distances of separation ($D_{Z}$) between the data points
and the correlation axis is then calculated with
the convention that $D_{Z}$ is positive for points below the $Z$-axis.
The distribution of $D_Z$ represents the relative fluctuations of
$\Ngl$ and $\Nc$ from the correlation axis at any 
given $\phi$ bin.

\begin{figure}[t]
\begin{center}
\epsfig{figure=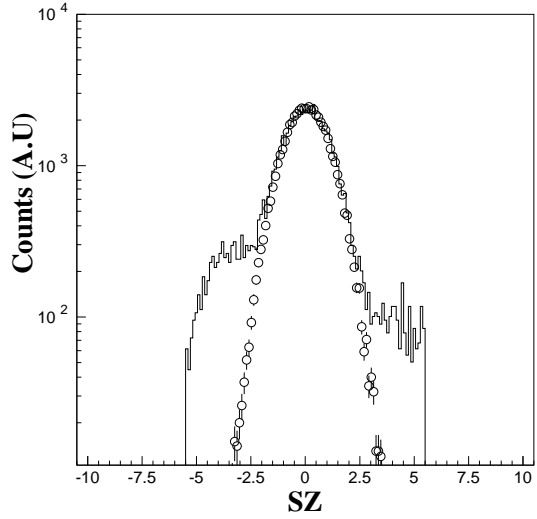,width=8cm}
\caption{\label{sz_result} \small
$S_Z$ distribution for normal events (open circles) and DCC-type of 
events (solid histogram) with $\Delta\phi=90^\circ$ DCC domain. From \cref{bedanga_thesis}.}
\end{center}
\end{figure}

One then applies this technique to simulated DCC events. The DCC
events considered corresponds to a domain size of $90^\circ$ in $\phi$. 
In order to compare the fluctuations for different $\phi$ bins in the 
same level, one introduces a scaled variable, $S_{Z} = D_Z/s(D_Z)$, 
where $s(D_Z)$ represents the root-mean-squared (rms) deviation of the 
$D_{Z}$ distribution for normal events. The presence of events with 
localized fluctuations in $\Ngl$ and $\Nc$ at a given $\phi$ 
bin is expected to result in a broader distribution of $S_Z$ compared to those 
for normal events at that particular bin. 
The typical $S_Z$ distributions for both class of events for 4 bins 
in $\phi$ are shown in \Fig{sz_result}.
One clearly observes a difference in width of the $S_Z$ distribution 
between the two class of events. So the rms deviation of the $S_Z$ 
distribution quantifies the amount of fluctuation in such an analysis method.

\begin{figure}[t]
\begin{center}
\epsfig{figure=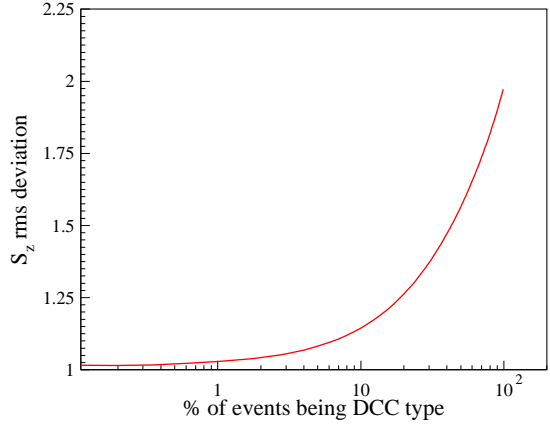,width=8cm}
\caption{\label{sz_sensitivity}\small
RMS deviation of $S_Z$ distribution for DCC-type of 
events with $90^\circ$ DCC domain in $\phi$ as a function of percentage of
events being DCC-type. This plot shows the sensitivity of the method. From \cref{bedanga_thesis}.}
\end{center}
\end{figure}
   
Figure~\ref{sz_sensitivity} shows the sensitivity of the method for picking 
up DCC-type fluctuations for a case of $90^\circ$ domain size. The width of the
$S_Z$ distribution has been plotted as a function of the number of events being 
DCC-type. As expected the rms deviation of $S_Z$ distribution increases with 
increase in $\%$ of DCC-type events in the sample. 

This technique has been applied to WA98 experimental charged-particle and
photon multiplicity data~\cite{wa98_global,wa98_local,wa98_local_cen} 
to look for extra DCC-type fluctuations as compared to normal events.

\subsubsection{\label{sec:dwt} Discrete wavelet analysis}

The discrete wavelet technique (DWT), described in \Sec{sec:pheno} above, has 
the beauty of analyzing a distribution of particles at different length scales with 
the ability of finally picking up the right scale at which there is a 
fluctuation. The method has been shown to be quite powerful to search 
for DCC-type of fluctuation~\cite{Huang:1995tv,Randrup:1997kt,strenght}.
Here we discuss how this method is actually appiled in analysing experimental
data.
As before, we analyze the simulated normal and DCC events using 
the multi-resolution DWT technique to search for bin-to-bin fluctuations
in the charged-particle and photon multiplicity distributions.
The analysis has been carried out in \cref{strenght} by making $2^j$ bins in $\phi$ 
where $j$ is the resolution scale. The input to the DWT analysis is a sample 
distribution function at the highest resolution scale $j_{max}$. In most 
analysis presented below, the sample function corresponds to the photon
fraction:\footnote{Neglecting detector effects, the photon fraction 
\eqn{gamfrac} can be approximately related to the neutral pion fraction
through (see \Eqn{new_f}): $f_\gamma\approx2f/(1+f)$. Assuming ideal DCC 
production, the corresponding probability distribution is given by:
$dP_{\rm DCC}(f_\gamma)/df_\gamma\approx1/\sqrt{f_\gamma(2-f_\gamma)^3}$.
}
\begin{equation}
\label{gamfrac}
    f_\gamma=\frac{\Ng}{\Ng+\Nc}
\end{equation}
where $\Ng$ and $\Nc$ are the multiplicities of photons and
charged particles respectively. The $\phi$-space is binned as discussed above 
and the analysis is carried out using the so-called $D-4$ wavelet basis \cite{refD4}.
The output of the DWT analysis consists of a set of wavelet or 
father function coefficients (FFCs) at each scale, from $j=0,\ldots,j_{max}-1$. 
The FFCs at a given scale carry information about the 
degree of fluctuation at higher scales.
Due to the completeness and orthogonality of the DWT basis, there is
no information loss at any scale.
For the present study the analysis has been carried out by 
taking $j_{max}=5$. 

\begin{figure}[t]
\begin{center}
\epsfig{figure=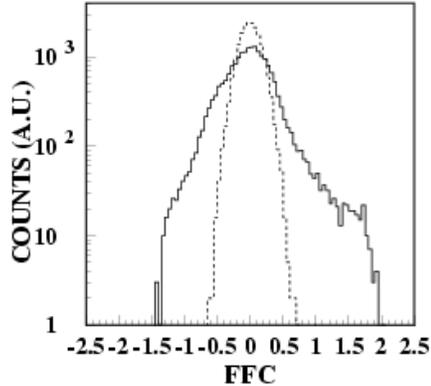 ,width=8cm}
\caption{\label{event_3}\small
FFC distribution for simulated normal events and pure
DCC-like events. The solid and the dotted lines correspond
to the DCC-like and normal events respectively. From \cref{eventshape}.}
\end{center}
\end{figure}

Typical FFC distributions at the scale $j=1$ are shown in 
\Fig{event_3}. The dashed histogram represents the FFC distribution
for normal events. The solid line histogram shows the corresponding 
distribution when a DCC domain of size given by $\Delta\phi=90^\circ$ and  
$3\le\eta\le 4$ is introduced in all the events. This constitutes an 
ensemble of DCC events having probability $\alpha=1$ of occurrence of 
a DCC. It is observed that the distributions for normal events is 
Gaussian,\footnote{The distribution for normal events is a perfect 
Gaussian with $\sigma=0.0908$ and $\chi^2/{\rm ndf}$ close to $1$ \cite{strenght}.}
whereas there is a broadening of the distribution in the presence of DCC. 
The presence of a DCC domain modifies the phase space distributions of 
charged particles and photons, resulting in an increase in the rms deviations 
of the FFC distribution. So in this analysis the rms deviation, $s$, of 
the FFC distribution quantifies the amount of fluctuation. 

\begin{figure}[t]
\begin{center}
\epsfig{figure=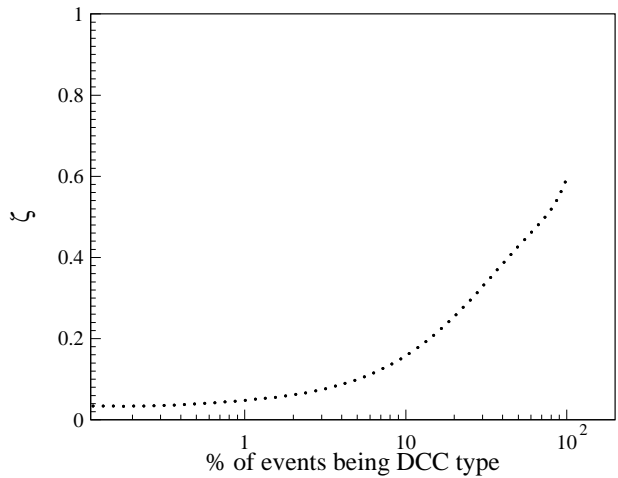,width=9.cm}
\caption{\label{ffc_sensitivity}\small
Variation of $\zeta$ as a function of percentage of events being DCC type. 
DCC events have $90^\circ$ DCC domain in $\phi$. 
This plot shows the sensitivity of the method. From \cref{bedanga_thesis}.}
\end{center}
\end{figure}

To quantify the DCC-type effect further, we introduce a strength parameter
$\zeta$, given by: 
\begin{equation}
\label{zeta}
 \zeta = \frac{\sqrt{(s_{\rm DCC}^2 - s_{\rm normal }^2)}}{s_{\rm normal}}\,,
\end{equation}
where $s_{\rm normal}$ is the FFC rms deviation for an ensemble of
normal events and $s_{\rm DCC}$ is the same for an ensemble of events
with DCC. 
Figure~\ref{ffc_sensitivity} shows the sensitivity of the method for picking 
up DCC-type fluctuations for a case of $90^\circ$ domain size. The strength 
parameter derived from the rms deviation of the FFC distributions of normal 
and DCC type events has been plotted as a function of the fraction of events 
being DCC-type. As expected $\zeta$ increases with the latter. 

This technique has been adopted by the WA98 Collaboration, which measures
multiplicities of charged particles and photons in overlapping parts of the 
detector phase-space \cite{wa98_local,wa98_local_cen}.

\subsubsection{\label{sec:robust} Robust observables}

\begin{figure}[t]
\begin{center}
\epsfig{figure=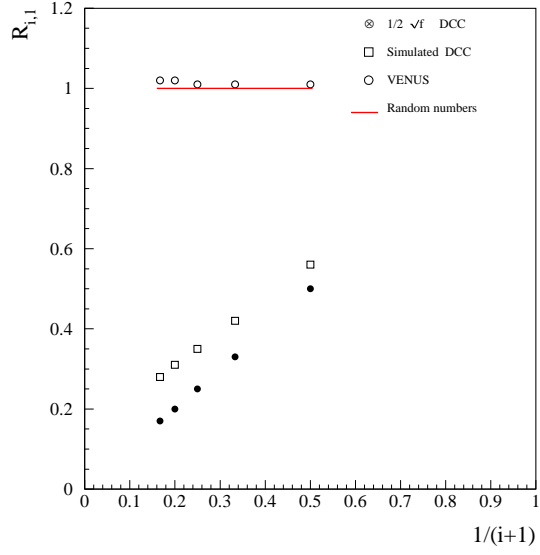,width=8cm}
\caption{\label{robust1}\small
Results of inclusive DCC analysis using robust observables.
Normal events (VENUS event generator) 
do not show any correlation whereas pure DCC events 
follow the relationship given in \Eqn{ri1DCC}. 
The simulated DCC events generated by taking detector effects
into account follow a similar behavior to that of pure DCC events.
From \cref{bedanga_thesis}.}
\end{center}
\end{figure}

This analysis uses ratios of factorial moments of the charged-neutral pion
distribution, as described at length in \Sec{sec:pheno}. It has been developed
for a dedicated DCC search at the Fermilab TEVATRON by the MiniMAX Collaboration 
\cite{minimax}. The set of robust observables is defined by Eqs.~\eqn{ri1}-\eqn{Fij},
where $\ng$ and $\nc$ are given by the photon and charged-pion multiplicities respectively 
in the studied region of phase-space.
For an inclusive analysis, the $n$'s denote the particle multiplicities in an event and
averages are taken over a large number of events. 
For event-by-event analysis, when the multiplicity is high, the phase space region is 
divided into bins and the $n$'s denote the multiplicities in a bin. In this case, averages 
are taken over the total number of bins. As emphasized previously, the variables $r_{i,1}$ 
are robust observables in the sense that the detection efficiencies, often difficult to 
estimate -- especially for photons, mostly cancel out and are thus expected to have little 
influence on the results. The main point \cite{Brooks:1996nu} is that $r_{i,1}=1$ for all 
$i\geq1$ in the case of generic pion production, whereas $r_{i,1}=1/(i+1)$ (see \Eqn{ri1DCC}) 
for pure ideal DCC production. Application of this technique to simulated normal and DCC-type 
events are shown in \Fig{robust1}.

This technique has been used by the MiniMAX Collaboration~\cite{minimax}  
and also to analyze cosmic ray experimental data at Pamir. 

\subsubsection{\label{sec:phi} $\Phi$-measure}

The $\Phi$-measure for a system of particles is defined as 
\cite{Gazdzicki:1992ri,Mrowczynski:1999un}:
\begin{equation}\label{phi}
 \Phi = \sqrt{\bra  Z^2 \ket \over \bra  N \ket} -
 \sqrt{\overline{z^2}} \,.
\end{equation}
where $z = x - \overline{x}$, where $x$ is 
the value of a given single-particle observable 
in a given event and $\overline{x}$ its average over all particles and all events.
The event variable $Z$ is a multi-particle analog of $z$ defined as 
$Z = \sum_{i=1}^{N}(x_i - \overline{x})$, where the summation runs over all 
particles from a given event. The brackets in \eqn{phi} represent averaging over 
all events. Note that, by construction, $\overline{z}=0$ and $\bra Z \ket = 0$.
As described in Refs.~\cite{Gazdzicki:1997gm,Mrowczynski:1999sf}, the $\Phi$-measure 
can be used to study the fluctuations in particle species. In that case, the single-particle 
variable $x=1$ if the particle is of a given sort (say a neutral pion) and $x=0$ otherwise 
(charged pion). This has been applied to study fluctuations of the neutral pion fraction
$f$ from a DCC in \cref{bedanga}.

Denoting by $N_\pi=N_{\pi^{0}} + N_{ch}$ the total pion multiplicity in a given event, 
one can write $N_{\pi^{0}} = fN_{\pi}$ and $N_{\pi^{\pm}}=(1-f)N_{\pi}$, where $f$ 
is the fraction of neutral pion in the event. Assuming that the relative fluctuations 
in $N_\pi$ are small, that is $\bra \delta N_\pi^2\ket\ll\bra N_\pi\ket^2$, one obtains
\cite{bedanga}:
\begin{eqnarray}\label{888}
 \Phi \simeq
 \sqrt{\bra  N_{\pi} \ket {\bra \delta f^{2}\ket}  } -
 \sqrt{\bra  f \ket \bra1 - f \ket} \,.
\end{eqnarray}
It is instructive to consider the properties of the $\Phi$-measure 
in the following simple cases:
\begin{itemize}
\item {\it Non-DCC case}: \\
  It is known from $pp$ experiments 
  \cite{pp_prod} that the produced pions have their
  charge states partitioned binomially with $\bra f\ket=1/3$. The
  fluctuation in $f$ is inversely proportional to the total 
  number of pions and is approximately given by 
  ${\bra \delta f^{2}\ket}\approx\bra f\ket\bra1-f\ket/\bra N_{\pi}\ket$. 
  In that case, one obtains, for the $\Phi$-measure: 
  \begin{eqnarray}\label{999}
    \Phi_{\rm non-DCC,~uncorr.} = 0\,.
  \end{eqnarray}
\item {\it No DCC, but $N_{\pi^{0}}$ and $\Nc$ have non-trivial correlations}:\\
  If $N_{\pi^{0}}$ and $N_{\pi^{\pm}}$ are assumed to be correlated in such 
  a way that there are {\it no} (DCC-type) event-by-event fluctuations of the neutral
  fraction $f$. Moreover, the latter is assumed to be strictly independent of the event 
  multiplicity. Then, $f = \delta$, where $\delta$ is a constant smaller than 
  unity, and ${\bra \delta f^{2}\ket} = 0$. One has:
  \begin{equation}\label{zero-DCC}
    \Phi_{\rm non-DCC,~corr.} =  - \sqrt{\delta (1- \delta)}\,.
  \end{equation}
  Notice that The $\Phi$-measure is negative in that case.\\
\item {\it Ideal DCC case}: \\
  For ideal DCC production, the probability distribution of the neutral fraction
  is given by the inverse square-root law, \Eqn{DCCsign}. One has $\bra f\ket = 1/3$
  and $\bra \delta{f}^2\ket=4/45$, so the $\Phi$-measure is given as:
  \begin{equation}\label{full-DCC}
    \Phi_{\rm DCC} =  
    \sqrt{\frac{4\bra N_{\pi}\ket}{45}}-\sqrt{\frac{2}{9}}\,.
  \end{equation}
  It is positive for $\bra N_\pi\ket>5/2$.
  For a typical case where the total pion multiplicity $N_{\pi}\approx 300$, 
  one finds $\Phi_{\rm DCC}\approx 4.7$. It is worth emphasizing that this assumes that
  all the observed pions are of DCC origin. A more realistic case would include pions 
  from non-DCC sources as well \cite{bedanga_thesis}. This, along with the decay of 
  neutral pions to photons as well as detector effects, have been discussed in 
  \cref{bedanga}. 
\end{itemize}

\begin{figure}[t]
\begin{center}
\epsfig{figure=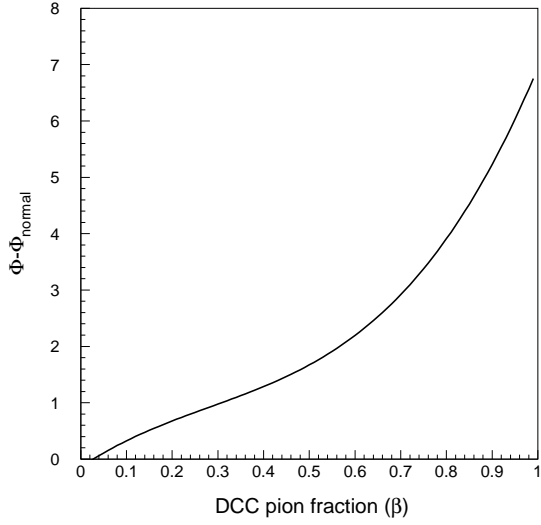,width=8cm}
\caption{\label{dcc_phi1}\small
Variation of $\Phi - \Phi_{\rm normal}$ as a function of the fraction
$\beta$ of DCC pions, obtained from simulated data using VENUS.
From \cref{bedanga}.}
\end{center}
\end{figure}

The $\Phi$-measure can be applied to simulated data (both normal and DCC-type)
in order to study the sensitivity of the measure to detection of DCC-type 
fluctuations. First we study the effect of varying the fraction $\beta$ of 
DCC pions in each event. To this aim, one calculates the quantity $\Phi-\Phi_{\rm normal}$, 
where $\Phi$ denotes the measure of fluctuation for a given DCC pion fraction $\beta$ and 
$\Phi_{\rm normal}$ is the measure of fluctuation for normal simulated events (detector 
and decay effects included). The variation of $\Phi - \Phi_{\rm normal}$ as a
function of $\beta$ is shown in \Fig{dcc_phi1}. One observes that the
size of anomalous fluctuations decreases with decreasing DCC pion fraction, which 
is as per expectation. The statistical error on $\Phi$ was calculated to be $0.006$,
from which one finds that such observable is a sensible probe of possible DCC-type 
fluctuations if the fraction of DCC pions in a typical event is above $\sim$ $3\%$.

\begin{figure}[t]
\begin{center}
\epsfig{figure=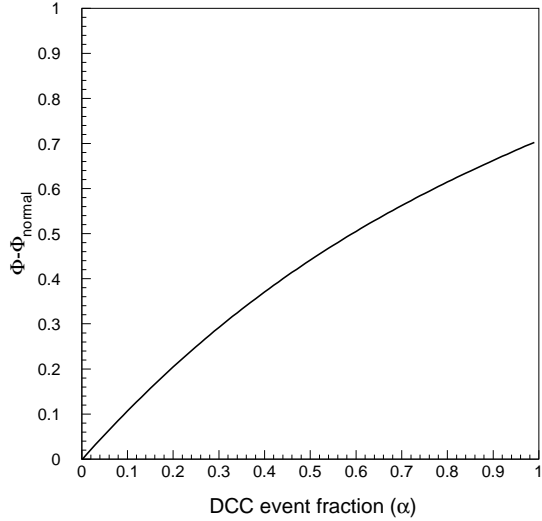,width=8cm}
\caption{\label{dcc_phi2}\small
Variation of $\Phi - \Phi_{normal}$  as a function of the DCC event fraction
($\alpha$) for a fixed DCC pion fraction ($\beta$ = 0.25), obtained from
simulated data using VENUS.
From \cref{bedanga}.}
\end{center}
\end{figure}

This analysis assumes that all events have DCC pions. Including the effect
of the probability of DCC formation being $\alpha\neq1$ has been studied for 
a fixed fraction of DCC pions in each event in \cref{poslim}. There, the authors 
assume that for DCC-type events the percentage of DCC pions is $\beta=0.25$ 
and vary the fraction $\alpha$ of events being DCC-type. The results are 
presented as $\Phi-\Phi_{\rm normal}$ vs. $\alpha$ in \Fig{dcc_phi2}. 
As expected, the value of $\Phi - \Phi_{\rm normal}$ decreases with decreasing 
$\alpha$. Keeping in mind the above mentioned statistical error on $\Phi$-measure, 
we find that for $\beta=0.25$, the measure is sensitive to DCC formation if the 
latter occurs in more than $1\%$ of the events.

\subsubsection{\label{sec:event_shape} Event shape analysis}

This method has been proposed in the context of DCC search in 
\cref{eventshape}. It is based on a technique that has been used 
successfully for flow analysis of heavy-ion data \cite{voloshin}. 
It uses the fact that localized DCC formation should lead to an 
event-shape anisotropy, which is expected to be out of phase for 
one detector (\eg charged pions) compared to the other (\eg photons). 
In other words, whenever there is large number of charged particles 
from a DCC region recorded in a given region of the first detector, 
there should be a depletion in the number of neutral particles recorded 
in the corresponding zone of the other detector. Therefore, from an 
event shape analysis using both detectors one can, in principle, look 
for DCC signals. In terms of the terminology used in flow analysis,  
this method is based on the fact that a simple redistribution of 
particles, with two detectors to detect charged particles and photons, 
would result in the same flow direction with the flow angle
difference peaking at zero. However, events where the neutral 
pion fraction has been modified according to the ideal DCC probability 
distribution, will show the flow angles in two detectors to be almost
$90^\circ$ apart \cite{eventshape}. 

In \cref{eventshape}, events with localized DCC domains were simulated 
along the lines described earlier in \Sec{sec:dcc_model} and event anisotropy 
(indicating flow) was introduced according to the following simple procedure.
First, a flow direction is selected at random, uniformly distributed between
$0^\circ$ and $360^\circ$. The directions of particles are then distributed about
this direction following a Gaussian distribution with a width of $10^\circ$.
Results were analyzed using the standard second order Fourier analysis of 
elliptic flow \cite{voloshin}.  
The particle hits in a given detectors can be labeled by their positions 
in the $\eta-\phi$ plane. For each event, one constructs the two sums:
\bea
 X&=&\sum_i \cos(2\phi_i)\\
 Y&=&\sum_i \sin(2\phi_i)\,,
\eea
where $\phi_i$ is the azimuthal angle of the particle $i$ and the sums run over all
particles in a given event. The flow angle $\Xi$ is defined as: 
\begin{equation}
  \Xi = {1\over 2}\arctan(Y/X)\,.
\end{equation}
In the absence of any detector imperfections and other geometrical
effects, the distribution of $\Xi$, taken over a large number of events,
is expected to be completely flat over $0^\circ$ to $180^\circ$. This is because 
the flow direction varies randomly from event to event. However, when the 
events are realigned, with respect to the flow angle in each event, as described
above, one can see the characteristic peaks (at $0^\circ$ and $180^\circ$) in the azimuthal 
distribution of particles.

In case of two detectors with the same phase space ($\eta-\phi$) coverage,
one detecting photons and other detecting charged particles, 
the situation is very interesting. If there is genuine flow in a particular 
event both detectors would show the effect in terms of their respective $\Xi$-angles 
getting aligned in the same direction. Therefore the 
distribution of $\Psi=\Xi_\gamma- \Xi_{\rm ch}$, the difference between the flow 
angles of the two detectors, is expected to be peaking at zero.
However, in case of DCC being prominent in a particular region, there
will be more photons detected in one detector. The other detector is 
expected to show less charged particles in the same region of phase space.
Therefore an event shape analysis is expected to show two flow angles
for both detectors which will be out of phase (for the present second order Fourier
analysis the angular difference is expected to peak at $90^\circ$).

\begin{figure}[t]
\begin{center}
\epsfig{figure=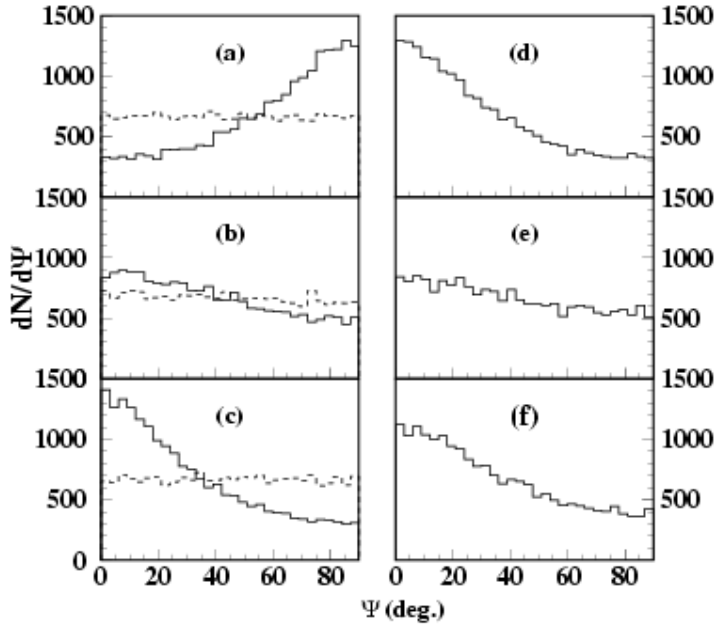 ,width=10.cm}
\caption{\label{event_1}\small
 The distribution of $\Psi$ for simulated DCC (a)-(c) and
 flow (d)-(f) events. Panel (a) shows anti-correlation between
 event planes determined for charged-particle and photon
 detectors. Panels (b) and (c) show correlations between event
 planes obtained considering two sub-events for the photon
 and charged-particle detectors respectively.
 The dotted lines correspond to generic VENUS events. In case 
 of flowy events, panels (d) shows correlation between event 
 planes obtained from the two detectors, whereas panels (e) and 
 (f) are analog to (b) and (c). From \cref{eventshape}.}
\end{center}
\end{figure}

The results of \cref{eventshape} are presented in \Fig{event_1}. The three left
panels (a)-(c) correspond to the case with only DCC in which one can notice (panel (a)) 
a clear peak at $90^\circ$ for the angle $\Psi$ between the event planes for the two 
detectors. Panels (b) and (c) show the angle between the event planes
as obtained for two sub-events in each of the detectors separately.
It is important to notice that both detectors show flow, but one
with respect to the other clearly shows an anti-flow type behavior.
In the same figure is also presented the data corresponding to pure VENUS 
events for comparison. One notices that the latter do not present neither any 
signature of flow nor DCC-like fluctuation. Finally, results shown in panels (d)-(f) 
show the same plots for flow type events which have no DCC-like fluctuation. Here, 
the individual detectors are seen to show the same effect as one with respect to the
other. 

A more detailed analysis \cref{eventshape} reveals that this method is 
sensitive if at least $10\%$ of the events are of DCC-type. It has been 
argued that one can improve upon this by combining with the DWT technique 
\cref{eventshape} . The event-shape analysis has not yet been applied to 
experimental data.

\subsection{\label{sec:expt_results} Experimental results}

We have discussed in the above section, the experimental set-up
for various experiments looking for DCC and the experimental techniques
used to look for DCC-type fluctuations in experimental data. Below we
present the results from various experiments which have looked for possible 
DCC formation in high energy collisions.

\subsubsection{\label{cosmic_result} Results from cosmic ray experiments}

\begin{figure}[t]
\begin{center}
\epsfig{file=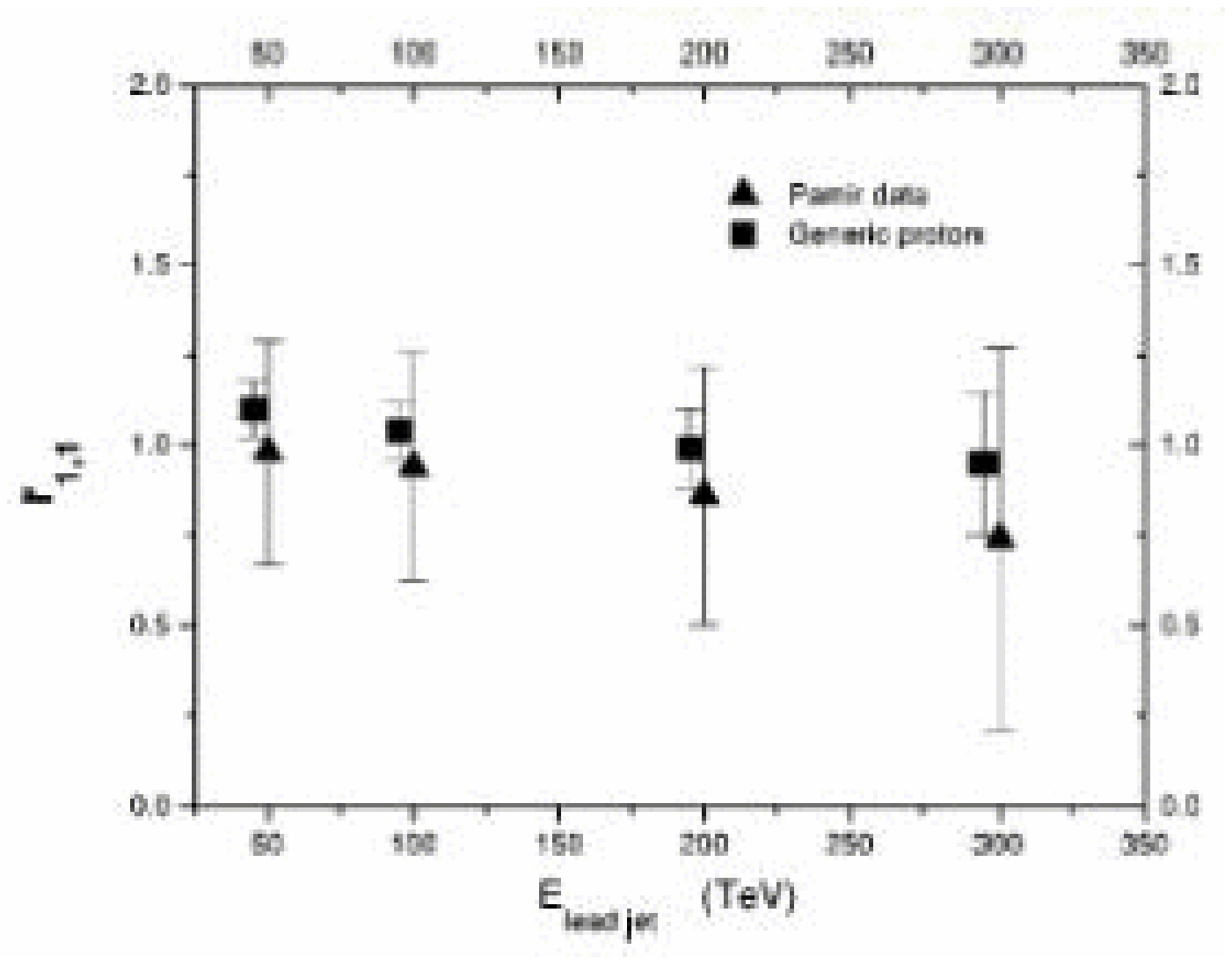,width=5.cm}
\epsfig{file=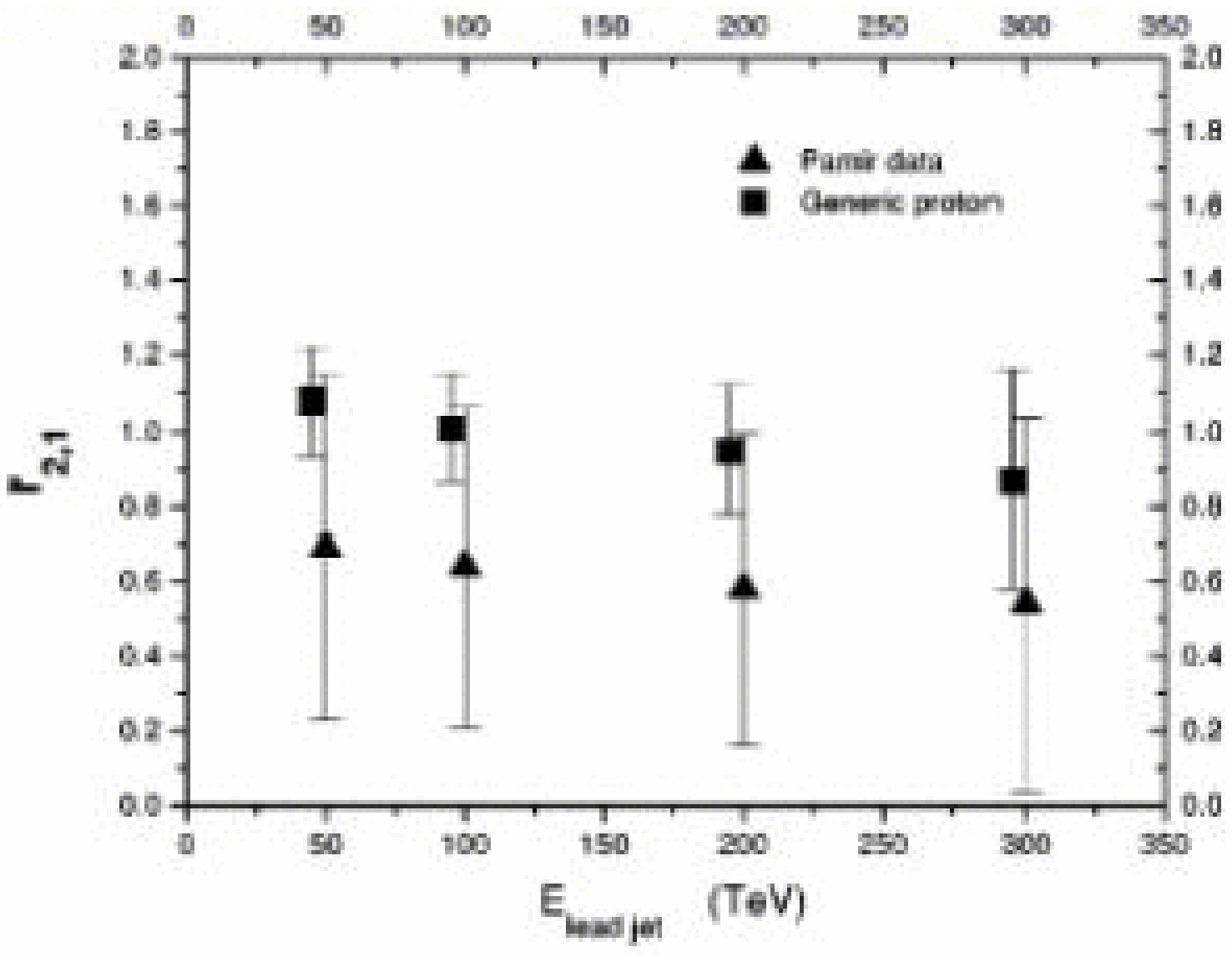,width=5.cm}
\epsfig{file=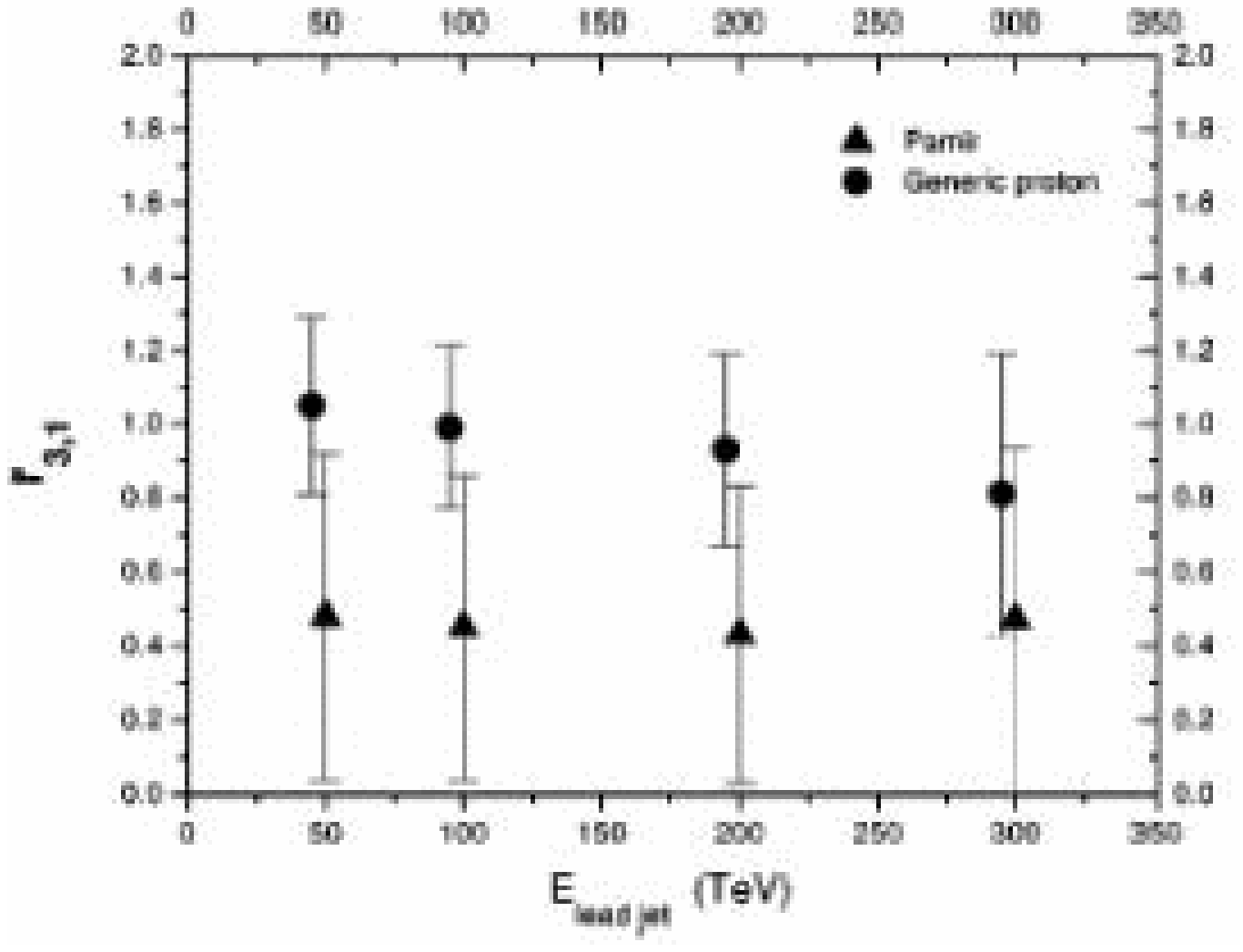,width=5.cm}
\caption{\label{pamir}\small
The robust observable $r_{1,1}$, $r_{2,1}$ and $r_{3,1}$ 
as a function of visible leading jet energy: experimental points are
represented by triangles, whereas squares are used for simulated data.
From \cref{pamir}.}
\end{center}
\end{figure}

There have been many cosmic rays experiments~\cite{jacee,pamir}. Here 
we discuss the most recent results from a systematic study of
the large asymmetries in the neutral pion fraction distribution in
high-energy cosmic-ray families (100 TeV $\le$ $E_{visible}$ $\le$
700 TeV) detected at high mountain altitudes at Pamir~\cite{pamir}. 
The experimental data has been analyzed using the technique of robust 
observables discussed earlier. The results for the robust ratios $r_{i,1}$ 
is presented as a function of leading jet energy $E_{lead~jet}$
in \Fig{pamir}. The pattern of energy flow in cosmic-ray interactions is
understood through the picture of jets: The stream of energy in the air 
cascade are represented by jets produced in the nuclear interactions. In
this picture the leading jet represents the pattern of atmospheric nuclear
and electromagnetic cascade having the largest energy flow.

From the figures one observes that the absolute values of $r_{i,1}$ are
decreasing with increasing energy of the jets and the order $i$ of the moments.
The results from data have been compared to simulated generic events.
One observes that with the increasing order of the robust observables the
difference between experiment and simulation becomes more important. Recall
that values of $r_{i,1}$ far below 1 can be an indication of events with DCC 
formation overlaying generic events. The experimental average values of $r_{3,1}$ 
that are most sensitive to DCC formation are clearly below $1$ but by less than 
a standard deviation. The averaged experimental values  for $r_{3,1}$ are presented 
in Table~\ref{tpamir}, together with the $95\%$ confidence interval. The latter does
not exceed $1$.

\begin{table}[h]
\caption{\label{tpamir}\small Experimental statistics and $95\%$ confidence interval of the
average $r_{3,1}$ in Pamir experiment. From \cref{pamir}.}
\vspace{.5cm}
\begin{center}
\begin{tabular}{|c|c|c|c|c|}
\hline
Jet energy (TeV) & $r_{3,1}$ & Number of events   & 95\% confidence interval\\
\hline
50  &  0.48 $\pm$ 0.44 & 78  & (0.38,0.58) \\
100 & 0.45  $\pm$ 0.41 & 61  & (0.38,0.55) \\
200 & 0.43  $\pm$ 0.40 & 30  & (0.28,0.58) \\
300 & 0.47  $\pm$ 0.49 & 10  & (0.12,0.82) \\
\hline
\end{tabular}
\end{center}
\end{table} 

The conclusion from Pamir experiment~\cite{pamir} is that, due to the limited
large error bars, the experimental data neither confirm nor rule out the 
possibility of a DCC formation mechanism in high-energy cosmic-ray interactions. 
A further increase in experimental statistics is needed to establish firm conclusions.

\subsubsection{\label{sec:nn_result} Results from nucleon-nucleon collision experiments}

Here we discuss results on DCC search from high-energy nucleon-nucleon
experiments UA1 and UA5 at CERN \cite{ua1,ua5_540,ua5_900} and MiniMAX
at Fermilab \cite{minimax}.

\paragraph{\label{sec:ua5_result} The UA5 Experiment}

\begin{figure}[t]
\begin{center}
\epsfig{figure=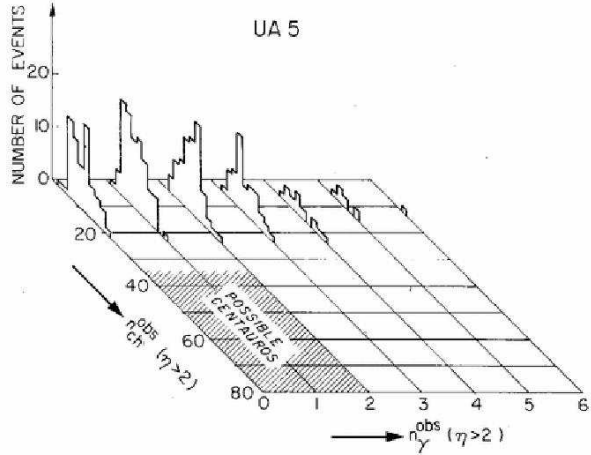,width=8cm}
\caption{\label{ua52}\small 
  Number of events plotted against multiplicities of charged particles
  observed in the pseudo-rapidities interval $-2\le\eta\le 2$. The 
  distributions are drawn separately for each value of observed
  photon multiplicity in the same range of pseudo-rapidity. The region
  which is populated by Centauro-like events if produced at the collider
  and observed by the detector is indicated.
  From \cref{ua5_540}.}
\end{center}
\end{figure}
 
Figure \ref{ua52} shows the distribution of events as a function of observed 
charged-hadron multiplicity in the pseudo-rapidity interval $-2\le\eta\le 2$ 
for each value of observed photon multiplicity in the same interval, for $p\bar{p}$ 
collisions at center of mass energy of 540 GeV~\cite{ua5_540}. This correlation 
plot does not support the existence of Centauro type of events, based on the 
characteristics of Centauros observed in cosmic ray interactions. 
One observes that there is no event with charged hadron multiplicity remotely 
comparable to the expected value\footnote{This number has been arrived at 
by making a comparative study of the collider data with the data obtained 
from the cosmic ray experiments \cite{ua5_540}.} of $30 - 40$ with associated 
photon multiplicity being close to zero.
The shaded portion represents the region where one would expect some
events in $p\bar{p}$ collisions at $\sqrt{s}$ = 540 GeV, had they similar 
characteristics as Centauro events in cosmic-ray experiments.

\begin{figure}[t]
\begin{center}
\epsfig{figure=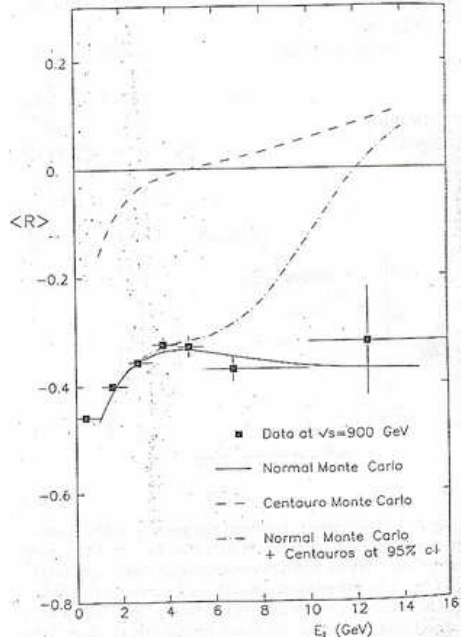,width=6cm}
\caption{\label{ua53}\small 
  The asymmetry in electromagnetic and hadronic energy ($R$) is plotted along
  the Y-axis and the transverse energy is plotted along the X-axis. Comparison to
  various model calculations are shown. From \cref{ua5_900}}
\end{center}
\end{figure} 

The UA5 Collaboration~\cite{ua5_900} also looked for Cen\-tau\-ro-\-like events in 
the $900$~GeV center of mass energy data, by studying the correlation of produced 
charged particles $\Nc$ versus photons ($\Ng$) and correlation between 
the electromagnetic ($E_{\rm em}$) and hadronic energies ($E_{\rm had}$) deposited 
in the calorimeters. Figure~\ref{ua53} shows the average value of the ratio 
\beq
 R=\frac{E_{\rm had}-E_{\rm em}}{E_{\rm had}+E_{\rm em}}
\eeq
as a function of the transverse energy $E_T$, for experimental data, simulated normal
and Centauro-like Monte-Carlo events, as well as a mixture of normal and Centauro events,
the percentage of Centauro events being $0.24\%$ of all events. The data clearly seems 
to follow the trend of normal events. Further the correlation of $\Nc$ with 
respect to $\Ng$ for data and Monte-Carlo simulation look similar, thereby ruling out 
the possibility of Centauro events or DCC-type events in the experimental data.

\paragraph{\label{sec:ua1_result} The UA1 Experiment}

The UA1 Collaboration \cite{ua1} have looked for Cen\-tau\-ro-\-type events in 
$p\bar{p}$ collisions at center of mass energy of $540$~GeV in $48,000$ 
low bias events. They have measured the correlation between deposited 
transverse electromagnetic and hadronic energies, where the energy deposited 
in the first four radiation lengths of material is termed as electromagnetic 
energy and that deposited beyond twelve radiation lengths is termed as hadronic 
energy. By comparing experimental data with Monte-Carlo simulations, they have 
excluded possible Centauros or global DCC-type events in their data sample.

\paragraph{\label{sec:minimax_result} The MINIMAX Experiment}

The analysis of the $p\bar{p}$ collisions data at center of mass energy of 
$1.8$~TeV for the MiniMAX experiment~\cite{minimax}, in the forward direction 
$3.2 \le \eta \le 4.2$, was carried out using the technique of robust observables,
which, as already emphasized, had been actually developed for this experiment. 
A total of $1.3\times10^6$ events was analyzed. 
A detailed Monte-Carlo simulation specific to the MiniMAX experiment,
with input being minimum bias PYTHIA events passed through full detector
simulation and analysis chain, yielded a value of $r_{1,1}\sim 1.02-1.13$
for the lowest robust ratio. A similar simulation with pure ideal DCC introduced 
by hand yielded a value of $r_{1,1}\sim 0.6-0.7$. This demonstrated the sensitivity
of the experiment to detect a single large ideal DCC. The analysis of the data 
reveals that the ratios $r_{i,1}$ are not smaller than one as would be 
expected for pion production from a DCC. The measured values of the ratios $r_{i,1}$ 
are given in Table~\ref{tminimax}.

\begin{table}
\caption{\label{tminimax}\small Values of $r_{i,1}$ for the whole event sample 
  (about $13.8\times10^5$ events) of the MiniMAX experiment \cite{minimax}.}
\vspace{.5cm}
\begin{center}
\begin{tabular}{|lcl|lcl|}
\hline
 &$i$ &&& $r_{i,1}$ &\\
\hline
 &$1$ &&& $1.026 \pm 0.004$ &\\
 &$2$ &&& $1.035 \pm 0.010$ &\\
 &$3$ &&& $1.059 \pm 0.027$ &\\
 &$4$ &&& $1.118 \pm 0.065$ &\\
 &$5$ &&& $1.310 \pm 0.151$ &\\
 &$6$ &&& $1.904 \pm 0.382$ &\\
\hline
\end{tabular}
\end{center}
\end{table} 

The MiniMAX Collaboration was able to put some limits on DCC production in
terms of the size of the DCC -- that is the number of DCC pions produced --
and the likelihood of DCC production. To this aim, two types of model for DCC 
production have been considered: 
\begin{itemize}
\item[(i)] Associated production, where the number of produced DCC pions in each 
event is proportional to the number of generically produced pions;
\item[(ii)] Exclusive production, where a given event is described by either 
pure DCC or pure generic production with an given probability.
\end{itemize}
Knowing the most probable lower limit of the ratios $r_{i,1}$ in data, they 
were able to extract an upper limit on DCC production for the above two 
scenarios. For the associated production model, the upper limit on DCC pion 
production has been set at
\beq
 \frac{\bra N\ket_{\rm DCC}}{\bra N \ket_{\rm generic}}\lesssim 0.21\,. 
\eeq
For the exclusive production model, the upper limit on DCC production is set
in terms of probability that an event is DCC type. This value must be less 
than $0.05$.

\subsubsection{\label{sec:AA_result} Results from nucleus-nucleus collision experiments}

Here we discuss results on DCC searches from the high-energy heavy-ion
collision experiments at the CERN SPS (WA98 \cite{wa98_local} and NA49 
\cite{na49} Collaborations) and at the RHIC (PHENIX Collaboration~\cite{phenixqm}). 

\paragraph{\label{sec:wa98_result} The WA98 Experiment}

DCC search in WA98 experiment involved three kinds of analysis.
One was to look for global DCCs \cite{wa98_global}, that is single 
large DCC domains, by measuring photon versus charged-particle correlation
in the full common $\eta-\phi$ phase-space of the photon (PMD) and 
charged-particle (SPMD) detectors. Second was to look for smaller
domains of DCC in common localized regions of $\eta-\phi$ phase space 
of the PMD and SPMD \cite{wa98_local,wa98_local_cen}. And the third
was to look for anti-Centauro--type events in the data by looking at
localized regions in the event displays \cite{mma}, as for example in 
the JACEE cosmic-ray experiment. 
Below we discuss the results on DCC search from the above three analysis
carried out by the WA98 Collaboration.

\begin{figure}[t]
\begin{center}
\epsfig{figure=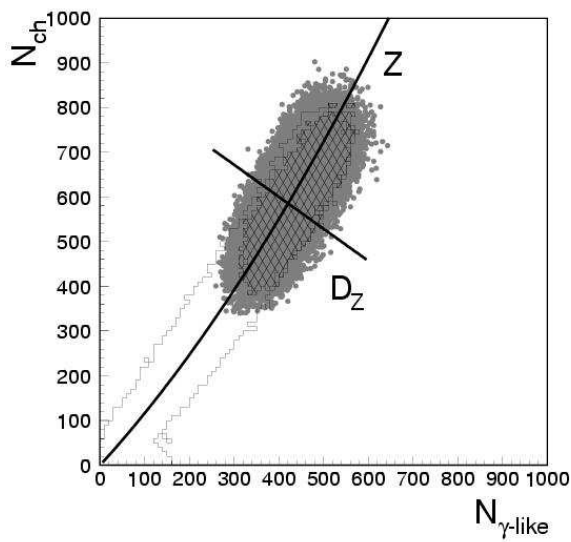,width=8cm}
\caption{\label{96data}\small 
  This is the scatter plot showing the correlation between $\Nc$
  and $\Ngl$. The solid outline shows the trend of the minimum
  bias data. The central sample (with $E_T>300$~GeV) is shown
  as points for the data and as a hatched region for VENUS (with much lower 
  statistics). Overlaid on the plot are the $Z$ axis and the $D_Z$ axis at 
  a particular value of $Z$ as explained in the text. From \cref{wa98_global}.
}
\end{center}
\end{figure}

\noindent{\it - Global DCC search:} \\
The WA98 Experiment attempted to look for DCC-type fluctuations
by looking at event-by-event correlation in the number of charged particles 
and photons detected over the full common $\eta-\phi$ coverage of the PMD 
and the SPMD \cite{wa98_global}.
The strong correlation between charged and neutral multiplicities shown 
in \Fig{96data} suggests to employ a coordinate system with one axis being 
the measured correlation axis and the other being perpendicular to it. If 
all detected particles were pions and the detectors were perfect and had identical
pseudo-rapidity acceptance, then the correlation axis would
be a straight line. The details of the correlation analysis has already
been described in \Sec{sec:corr}. Here we present the results of such analysis
applied to the WA98 data.

\begin{figure}[t]
\begin{center}
\epsfig{figure=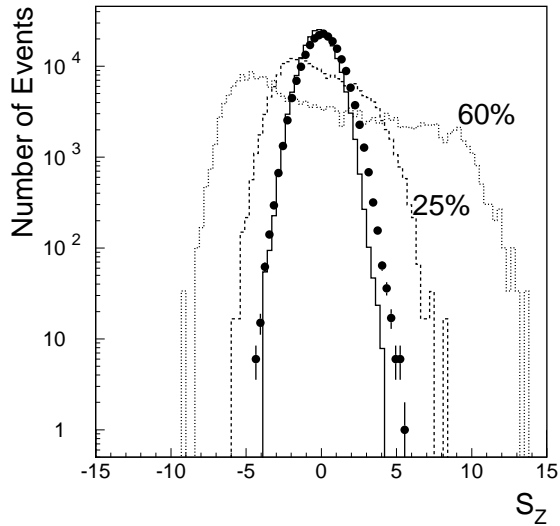,width=8cm}
\caption{\label{dcc_hyp}\small
  $S_Z$ distribution for the experimental data is shown, 
  overlaid with VENUS simulations incorporating 
  $0\%$, $25\%$ and $60\%$ DCC-type events.
  All of the distributions are normalized to
  the total number of data events. From \cref{wa98_global}. 
}
\end{center}
\end{figure}

The $S_Z$ distribution for both data and simulated events are shown in
\Fig{dcc_hyp}. The simulated events have been generated using the VENUS 
4.12 event generator, with output processed through a detector simulation 
package in the GEANT 3.21 framework \cite{geant}. This incorporates the 
full WA98 experimental setup.
The centrality selection for simulated data has been made in an
identical manner as in the data, determined from the
simulated total transverse energy of mid-rapidity calorimeter used
in the experiment. The discrepancy between VENUS and the data 
can be seen more clearly by measuring the width of the $S_Z$ distribution.
The simulated and measured distributions are both Gaussian with respective
widths $\sigma_{\rm simul.}=0.998\pm0.002$ (fit error only) and 
$\sigma_{\rm meas.}=1.13\pm0.07$ (error from relative scale uncertainties 
included).

\begin{figure}[t]
\begin{center}
\epsfig{figure=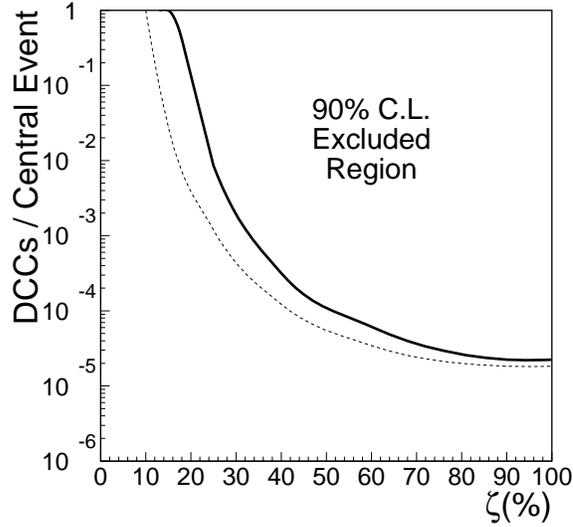,width=8cm}
\caption{\label{limits} \small The $90\%$ C.L.
  upper limit on DCC production per central event as a function of
  the fraction of DCC pions under two assumptions.  The thick line
  gives the upper limit obtained by assuming the $\sigma_{D_Z}$ in 
  $S_Z$ is completely given by the VENUS calculation requiring to
  make a cut at $6\sigma$. The dashed line shows a less conservative 
  limit obtained by using the $\sigma_{D_Z}$ measured in
  the data itself. This allows one to make a tighter cut
  at $5\sigma$, increasing the DCC detection efficiency. From \cref{wa98_global}.
}
\end{center}
\end{figure}

One expects that DCC events would show up as non-statistical tails on the 
$S_Z$ axis. No such events were seen by the WA98 Collaboration in their global 
analysis of the data sample. They concluded that a possible single DCC domain 
is either very rare, very small, or both. To check which hypothese is consistent 
with their data, they determined upper limits on the frequency of DCC production 
as a function of its size, as represented by the fraction $\beta\equiv\zeta=
N_{\rm DCC}/N_{\pi}$ of DCC pions. For this they have computed $S_Z$ distributions
for several values of $\zeta$, ranging from $15\%$ to $90\%$, and they have considered 
two possible scenarios. The first is based upon the conservative assumption that 
VENUS should describe the data perfectly in the absence of a DCC signal.
The corresponding $90\%$ C.L. excluded region for possible DCC formation is 
presented in \Fig{limits} as a solid line. The second scenario assumes that the 
difference between the data and VENUS seen in \Fig{dcc_hyp} is due to detector 
effects and that the widths should in fact be the same. The resulting excluded 
region is shown in \Fig{dcc_hyp} as a dashed line. The boarders of the two 
regions are quite different at $\zeta=15\%$, but get closer at $\zeta >30\%$.

\noindent{\it - Search for localized domains of DCC:} \\
Photon and charged-particle multiplicity correlation studies have also been carried 
out in localized regions of the $\eta-\phi$ phase-space to look for the possible existence 
of small localized DCC domains. This analysis uses both the techniques of $\Ng-\Nc$ 
correlation and DWT in various bins in azimuthal angle~\cite{wa98_local,wa98_local_cen}, 
discussed previously under sections~\ref{sec:corr} and \ref{sec:dwt}. 
The WA98 Collaboration have also studied the centrality dependence of this analysis.
For this, four different centrality bins were considered: top $5\%$ 
(henceforth referred to as centrality-1); $5\%-10\%$ (centrality-2); $15\%-30\%$ 
(centrality-3); and $45\%-55\%$ (centrality-4) of the minimum bias cross-section.

In looking for non-statistical DCC-type fluctuations 
in experimental data, it is necessary to understand all
detector related effects and to try to filter
out the various components contributing to the total observed 
fluctuations. For this, results from experimental data need to be 
compared to some base line results to draw proper conclusions. 
Due to the inherent uncertainties in the description of ``normal'' 
physics and detector response in simulations, the observation of an 
experimental result which differs from the case with statistical 
fluctuation, given by simulations, cannot be taken alone as evidence 
of presence of non-statistical fluctuations. 
This problem can be solved by generating different sets of mixed 
events from data, keeping specific physics goals and detector effects 
in mind. This remarkable ability of mixed events to probe different 
physics goals is possible through construction of suitable ensembles of
mixed events by removing various correlations in a controlled manner while
preserving the characteristics of the measured distributions as
accurately as possible. 
Fluctuations in the ratio of $\Ng$ to $\Nc$ can have three different origins:
They can arise due to independent fluctuation in either $\Ng$ only, $\Nc$ only,
or both. Further, in the latter case the fluctuations in $\Ng$ and $\Nc$ may be
correlated event by event. In particular, such a correlation is expected to arise
in the case of DCC production. By intelligently constructing different types of
mixed events one can probe possible non-trivial fluctuations in the $\Ng$-to$\Nc$ 
ratio and their origin.

A first set of mixed events, referred to as M1-type, is constructed to provide a 
base line for comparison to real event. These are generated by mixing hits in both 
photon (PMD) and charged-particle (SPMD) detectors separately, satisfying the 
$\Ngl-\Nc$ correlation as in the real event. This is done by producing one mixed 
event for each real event with the same multiplicity of $\Ngl$ and $\Nc$ pair as 
in the VENUS event.
The idea is to globally keep the distributions similar to real events and look 
for event-by-event fluctuations in localized regions of phase-space. In the construction
of such mixed events, care is taken such that no two hits come from the same real event. 
Moreover, hits within a detector in the mixed events are not allowed to lie within the 
two track resolution of the detector. Results from such mixed events when compared to 
those of real events were expected to reveal the presence of localized
fluctuations. However, they do not give any information neither regarding 
the origin of such fluctuations nor concerning possible $\Ng-\Nc$ correlations.

A second set of mixed events, referred to as M2-type events, has been
constructed to provide information regarding the possible presence of
localized event-by-event correlated fluctuation in $\Ngl$ and
$\Nc$. They are generated by mixing photon events with different charged-particle 
events so that no two events were repeated. Here also, care is taken that the 
original $\Ngl-\Nc$ correlation is maintained. The original 
event-by-event hit structure in each individual detector being left unchanged,
such mixed event set keeps identical the individual localized fluctuations 
due to $\Ngl$ and $\Nc$, but removes possible event-by-event 
correlations. Therefore, by comparing the results from such mixed events to those 
from real events reveals the presence of localized event-by-event correlated
fluctuations in $\Ngl$ and $\Nc$.

Finally, two other sets of mixed events, referred to as M3-$\gamma$ and M3-ch type, 
have also been considered to provide information regarding contribution to 
localized fluctuations in the $\Ng$-to-$\Nc$ ratio from either $\Ng$ fluctuations
or $\Nc$ fluctuations separately. Such events are generated from real events by
mixing hits in one of the detectors (following the procedure of construction of 
M1-type events) and keeping the hit structure of the event in the other detector 
intact. M3-$\gamma$ events corresponds to the case where the hits within the photon 
detector are unaltered while the charged-particle hits from different events are
mixed, whereas M3-ch events correspond to the opposite case.
The total number of mixed events generated is the same as the number of real events. 
Results from such mixed events, when compared to those of real and M1-type events, 
are expected to reveal the presence of event-by-event localized fluctuations in 
$\Ngl$ and $\Nc$ individually. 
The various sources of fluctuation in the $\Ng$-to-$\Nc$ ratio which are preserved
by the different types of mixed events described above are summarized in Table~\ref{tab:table1}.

\begin{table}
\begin{center}
\caption{\label{tab:table1}\small Sources of fluctuations preserved by
  various mixed events.}
\vspace{0.5cm}
\begin{tabular}{|c|ccccc|}
\hline
 Type of & &  &Mixed & Event & \\ 
 fluctuation& &M1&M2&M3-$\gamma$&M3-ch \\
\hline
$\Ng$ only& &No & Yes & No & Yes \\
$\Nc$ only& &No & Yes & Yes &No \\
correlated $\Ng-\Nc$& &No & No & No &No \\
\hline
\end{tabular}
\end{center}
\end{table}

Mixed events constructed from experimental data were compared to simulated 
events, generated as for the global DCC case discussed previously, and were
analyzed using both the correlation and the DWT analysis techniques.
Since the widths (rms deviations) of the $S_{Z}$ and FFC distributions
carry the information concerning fluctuations, below we compare these
widths as obtained from data as well as mixed and simulated events.

\begin{figure}[t]
\setlength{\unitlength}{1mm}
\begin{center}
\begin{picture}(100,100)(10,10)
\put(0,60){
\epsfxsize=6.cm
\epsfysize=5.cm
\epsfbox{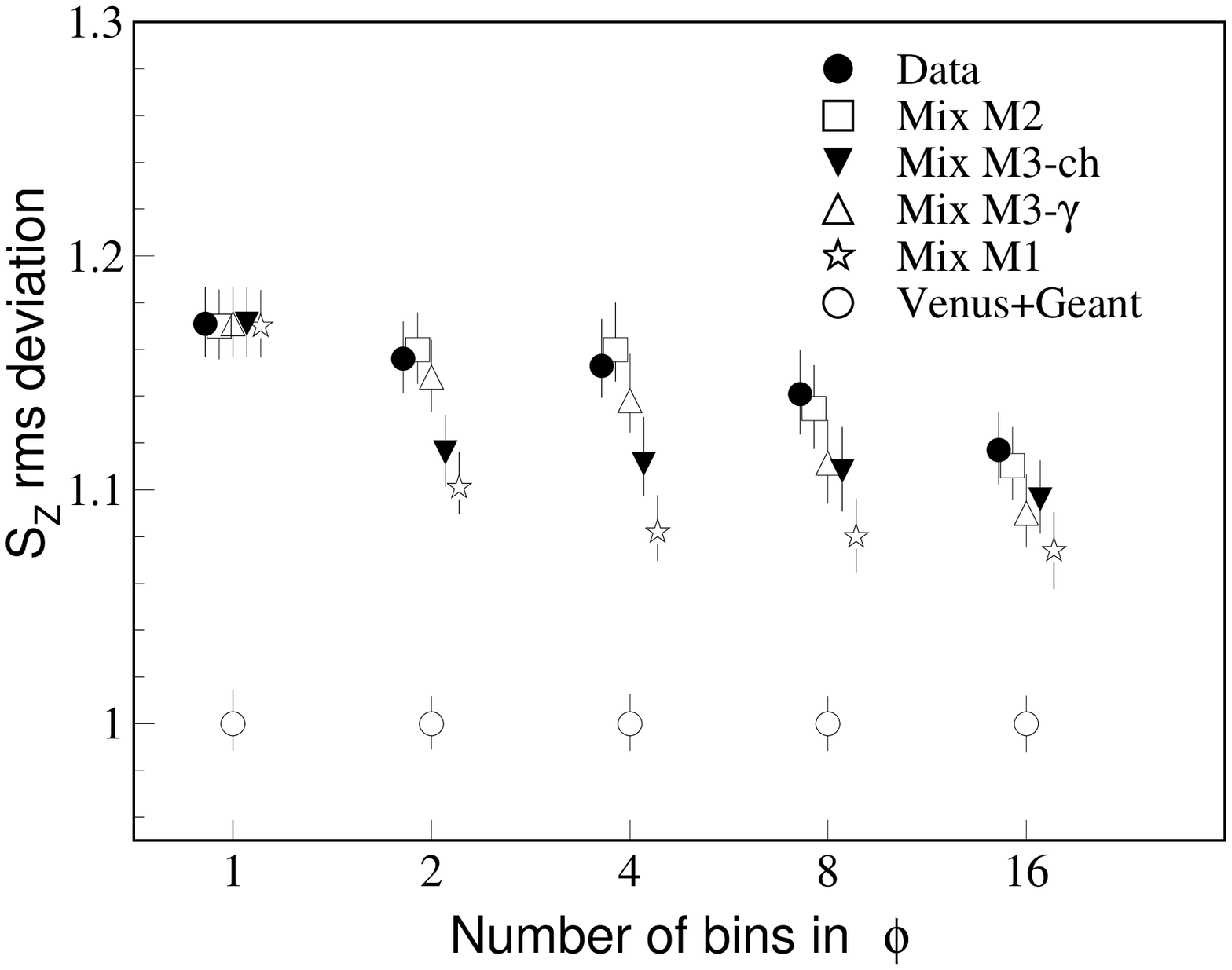}
}
\put(60,60){
\epsfxsize=6.cm
\epsfysize=5.cm
\epsfbox{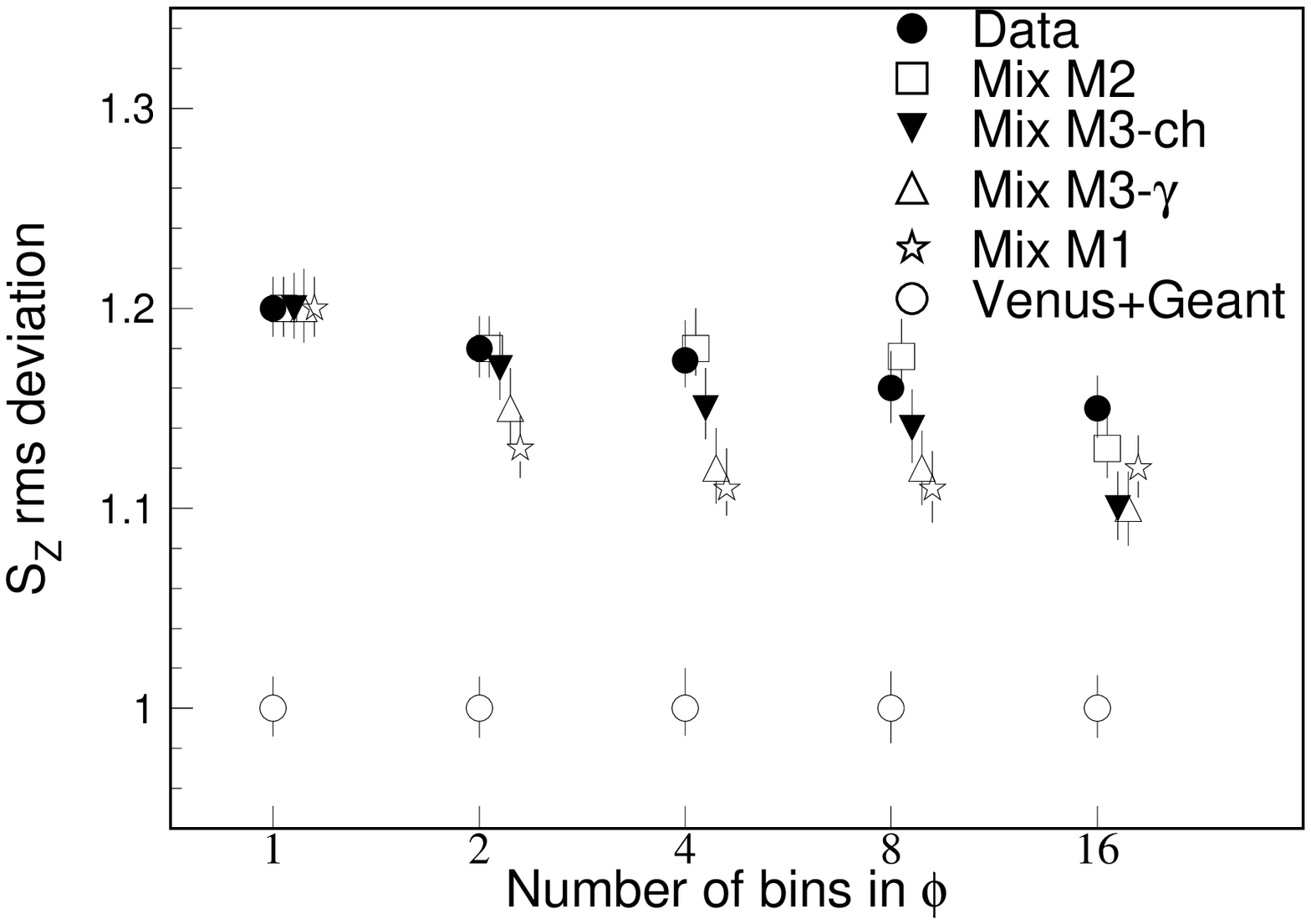}
}
\put(0,10){
\epsfxsize=6.cm
\epsfysize=5.cm
\epsfbox{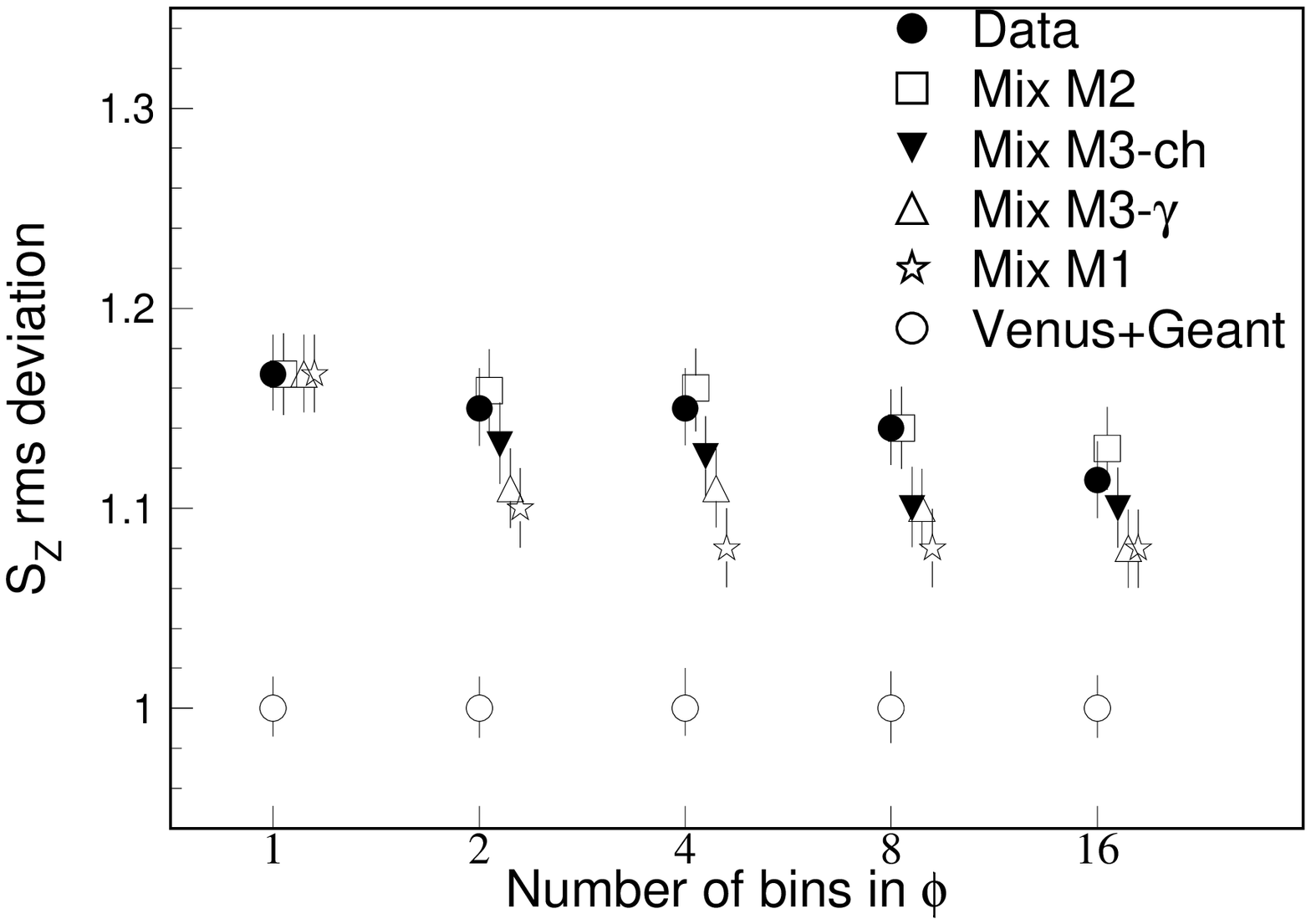}
}
\put(60,10){
\epsfxsize=6.cm
\epsfysize=5.cm
\epsfbox{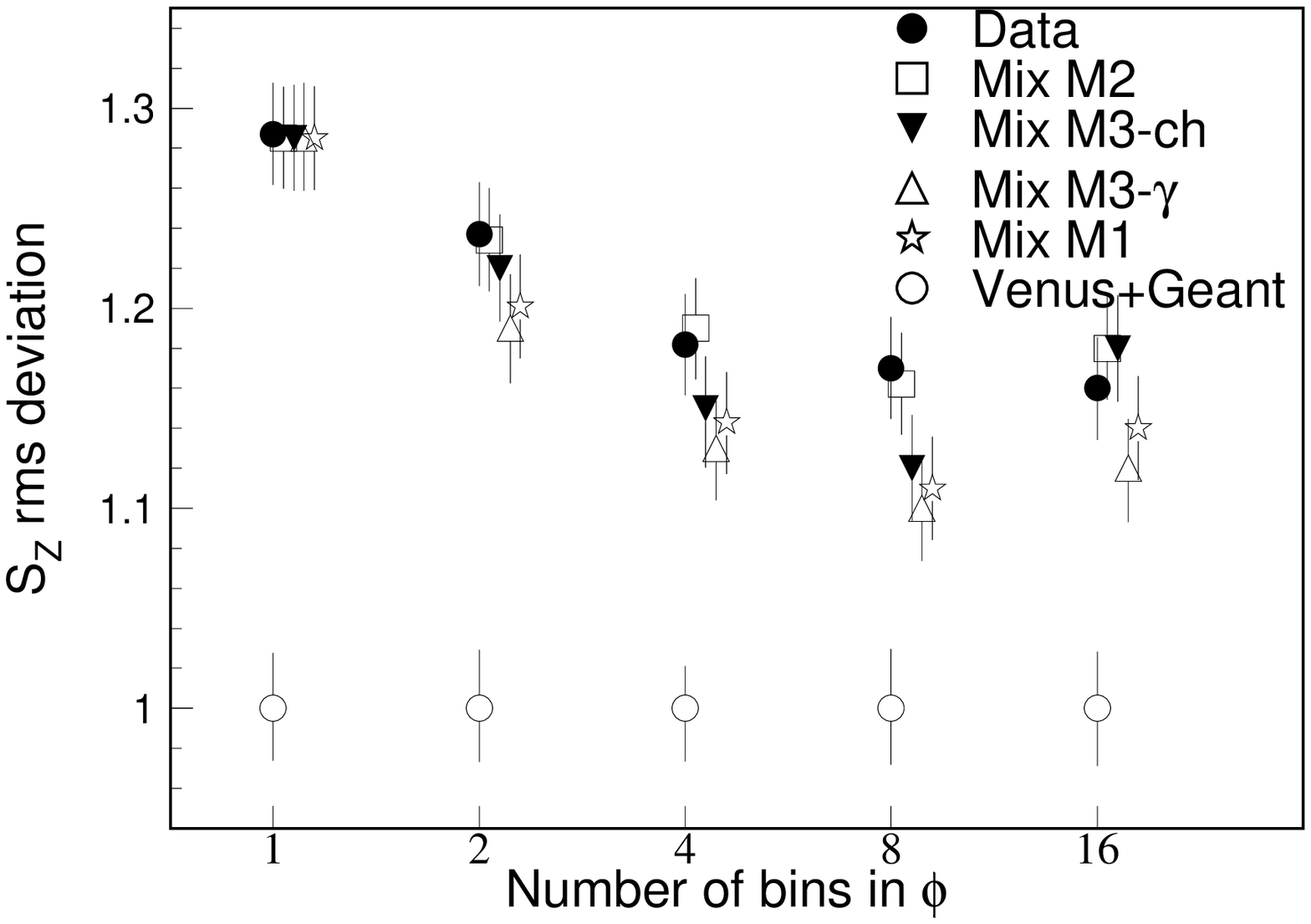}
}
\end{picture}
\end{center}
\caption{\label{sz_rms}\small
  RMS deviations of the $S_Z$ distributions for data, mixed and simulated events 
  for the four centrality bins: top-left for centrality-1, top-right for centrality-2, 
  bottom-left for centrality-3 and bottom-right for centrality-4. From \cref{wa98_local_cen}.
}
\end{figure}

The rms deviations of the $S_Z$ distributions calculated at different
$\phi$ bins are shown in \Fig{sz_rms} for data, mixed (M1, M2, M3-$\gamma$ and M3-ch)
and VENUS events, in various centrality bins. The statistical errors on the values
are small and are within the sizes of the symbol.
The bars represent statistical and systematic errors added in
quadrature. The total error is not more than
$3\%$ for the most central events and goes to not more than
$10\%$ for the most peripheral events.
From \Fig{sz_rms}, one observes that the nature of the rms 
deviations of the mixed events closely follow those of the data.
The mixed events have been constructed such that
the $\Ngl$ vs. $\Nc$ correlations are maintained for full azimuth 
-- that is one single bin, so the rms deviations of data and mixed 
events are identical by construction in that case. 
The rms deviations of M2-type events are found to agree with those 
of the experimental data within errors for all
four centrality classes, thereby suggesting the absence of localized
correlations in $\Ngl$ and $\Nc$. The same for
M1-type events are found to be lower than those
obtained for data for $2$, $4$ and $8$ bins in $\phi$ for centrality
bins $1$, $2$ and $3$. The results form M3-type events are found
in between those obtained from data and M1-type events. These
results indicate the presence of localized fluctuations in data, due to
both photons and charged particles. However for the case of centrality-4,
the rms deviations of $S_{Z}$ distributions from data and the mixed
events are found to more or less agree  with each other within the quoted
errors. The rms deviations for the simulated events are $1$ for all
bins in azimuth by definition of $S_{Z}$.

\begin{figure}[t]
\setlength{\unitlength}{1mm}
\begin{center}
\begin{picture}(100,100)(10,10)
\put(0,60){
\epsfxsize=6.cm
\epsfysize=5.cm
\epsfbox{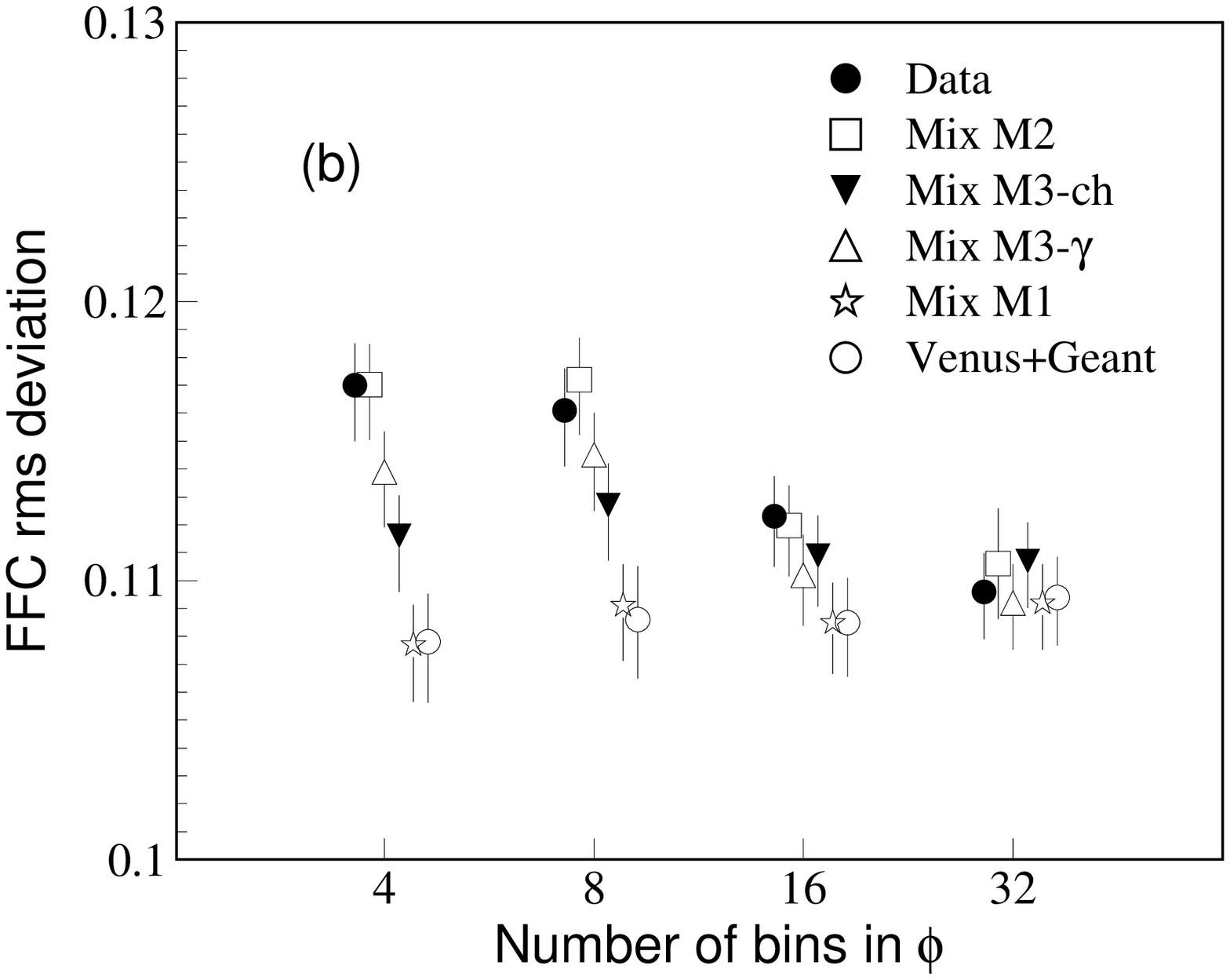}
}
\put(60,60){
\epsfxsize=6.cm
\epsfysize=5.cm
\epsfbox{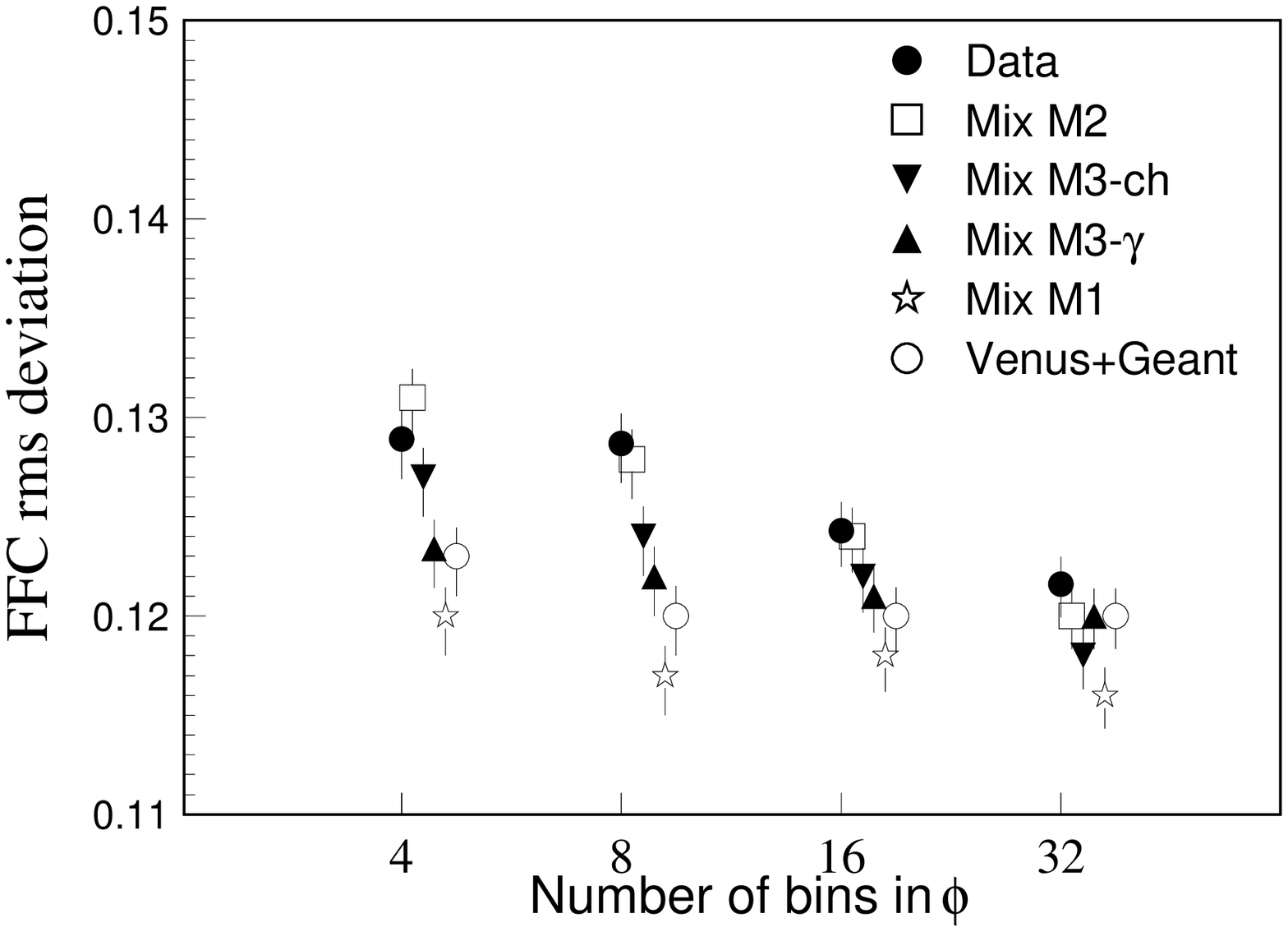}
}
\put(0,10){
\epsfxsize=6.cm
\epsfysize=5.cm
\epsfbox{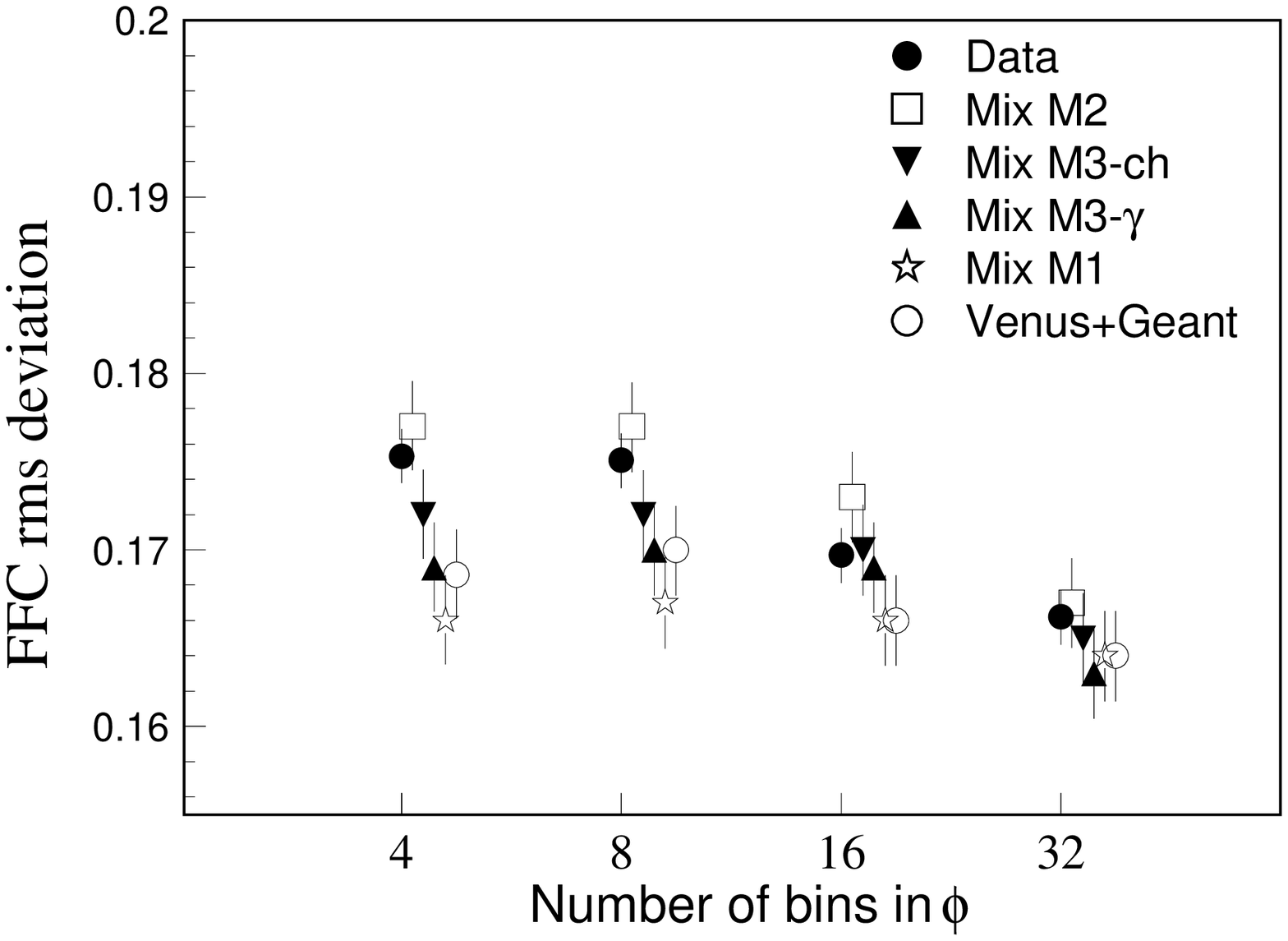}
}
\put(60,10){
\epsfxsize=6.cm
\epsfysize=5.cm
\epsfbox{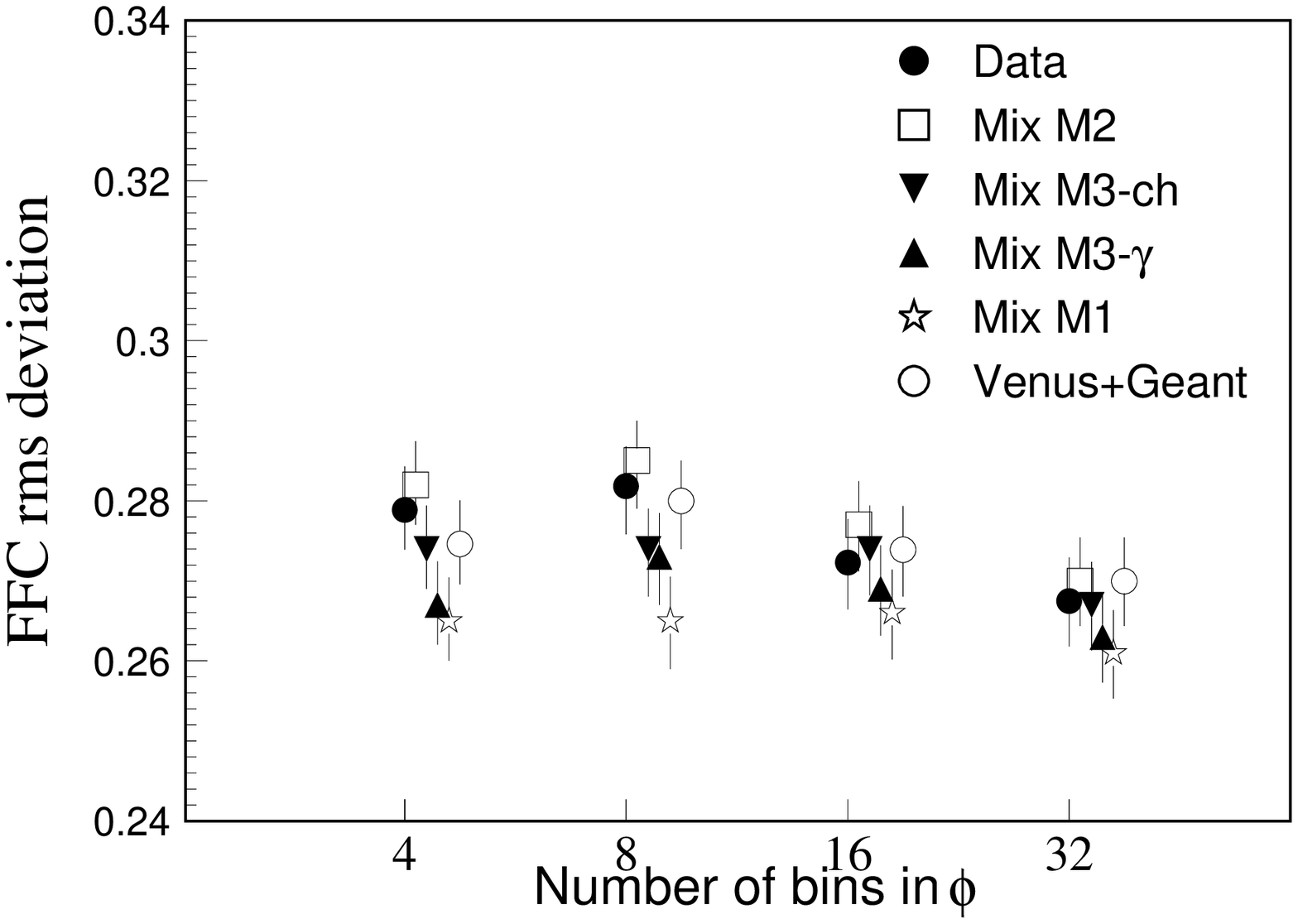}
}
\end{picture}
\end{center}
\caption{\label{ffc_rms}\small
  RMS deviations of the FFC distributions for data, mixed and simulated events 
  for the four centrality bins: top-left for centrality-1, top-right for centrality-2, 
  bottom-left for centrality-3 and bottom-right for centrality-4. From \cref{wa98_local_cen}.}
\end{figure}

The rms deviations of the FFC distributions calculated at different
$\phi$ bins are shown in \Fig{ffc_rms} for data, all four types
of mixed events, as well as generic VENUS events. 
We observe that the rms deviations of data, VENUS and mixed events
match with each other (within quoted errors) for $32$ bins
in $\phi$ for all the the four centrality classes. 
The VENUS results are close to the M1-type events, for centrality
classes 1, 2 and 3 and are slightly higher for centrality-4.
The rms deviations of the FFC distributions for 
M2-type events are found to agree closely with those for
data for all centrality classes and all bins in $\phi$. 
The rms deviations for M3-type events lie
in between those obtained from data and M1-type mixed events. 
These results are consistent with those obtained from the $S_{Z}$
distributions. They indicate the presence of individual localized 
fluctuations in photons and charged particles for certain bins in 
azimuth (resulting in fluctuations of the ratio, as seen).
However, they show no clear evidence for localized correlated 
fluctuations (DCC-type) between $\Ngl$ and $\Nc$.

\begin{figure}
\begin{center}
\vspace{-1cm}
\epsfig{file=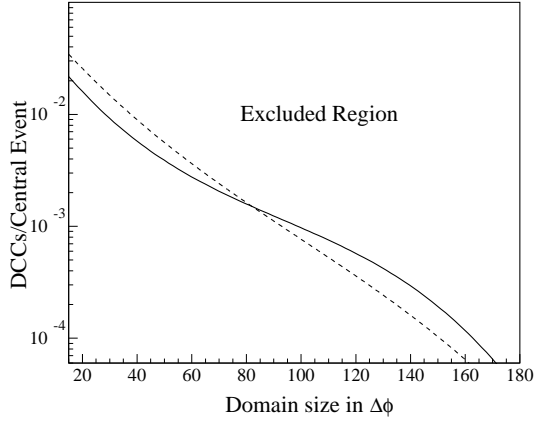,width=8.cm}
\caption{\label{upper_limit}\small 
  $90\%$ confidence level upper limit on DCC production per central
  event at SPS as a function of DCC domain size in azimuthal angle
  within the acceptance of the detectors.
  The solid line is for top $5\%$ central events and dashed line
  for  $5\%-10\%$ centrality class. From \cref{wa98_local_cen}.
}
\end{center}
\end{figure}

Comparing these results with those obtained from simulated DCC events
described in \Sec{sec:dcc_model}, one can extract an upper limit on the 
occurrence probability of DCC-like fluctuations at the $90\%$ confidence 
level, following the standard procedure~\cite{feld} as described below. 
This has been done, in particular, for central Pb-Pb events with $E_{T}\ge
341$~GeV at $158$~AGeV collision energy, as DCC formation is expected to be
more likely in these events. First, a quantity $\chi$ is defined as:
\begin{equation}
\label{xi_eqn}
    \chi =  \frac{\sqrt{(s_1^2 - s^2)}}{s}\,,
\end{equation}
where $s$ and $s_1$ correspond to the rms deviations
of the FFC (or $S_Z$) distributions of M2-type mixed events and
real data respectively. It is assumed that errors are Gaussian-distributed. 
Since errors are asymmetric one considers the largest error for the calculation 
to be on the conservative side.
The errors on the rms deviations are propagated to get errors
on $\chi$.
Upper limit contours with $90\%$ C.L. have been calculated
by taking the limit at $\chi + 1.28e_\chi$ where $e_\chi$
is the error in $\chi$ from $S_Z$ and FFC analysis.
The two upper limit contours set by the two different analysis were 
found to be consistent  with each other.  It must be mentioned that,
for calculating the upper limits, it has been assumed that the total
difference in rms values of data and M2-type events is
due to DCC-like fluctuations.
Now, to relate them to DCC domain size and frequency of occurrence of
DCC, one proceeds as follows: Obtain the rms deviations of FFC distributions
for various domain sizes (from $15^\circ$ to $180^\circ$ in $\phi$ with $15^\circ$ increments) 
for different values of the frequency of occurrence of DCC (varying from $0\%$ to $100\%$),
where simulated DCC events are generated from the simple model discussed earlier. 
Then carry out the entire analysis as done for data to get $\chi$ values for each 
set of events. One then finds for which frequency of occurrence 
for a fixed DCC domain size, the $\chi$ value from simulation
matches with that of the $\chi + 1.28e_\chi$ for data. This
is done for each of the above domain sizes. 
This is then used to set the upper limits 
in terms of domain size and frequency of occurrence of DCC, as shown on
\Fig{upper_limit}. It may be mentioned that results from both the $S_Z$ and 
FFC analysis methods gives similar upper limits.

Thus, within the context of this simple DCC model, an upper limit on the
presence of localized non-statistical DCC-like fluctuations is found to be
of the order of $10^{-2}$ for an azimuthal domain size $45^\circ\lesssim\Delta\phi
\lesssim 90^\circ$ and about $3\times 10^{-3}$ for $90^\circ\lesssim\Delta\phi\lesssim 
135^\circ$. This upper limit is the most strict limit set so far at SPS. Compared 
to earlier limits set in global analysis of WA98 and NA49, this analysis is 
more sensitive to possible presence of small localized domains of DCC in 
$\eta-\phi$ phase-space. 

\noindent{\it -Search for anti-Centauro events:}
Finally, a new method to look for anti-Centauro events in WA98 data
has been proposed \cite{mma}, where one studies events having photon excess
in azimuthal patches within the overlap zone of the PMD and the SPMD. 
One of the advantage of searching for anti-Centauro type events is that
the purity of the photon sample observed by the PMD will be higher
than those for normal patches due to the depleted flux of charged particles.
The WA98 Collaboration have analyzed 196K events corresponding to the top $15\%$ 
of the minimum bias cross-section. One looks for fluctuations in the 
neutral pion fraction in localized regions of the $\eta-\phi$ phase-space on 
an event-by-event basis. A particular azimuthal window $\Delta\phi$ is selected 
in the pseudo-rapidity range $ 2.9<  \eta \le 3.75$ and the entire azimuthal 
range from $0^\circ$ to $360^\circ$ is scanned in order to find the patch 
having the maximum value of $f$, referred to as $f_{max}$.
This scan is performed by successive $2^\circ$ rotations of the $\Delta\eta-
\Delta\phi$ patch in the $\eta-\phi$ plane.
To minimize the statistical fluctuations, a patch with maximum $f$ value
in an event was required to have at least $40$ photons corresponding to a  
$15\%$ statistical error.

\begin{figure}[t]
\begin{center}
\epsfig{figure=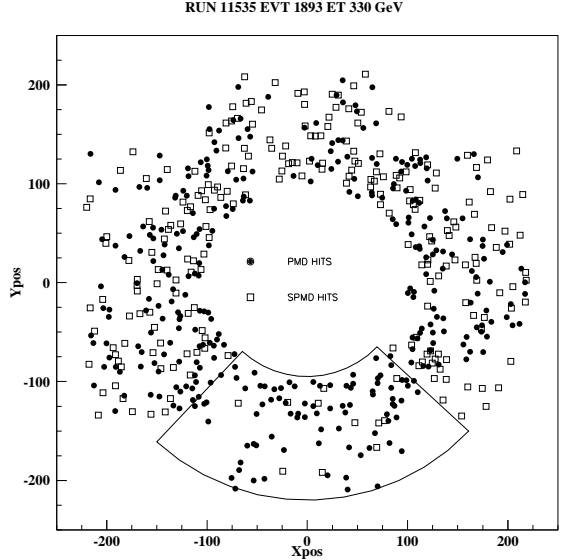,width=8cm}
\caption{\label{mma1}\small 
  Plot showing photon hits (PMD) and charged particle hits (SPMD)
  in an azimuthal plane. The marked $90^\circ$ patch corresponds to 
  $f_{max}$ = 0.77. From \cref{mma}.}
\end{center}
\end{figure}

\begin{figure}[t]
\begin{center}
\epsfig{figure=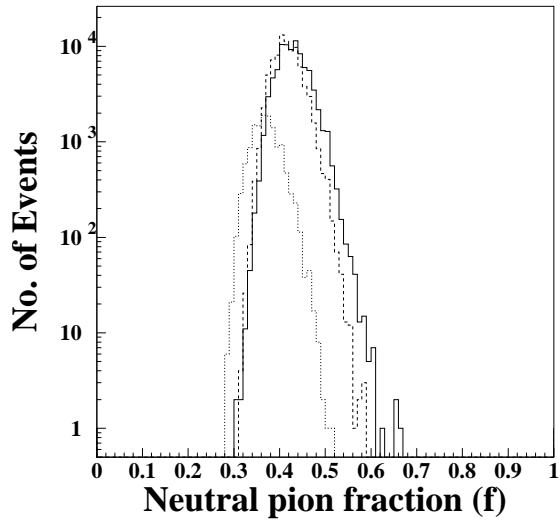,width=8cm}
\caption{\label{mma2}\small 
$f_{max}$ distributions for 60$^\circ$ patch 
for data (solid histogram), mixed Events (dashed histogram) and V+G 
(dotted histogram). From \cref{mma}.}
\end{center}
\end{figure}

Results from such an analysis have been compared with simulated events obtained 
using the VENUS 4.12 event generator and processed through the WA98 detector setup
using GEANT 3.21. These are referred to as V+G events. The statistical significance 
of these results is obtained by comparing them with those obtained from a similar
analysis applied to the various types of mixed events discussed above. 
An event displaying PMD hits (filled circles) and SPMD hits (open squares) within 
the overlap $\eta-\phi$ zone (obtained by using the method described above) is shown
in \Fig{mma1}. The patch of size $\Delta\phi=90^\circ$ in azimuth having the highest neutral
ratio $f$ is also marked. The number of charged particles in this patch is only $12$ 
as compared to $84$ photons, which corresponds to $f_{max} = 0.77$.
Figure~\ref{mma2} shows histograms of the maximum value $f_{max}$ of the neutral 
fraction in each event, for data, mixed events, and V+G events. This corresponds
to a patch size of $\Delta\phi=60^\circ$. It is seen that the $f_{max}$ distribution 
for the data extends to much larger values than for mixed and V+G events. 

\begin{figure}[t]
\begin{center}
\epsfig{figure=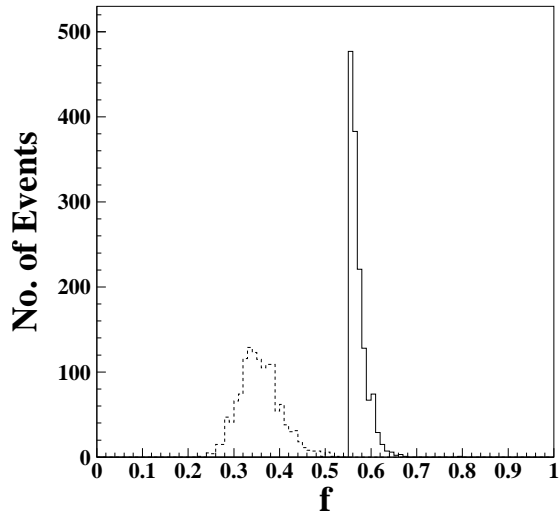,width=8cm}
\caption{\label{mma3}\small
  Neutral pion fraction ($f$) distributions
  for events having $f_{max}>0.55$
  (solid line) and for preceding and succeeding
  events (dashed line) for
  $60^\circ$ patch having $\Ng>40$. From \cref{mma}.}
\end{center}
\end{figure}

To test the authenticity of these events, the neutral pion fraction $f$ 
has also been calculated for the immediate preceding and succeeding events 
in the data in the same patch in $\eta-\phi$. The distribution of these $f$ 
values is shown in \Fig{mma3}. The solid line corresponds to the neutral pion 
fraction for patches having  $f_{max}> 0.55$ with $\Ng>40$ and the dashed 
histogram represents the $f$ distribution for preceding and succeeding events
in the data sample. The latter has its peak at about $0.35$, which is characteristic
of generic (normal) events. This shows that events with $f_{max}>0.55$ are indeed
special events, having large non-statistical charged-to-neutral fluctuations. 
In this sense , these events \cite{mma}
may resemble the anti-Centauro events found in cosmic ray experiments
\cite{jacee}.

In summary, using this new event-by-event analysis the WA98 have found 
events with large charged-to-neutral fluctuations in their data, which exhibit 
$f_{max} >0.55$ in certain $\Delta\phi$ domains. These domains are uniformly 
distributed in azimuth. The fraction of such events are much larger in the data
as compared to those seen in the mixed events and V+G events. The percentage of
these events increases significantly with decreasing patch size. Also 
the fraction of such events increases significantly as one decreases
the centrality. 
The special events selected by this procedure appear anti-Centauro like~\cite{mma} similar to those found in cosmic ray experiments \cite{jacee}.
A more careful analysis of the statistical significance of these events 
is needed to conclude about observation of DCC.

\paragraph{\label{sec:na49_result} The NA49 Experiment}

\begin{figure}[t]
\centerline{\epsfig{file=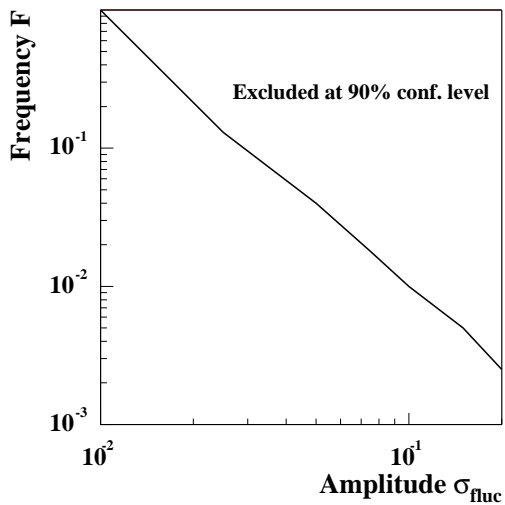,height=7.5cm}}
\caption{\label{na49dcc}\small Limit on the amplitude of fluctuations in the 
  $p_T$ parent distribution as a function of the frequency of 
  occurrence of the event showing the fluctuation. From \cref{na49}.}
\end{figure}

The NA49 Experiment, which detects only charged hadrons, has tried to
put an upper limit on DCC production at SPS from the measurement
of fluctuations in mean transverse momentum $M(p_T)$ \cite{na49}.
They argue that fluctuations in $M(p_T)$ are also relevant for models 
of processes that lead to non-statistical fluctuations localized 
in transverse momentum, like the formation of DCCs. As we have seen 
previously in the theory review, pions emitted from DCC domains
are expected to be preferentially produced at low transverse
momenta. This provides a translation of the number-fluctuations, 
expected from dynamical DCC simulations (see \Sec{sec:pheno}), into 
$p_T$ fluctuations accessible to the NA49 experiment. 
For comparison purposes NA49 used the same type of simulated DCC events
as the WA98 Collaboration \cite{wa98_global}, which we described in detail 
above. The DCC production is characterized by the probability $\alpha\equiv F$
to form a single DCC domain in an event and the fraction $\beta\equiv\xi$
of pions coming from the DCC. They made the additional assumption
that the DCC pions are produced with $p_T <  p_T^{max} =  m_{\pi}$.
The ratio of neutral to charged pions was chosen randomly according 
to the ideal DCC probability distribution of the neutral pion fraction,
\Eqn{DCCsign}. 
The isospin fluctuations of pion production from DCCs then
lead to multiplicity fluctuations of charged pions at low
transverse momenta and, therefore, to non-statistical fluctuations in 
$M(p_T)$. For DCCs occurring in every event ($F = 1$) the fluctuations 
observed in the data rule out DCC sizes of $\xi > 0.35$.  
The limit on the amplitude of fluctuations in the $p_T$ parent distribution 
as a function of the frequency of occurrence of the events showing the 
fluctuation is shown in \Fig{na49dcc}.

\paragraph{\label{sec:rhic_results} RHIC experiments}

\begin{figure}[t]
\centerline{\epsfig{file=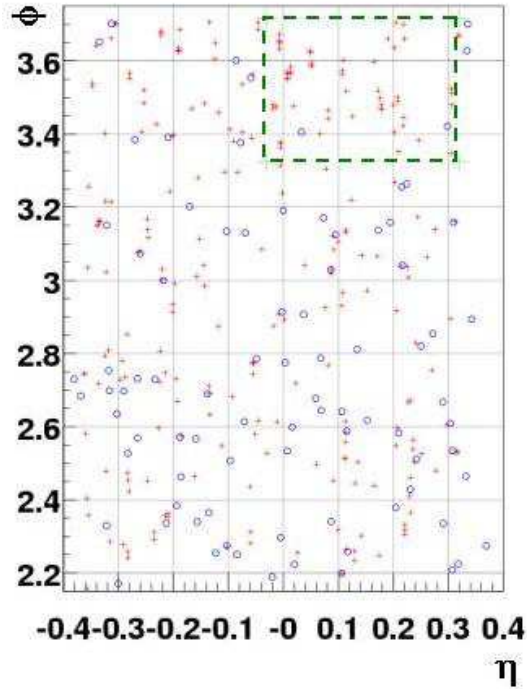,width=7.1cm}}
\caption{\label{phenix_dcc}\small Centauro-type event seen in the PHENIX Experiment at RHIC.
  The solid circles are photon hits and plus signs are charged-particle hits.
  The box shows the Centauro region in $\eta-\phi$ phase-space.
  From \cref{phenixqm}.}
\end{figure}

Among the four RHIC experiments, both STAR \cite{starnim} and PHENIX \cite{phenixnim} have 
the capability to look for DCC-type fluctuations. So far the PHENIX Collaboration
have reported \cite{phenixqm} some results on search for Centauro and anti-Centauro 
type of events at RHIC energies. They use the photon information from the
electromagnetic calorimeter and charged-track information using
drift chamber and pad chamber detectors. One such typical possible
Centauro event is shown in \Fig{phenix_dcc}. The analysis is
still in preliminary stage. In addition to looking for such
events, they also carry out a multi-resolution analysis, with input 
being the asymmetry between the number of charged tracks and
neutral clusters, on an event-by-event basis as a function of 
subdivided $\eta-\phi$ phase-space.

\subsection{\label{sec:poslim_expt} Possibility and limitations of DCC search}

For detecting a DCC signal with very low probability of occurrence,
it is necessary to understand, from an experimental point of view,
all the factors that affect the detection of DCC domain during its 
transition from the time of formation to the time of detection. 
This has been investigated in \cref{poslim} using the simple model
for simulating DCC events described earlier. The analysis is carried 
out using the method based on the DWT technique and looking at the 
signal-to-background ratio.
Possible factors that can affect the DCC signal are: effect of the 
$\pi^{0}$ decay; the presence of multiple DCC domains; detector related 
effects, such as efficiency and purity of particle detection; use (or lack) 
of $p_{T}$-information of the detected particles; increase in particle 
multiplicity as one goes from SPS to RHIC and LHC energies. 

\subsubsection{\label{sec:pi0_decay} Effect of decay of $\pi^{0}$}

\begin{figure}[t]
\begin{center}
\epsfig{figure=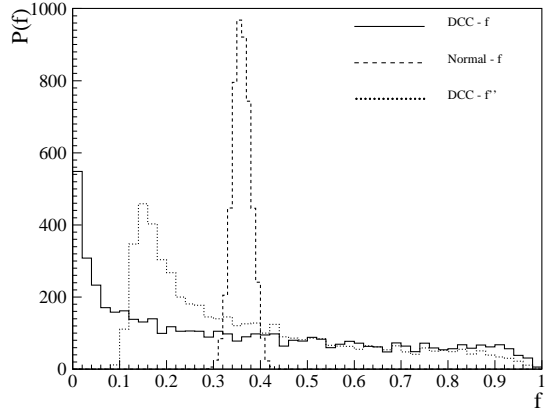,width=8cm}
\caption{\label{domain_deccay}\small
  Typical $f$ distributions inside a DCC domain before (solid line) 
  and after decay (dotted line) of $\pi^{0}$ ($f^{\prime \prime}$). 
  Also shown for comparison is the $f$ distribution for normal
  events (dashed line). From \cref{poslim}.}
\end{center}
\end{figure}

The formation of DCC leads to large event-by-event fluctuation
in the neutral pion fraction $f$. To observe this in an ideal situation 
one has to count the number of neutral pions $\pi^{0}$ event-by-event. 
But the $\pi^{0}$'s decay into photons by the time they reach the detector. 
As a result, photons coming from DCC $\pi^{0}$'s inside a given DCC 
domain may move out of the $\eta-\phi$ phase-space of the domain -- depending 
on their momentum -- before they are detected. Also, photons from outside 
the DCC domain may enter the $\eta-\phi$ phase-space of the domain. 
Both effects result in dilution of the strength of the signal. Considering 
the effect of $\pi^{0}$ decay, one modifies \Eqn{new_f} as:
\begin{equation}
  \label{decay_eqn}
  f^{\prime\prime}=\frac{\Ng/2\pm\delta_{\gamma}}{\Ng/2\pm\delta_{\gamma}+\Nc}\,,
\end{equation}
where $\delta_{\gamma}$ is the resultant number of photons that get removed
from or added into the  DCC domain and $\Nc$ is the number of charged pions
in the region of $\eta-\phi$ phase-space under consideration. For multiplicity 
detectors (\eg SPMD in WA98, or FMD in ALICE), which do not have the capability 
of particle identification, $\Nc$ denotes the multiplicity of all charged particles
in the considered phase-space region. Since $\delta_{\gamma}$ is a non-zero number, 
the possibility of observing a pure Centauro-type event ($f\ll1$) is low. 
In fact the distribution of the ratio $f^{\prime \prime}$ will be shifted away 
from zero, as illustrated in \Fig{domain_deccay}, which shows the  
probability distributions for $f$ and $f^{\prime \prime}$  inside a DCC domain 
in the model calculation. 
Here $\Nc\equiv N_{\pi^\pm}$. The effect of detecting other charged particles along 
with pions will be discussed later. Clearly one can notice that, as a result of 
$\pi^{0}$ decay, the peak of the original DCC distribution is shifted to a non-zero
value $\simeq 0.15$. 
Also shown in the figure is the  $f$ distribution for normal events. In that case, the
$f$ and $f^{\prime \prime}$ distributions are not very different because the relative 
population of $\pi^{0}$ is the same within and outside the domain and hence the loss 
due to decay is compensated by the gain due to decay from other phase-space regions. 

\begin{figure}[t]
\begin{center}
\epsfig{figure=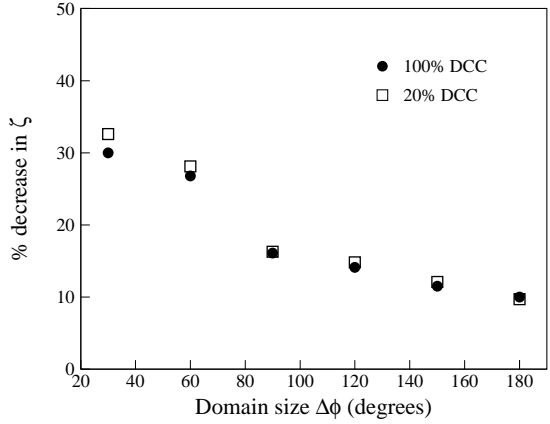,width=8cm}
\caption{\label{deccay}\small
  Percentage decrease in the strength of DCC signal
  due to the effect of decay of $\pi^{0}$ as a function of domain size 
  $\Delta \phi$. From \cref{poslim}.}
\end{center}
\end{figure}

In order to quantify the decrease in the detectability of the DCC signal, the 
authors of \cite{poslim} considered the following two cases:
\begin{itemize}
\item[(i)]{\it Analysis using a $\pi^{0}$ detector:}\\
  In the first case one puts two hypothetical 
  detectors, one for detecting $\pi^{0}$'s
  and other for detecting charged pions, 
  with $100\%$ efficiency, within one unit of $\eta$ 
  and full $\phi$ coverage. 
  The detector effects will be discussed later taking realistic efficiency 
  and other parameters. The goal here is to study the $\pi^{0}$ 
  decay effect only. 
  Then one introduces DCC domains in the simulated events with 
  domain size $\Delta \phi$ varying from $30^\circ$ to $180^\circ$. 
  These events without $\pi^{0}$ decay and having DCC domain of
  a particular size are then analyzed using the DWT method to 
  obtain the FFC distribution. The strength parameter $\zeta$ is 
  calculated from \Eqn{zeta}.
\item[(ii)]{\it Analysis using an ideal photon detector:}\\
  In the second case one replaces the $\pi^{0}$ detector 
  with an ideal photon detector having $100\%$ detection efficiency
  and no contamination from charged particles. 
  The $\pi^{0}$'s are allowed to decay and a similar analysis as 
  mentioned above is carried out to obtain the strength values. 
\end{itemize}  
From the two such obtained strength values for each DCC domain size, 
one calculates the percentage decrease in strength value of the second 
case (ii) compared to the first case (i). This is shown in
\Fig{deccay} for various domain sizes in $\Delta \phi$.
The results show that the decrease in strength of the signal is 
more for smaller domains of DCC as compared to larger ones. 
We mention that for each DCC domain size, two cases have been studied: one in
which all the events have DCC-type fluctuations and another where only $20\%$ 
of the events have DCC-type fluctuations introduced. It is found that the 
results are similar for both cases.

\subsubsection{\label{sec:multi_domain} Multiple domains of DCC}

In heavy-ion collisions there is a possibility that more than one domain
of DCC be formed, each with a chiral condensate pointing in independent 
directions in isospin space. Of course, the contribution to the neutral
fraction from different independent domains is expected to reduce the 
DCC signal in isospin fluctuations~\cite{amado}. In particular, in case 
of a large number of independent domains of equivalent size,\footnote{Clearly, 
the situation is better if one of the domains is predominant~\cite{Li}.} 
one expects the distribution of the total neutral fraction $f$ to approach a 
Gaussian centered at $\bra f\ket=1/3$ and of width inversely proportional to the 
square root of the number of domains, as per the central limit theorem.
This makes the dis-entanglement of DCC signal from that of normal events 
difficult.

These effects have been studied in \cite{poslim} 
using the multi-resolution DWT method. For simplicity, and in view of limitations 
of the model in terms of introducing DCC at the freeze-out stage, these authors have 
placed varying number of DCC domains in each event, depending on the domain 
size in $\Delta\phi$. All the domains are assumed to extend over one unit in 
pseudo-rapidity. For example, we can place up to four domains of $\Delta\phi=90^\circ$, 
without any overlap among themselves. These authors have also carried out the 
analysis by placing up to six DCC domains with $\Delta\phi=30^\circ$ in each event.
Care has been taken in placing the domains randomly so that no two domains overlap.
The $f$ value of each domain in each event is randomly chosen following the 
$1/2\sqrt f$ probability distribution. The aim is to see how the signal changes in 
terms of the strength parameter as given in \Eqn{zeta}. 

\begin{figure}[t]
\begin{center}
\epsfig{figure=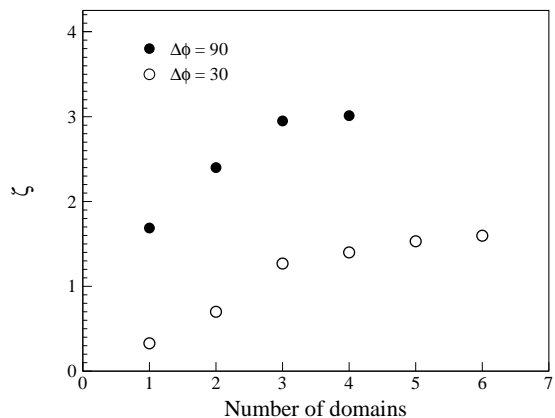,width=8cm}
\caption{\label{multidomain}\small
  Variation of the strength of DCC signal
  with increase in number of DCC domains in an event. 
  All the events are DCC events. From \cref{poslim}.}
\end{center}
\end{figure}

Figure~\ref{multidomain} shows the variation of strength of the DCC signal 
with the number of domains in a given event for two domain sizes of 
$\Delta\phi=90^\circ$ and $30^\circ$. One observes that the strength of the DCC 
signal increases as the number of domains increases and saturates for 
larger number of domains. 
The increase in strength value with increase in number of domains 
is because the multi-resolution event-by-event analysis
based on the DWT method has been able to pick up signals by looking 
at bin to bin (in $\phi$) fluctuations in each event. 
It is worth emphasizing that this result is not in contradiction with the 
theoretical considerations described above~\cite{amado,Li}. Indeed, if the 
multiple DCC domains formed in the initial stages, in the course of their 
evolution, move towards the same part of phase space covered by a detector, 
then the strength of the signal will reduce. However, if the phase space 
separation is maintained in evolving DCC domains, then the strength will 
increase as found in the present analysis.

\subsubsection{\label{sec:detector_effect} Detector effects}

The detector limitations are basically those related to efficiency and purity
of particle detection and to the acceptance in $\eta-\phi$. 
The effect of efficiency of particle detection 
is trivial, higher the efficiency of particle detection, more reliable is 
the search for DCC. Similarly the role of higher acceptance in $\eta$ and 
full azimuthal coverage can be hardly over-emphasized (lower acceptance 
effectively reduces the total multiplicity of observable DCC pions and hence
reduces the strength of the DCC signal). If the acceptance in $\eta$ is 
sizable, one can attempt DWT analysis using bins in $\eta$ for limited 
azimuthal coverage, which would be complementary to the various analysis 
reviewed here. Such an analysis would certainly be of great interest.

The purity of the photon sample, as measured in a detector like
PMD, is around $60\% - 70\%$, whereas that of charged particle samples 
measured  using detectors like FTPC or FMD is quite high, larger than $95\%$.
Hence we discuss here only the effect of purity of photon sample.

\paragraph{Effect of neutral pion fraction on the purity of photon sample}

\begin{figure}[t]
\begin{center}
\epsfig{figure=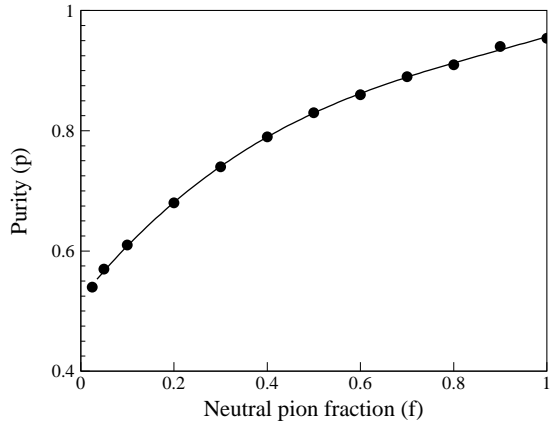,width=8cm}
\caption{\label{pur_f}\small
 Variation of purity of the photon sample 
 as a function of neutral pion fraction. For $f = 0.33 $ (generic event)
 purity is $\sim 0.74$. From \cref{poslim}.}
\end{center}
\end{figure}

It is of interest to study the effect of the DCC neutral pion fraction on the
purity of the photon sample. Consider the case where the neutral pion fraction 
for a DCC domain is $f=1$. This corresponds to a situation where there are 
no charged pions. This will lead to greatly reduced number of 
charged particles  falling on the photon detector.
So the purity of the photon sample will be high. 
The variation of purity with neutral pion fraction can be investigated in the
following manner \cite{poslim}: 
The number of $\gamma$-like hits on the 
photon detector can be written as
\begin{equation}
\label{a_def}
 \Ngl = \epsilon\Ng + c \Nc\,,
\end{equation}
where the first term denotes the contribution of actual photons and the
second term denotes charged-particle contamination. The factor $c$ is the 
fraction of charged-particles on the PMD acceptance treated as
contamination to the detected photon sample.
One assumes that the latter is given by a normal distribution with mean 
of $15\%$ and width of $5\%$, the percentage being taken with respect to 
the total charged particles within 
the acceptance of the photon detector. This is a reasonable number, 
considering that the converter thickness of the preshower detector is $\sim$
$10\%$ of an interaction length and some interactions lead to 
multiple clusters (the effect of overlapping clusters is ignored).
DCC-type fluctuations are introduced for fixed $f$ values
and domain size $\Delta \phi = 360^\circ$ and $\Delta \eta = 1$ 
in a set of event. Several such sets of events were generated for $f$ values 
varying from $0.025$ to $1.0$. The purity of the photon sample is then calculated for 
each set of events with a fixed neutral pion fraction value.

The results are shown in \Fig{pur_f}, where one clearly sees that the
purity of photon sample increases with the increase in the neutral pion
fraction value. However the purity does not reach a value of $1$ for 
$f$ = $1$,  because there are charged particle 
other than $\pi^{\pm}$ falling on the detector. Similarly it never reaches a 
value of zero for $f$ = 0, 
as there are $\gamma$'s from $\pi^0$'s decaying outside the 
detector acceptance and from other sources.

\paragraph{Effect of charged-particle contamination in PMD}

\begin{figure}[t]
\begin{center}
\epsfig{figure=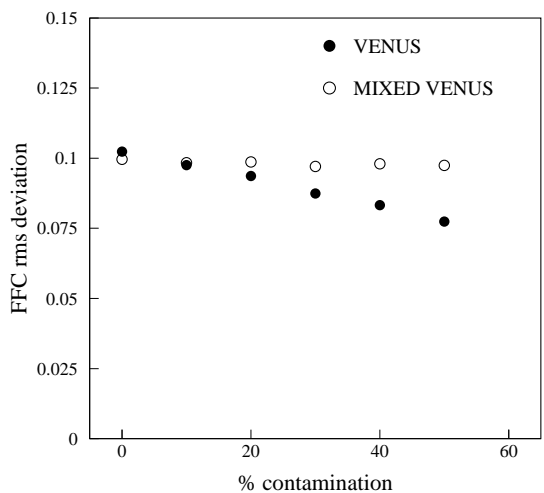,width=8cm}
\caption{\label{conta_var}\small
  Variation of rms deviation of FFC distribution for pure
  VENUS events with increase in the percentage of contamination in the photon
  detector. The variation of rms deviation of FFC distributions of
  mixed events constructed form VENUS are also shown. From \cref{poslim}.}
\end{center}
\end{figure}

Photon measurement in heavy-ion collisions has the problem of
charged-particle contamination. 
The charged particles detected as photons have a
correlation with those detected in the charged-particle 
multiplicity detector. This additional correlation suppresses 
the possible anti-correlation between photons and charged particles that 
may arise due to DCC formation.

The effect of such a correlation has been studied in \cite{poslim} with the 
help of VENUS (no DCC) events and a set of mixed events generated from them
as described above. 
The percentage of charged-particle contamination was varied 
from $0\%$ to $50\%$. 
The event-by-event $\Ngl$ was kept the same for all the 
cases. The result of the DWT analysis on these events is shown in 
\Fig{conta_var}. There is a decrease in the 
rms deviation of the FFC distribution for pure VENUS events with 
increase in the percentage contamination in the photon sample. 
Thus the presence of any additional anti-correlation due to DCC-like 
effect will have to overcome the opposing correlation effect due to 
contamination in order to be observed. 
Figure~\ref{conta_var} also shows the effect of contamination on
M1-type mixed events generated from the VENUS events. 
It is found that the rms deviations of the FFC distributions for the 
mixed events remain independent of the level of contamination. The 
mixed events appropriately break the additional 
correlation due to charged-particle contamination.

\begin{figure}[t]
  \begin{center}
    \epsfig{figure=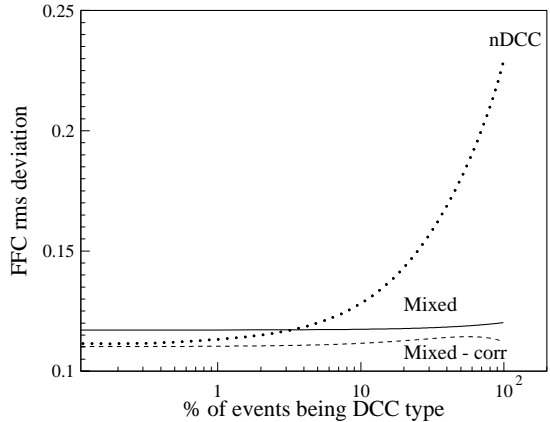,width=8cm}
    \caption{\label{conta_var2}\small
      Variation of rms deviation of FFC distribution of DCC-type
      events as a function of the percentage of events being DCC-type. Also shown
      are the results for corresponding mixed events and mixed events
      corrected for contamination effect. From \cref{poslim}. 
    }
  \end{center}
\end{figure}

The effect of charged-particle contamination can be corrected by 
knowing the level of contamination in photons for a given detector. 
Figure~\ref{conta_var2} shows the variation of
rms deviation of FFC distribution for DCC-like events with increase in
percentage of events being DCC-type in a given ensemble of nDCC events,
described previously.
It also shows the variation of rms deviation of 
FFC distribution of corresponding mixed events. 
The rms deviations of the FFC distribution of mixed events  are
found to be independent of the percentage of events being DCC-type. 
This is along the expected lines. But for lower fraction of events 
being DCC-type it is above that of the parent sample of nDCC events. 
This is because the $\Ngl-\Nc$ anti-correlation due  
to DCC-type effect is not sufficient 
to overcome the correlation between $\Nc$ and
charged particles detected as contamination in the photon detector. 
However as the fraction of events being DCC-type increases, the DCC-type 
effect dominates. 
The effect of the contamination can be corrected.  
This is done by taking into account the difference in 
rms deviation of normal events and mixed events. 
The rms deviation of FFC distribution of corrected mixed events are 
shown in \Fig{conta_var2}. 

The results discussed in the above two sections indicate that it is better 
to look for anti-Centauro (photon excess) events in studies using 
photon and charged-particle multiplicity detectors.
This is because the effect of decay is primarily to shrink the $f$ 
distribution from the lower $f$ side and high value of $f$ leads to a
reduced effect of charged particle contamination on DCC search.

\subsubsection{\label{sec:low_pt} Effect of low-$p_T$--information of particles}

It is expected that $p_{T}$-information of particles would be very
helpful in DCC search. This would also enable one to verify various 
expected features of DCC production, such as the fact that DCC-pions 
have low transverse momenta. 
In order to show the utility of $p_{T}$-information, the following study
has been carried out in \cref{poslim}. One assumes that all charged particles 
with $p_{T}$ greater than $50$~MeV/c are detected (the $p_T$ acceptance of
preshower detector like PMD extends down to $30$~MeV/c~\cite{pmdnim}). 
The $p_{T}$-resolution for the charged-particle detector was taken as 
$\Delta p_{T}/p_{T}=0.2$~\cite{ftpc}. These realistic parameters
correspond to a typical experiment for DCC search, as in STAR.
DCC events are simulated as described above. 
Since DCC pions are believed to have low $p_{T}$, one introduces DCC-type
fluctuations in pions with $p_{T}\le150$~MeV/c. Then a DWT analysis is 
carried out. The sample function in the DWT analysis is modified such that  
$\Nc$ taken corresponds to charged pions having $p_{T}\le150$~MeV/c.

\begin{figure}[t]
  \begin{center}
    \epsfig{figure=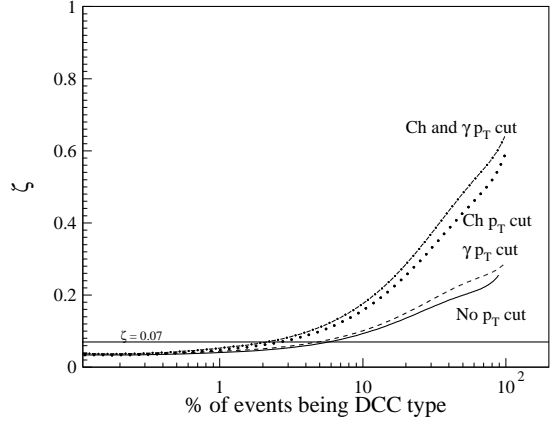,width=8cm}
    \caption{\label{lowpt}\small
      Variation of $\zeta$ as a function of the fraction of events being
      DCC-type, for charged particle detector with $p_{T}$-information 
      and  photon detector without $p_{T}$-information (dotted line), 
      for photon detector with $p_{T}$-information and
      charged particle detector without $p_{T}$-information (dashed line)
      and for both charged particle and photon detector with $p_{T}$-information 
      (dot-dashed line).
      Also shown are the corresponding results with both charged particle
      detector and photon detector without $p_{T}$ information (solid line).
      The horizontal straight line indicates the statistical error on $\zeta$.
      From \cref{poslim}.
    }
  \end{center}
\end{figure}

Figure~\ref{lowpt} shows the strength $\zeta$ of the DCC signal as a 
function of the percentage of events being DCC-type. For comparison, 
also shown is the corresponding $\zeta$ where the charged particle 
detector has no $p_{T}$-information. We note that the absolute 
value of $\zeta$ is lower compared to those presented in previous sections.
This is because here we have taken only low-$p_{T}$ pions as DCC pions in 
a given domain, whereas earlier all the pions in the domain were 
considered to be
DCC-type. This reduces the strength $\zeta$ as a result of reduction in 
multiplicity. The statistical error on $\zeta$ for the present case is $0.07$.

Clearly one can see the increase 
in strength of the signal with the use of $p_{T}$-information. A similar
analysis was carried out assuming that all photons have $p_T$-information
while the charged particles do not have any.
The results are also shown in \Fig{lowpt}. One sees that,
although there is an increase in signal strength compared to the case with no 
$p_T$-cut, it is much less compared to the case with
$p_T$-information for charged particles. For the case where the analysis
is carried out assuming that both photons and charged particles have 
$p_{T}$-information, one finds that the results are close to those obtained 
for charged-particle detector with $p_{T}$-information, except for events
having higher percentage of DCC type events.

\begin{figure}[t]
\begin{center}
\epsfig{figure=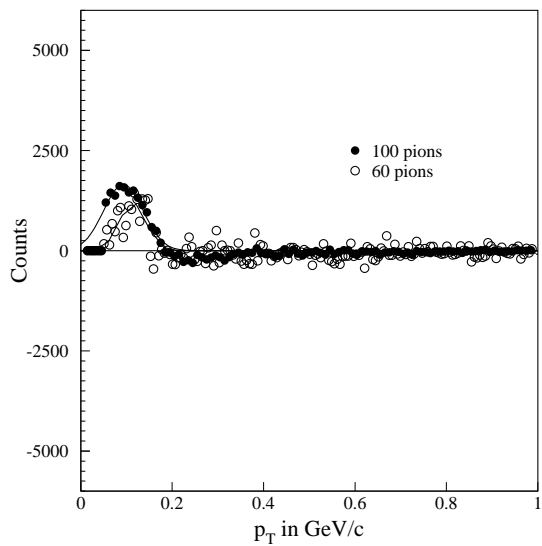,width=8cm}
\caption{\label{pt_enhance}\small
  Resultant $p_{T}$ distribution of charged particles with 
  $20\%$ of the events being DCC-type, obtained by subtracting the
  $p_{T}$ distribution of charged particles within $2~\sigma$ of the
  FFC distribution from those beyond. Two cases are presented, additional
  low $p_{T}$ DCC pions being $60$ and $100$.
  The statistical errors are within the symbol size. From \cref{poslim}.}
\end{center}
\end{figure}

It is also interesting to look whether one can detect the enhancement of
low-$p_T$ pions due to the presence of a DCC.
This can be done if the transverse momentum of charged particles is 
measured. In order to incorporate this effect in simulation, a number of 
low-$p_{T}$ pions ($p_{T}\le150$~MeV/c), with neutral fraction distributed 
according to the ideal DCC inverse square root law
and having uniform $p_{T}$ distribution were added within 
the chosen domain  on top of the existing pions of a normal 
event~\cite{strenght}. 
The number of pions to be added depends on the size and
energy density of the DCC domain. Two cases, with $100$ and $60$
additional low-$p_{T}$ pions were considered, corresponding to an energy 
density within a DCC domain of about $50$~MeV/fm$^{3}$ and a domain radius 
of the order of $3-4$~fm. As a test case a event sample with $20\%$ of the events 
having a DCC domain of size $\Delta\phi=90^\circ$ and 
$\Delta \eta=1$ was generated. 

The event sample has been analyzed by the DWT method to 
obtain the FFC distribution, which was found to be near Gaussian for 
the present case \cite{poslim}. One divides the FFC distribution into 
two parts: one within $\pm 2 \sigma$ of the mean and the other beyond 
this. One then obtains the separate $p_{T}$ distributions of these two 
subsets of events. Finally, the histogram corresponding to events within 
$2\sigma$ of the FFC distribution is subtracted from those beyond, 
after proper normalization. The resultant spectrum is shown in \Fig{pt_enhance}.  
Clearly one sees that the above analysis technique should provide a useful
tool to able to observe the low-$p_{T}$ pions enhancement due to possible DCC 
formation. 

\subsubsection{\label{sec:cms_energy} Effect of increase in center of mass energy}

\begin{figure}[t]
\begin{center}
\epsfig{figure=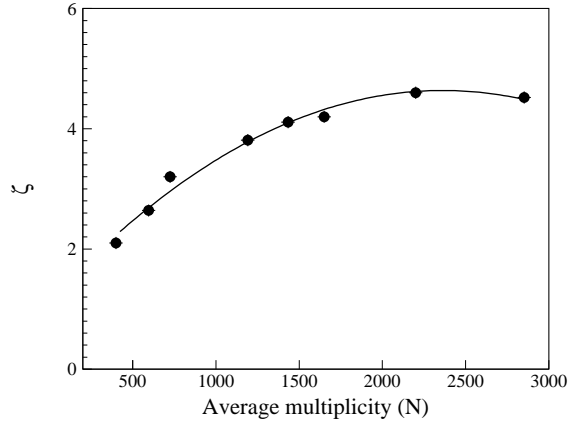,width=8cm}
\caption{\label{multi}\small
  Variation in the strength of DCC signal
  with multiplicity in a domain of $\Delta\phi=90^\circ$ and $\Delta\eta=1$. 
  Here, $N$ is the average of the sum of mean photon and mean charged particles 
  in a set of events. All the events are DCC events. From \cref{poslim}.}
\end{center}
\end{figure}

The effect of higher multiplicity is to reduce the event-by-event
statistical fluctuation associated with the multiplicity of photons 
and charged particles. A simple estimate of the strength of DCC-type
fluctuations shows that for a DCC domain of azimuthal size $\Delta\phi=90^\circ$  
and $\Delta\eta=1$, the strength of DCC signal increases 
with increase in average multiplicity of photons and charged particles. 
Figure~\ref{multi} shows the variation 
of the strength $\zeta$ of the DCC signal with increasing multiplicity, where
$N$ is the average of the sum of mean photon and mean charged particles 
in a set of events for one unit in $\eta$. 
One observes that the increase is almost like a $\sqrt{N}$ effect.
This is probably a consequence of going
from a narrow Gaussian distribution to a wider $1/2\sqrt{f}$ distribution.
It may be mentioned that typical values of $N$ in central 
collisions for $\Delta \eta\sim 1$, for the WA98 Experiment at SPS 
(combination of PMD and SPMD detectors for charged-particle detection) was 
about $320$. It is expected to be about $500$ for the STAR Experiment at RHIC 
(PMD + FTPC) and about\footnote{This assumes a pseudo-rapidity density of $\sim8000$ 
at mid-rapidity \cite{poslim}.} $2800$ at the LHC, with the ALICE detector (PMD + FMD).

\section{Summary and conclusions}

Since it has been proposed in the early 1990's, the idea that a DCC might be 
formed in high-energy collision experiments has triggered an intense activity,
both theoretical and experimental. Attempts to detect this phenomenon include
various cosmic ray experiments, the T-864 MiniMAX experiment -- a dedicated search 
in proton-antiproton collisions at the Fermilab Tevatron, or event-by-event analysis 
of Pb-Pb collisions data by the WA98 and NA49 collaborations at the CERN SPS.
No clear experimental evidence for DCC formation has been reported so far.
This is a rather disappointing result in view of the fact that the phenomenon seems 
quite natural from a theoretical point of view. In particular, the growth of 
long-wavelength fluctuations due to a sudden cooling is a generic phenomenon, which, 
for instance, is commonly observed in ferromagnets in condensed matter physics. 
Numerical simulations of the non-equilibrium chiral phase transition in heavy-ion 
collisions suggest that this phenomenon might actually occur in existing or planned 
experiments at RHIC and LHC.

The absence of experimental evidence may indicate that the appropriate conditions 
for DCC formation have not been met yet. For instance, it has been suggested that
non-central collisions of large nuclei could be a more interesting place than central
ones to look for DCC signatures. Another interesting suggestion is to look
at very forward rapidities of high energy hadronic or nuclear collisions. 
But it should also be stressed that the present experimental situation is 
actually consistent with present theoretical expectations: The estimated upper 
bound for the probability of DCC formation in heavy-ion collisions at SPS energies
is at the edge of the current experimental limit. Moreover, present 
theoretical calculations suggest that the phenomenon we are seeking may actually be more 
difficult to observe than originally thought. In particular, numerical simulations
of the chiral phase transition indicate that the DCC might not have the ideal 
isospin structure originally proposed. This should be taken into account in future 
theoretical investigations as well as experimental searches at RHIC and LHC. 

Besides providing a useful tool for the study of the chiral structure of 
QCD, the detection of a DCC is of fundamental interest on its own. Indeed, 
the possible formation of a coherent, classical pion field is to be expected 
on very general grounds as a direct consequence of the bosonic nature of pions. 
In analogy with the -- nowadays commonly observed -- phenomenon of Bose-Einstein 
condensation for non-relativistic systems, the DCC can be thought of as a macroscopic 
wave of QCD matter. The possibility that the latter might indeed be produced in 
existing experiments is very exciting. It is certainly worth have a look!

\section*{Acknowledgments}

We would like to thank J.~Alam, T.~Awes, J.D.~Bjorken and A.~Krzywicki for 
reading the manuscript and for their useful sugestions.
J.S. would like to thank A.~Krzywicki for introducing him to this subject as 
well as J.D.~Bjorken, J.-P.~Blaizot, K.~Rajagopal and C.~Volpe for interesting 
discussions.
B.M. would like to thank B.K.~Nandi, P.~Steinberg,
G.C.~Mishra, T.K.~Nayak, D.P. Mahapatra, B.~Wyslouch,
M.M.~Aggarwal and Y.P. Viyogi for fruitful
collaboration on this topic.
J.S. would like to dedicate this article to D. Vautherin.

\end{document}